\documentclass[aps,prb,reprint,superscriptaddress,amsmath]{revtex4-1}
\usepackage{graphicx}
\usepackage{enumerate}
\usepackage{refcount}
\usepackage{fmtcount}
\usepackage{amssymb}
\usepackage{dsfont}
\usepackage{hyperref}
\usepackage[usenames,dvipsnames]{color}
\usepackage[normalem]{ulem}
\usepackage{pdfpages}

\makeatletter
\AtBeginDocument{\let\LS@rot\@undefined}
\makeatother

\hypersetup{pdfstartview=FitH,pdfpagemode=UseOutlines,colorlinks,
citecolor=blue,linkcolor=blue,urlcolor=blue}

\newcommand{\s}{\sigma}

\newcommand{\dg}{\dagger}
\newcommand{\sd}{\downarrow}
\newcommand{\su}{\uparrow}

\definecolor{blue(munsell)}{rgb}{0.0, 0.5, 0.69}

\definecolor{aaron}{rgb}{0.6, 0.6, 0.8}

\begin{document}


\title{Chiral spin liquid phase of the triangular lattice Hubbard model: a density matrix renormalization group study}

\author{Aaron Szasz}
	\email[]{aszasz@berkeley.edu}
	\affiliation{Department of Physics, University of California, Berkeley, California 94720, USA}
	\affiliation{Materials Sciences Division, Lawrence Berkeley National Laboratory, Berkeley, California 94720, USA}
	\affiliation{Perimeter Institute for Theoretical Physics, Waterloo, Ontario N2L 2Y5, Canada}
\author{Johannes Motruk}
	\affiliation{Department of Physics, University of California, Berkeley, California 94720, USA}
	\affiliation{Materials Sciences Division, Lawrence Berkeley National Laboratory, Berkeley, California 94720, USA}
\author{Michael P. Zaletel}
	\affiliation{Department of Physics, University of California, Berkeley, California 94720, USA}
	\affiliation{Materials Sciences Division, Lawrence Berkeley National Laboratory, Berkeley, California 94720, USA}
	\affiliation{Department of Physics, Princeton University, Princeton, New Jersey 08540, USA}
\author{Joel E. Moore}
	\affiliation{Department of Physics, University of California, Berkeley, California 94720, USA}
	\affiliation{Materials Sciences Division, Lawrence Berkeley National Laboratory, Berkeley, California 94720, USA}

\date{\today}

\begin{abstract}
Motivated by experimental studies that have found signatures of a quantum spin liquid phase in organic crystals whose structure is well described by the two-dimensional triangular lattice, we study the Hubbard model on this lattice at half filling using the infinite-system density matrix renormalization group (iDMRG) method. On infinite cylinders with finite circumference, we identify an intermediate phase between observed metallic behavior at low interaction strength and Mott insulating spin-ordered behavior at strong interactions.  Chiral ordering from spontaneous breaking of time-reversal symmetry, a fractionally quantized spin Hall response, and characteristic level statistics in the entanglement spectrum in the intermediate phase provide strong evidence for the existence of a chiral spin liquid in the full two-dimensional limit of the model.
\end{abstract}

\maketitle


\section{Introduction:}

Quantum spin liquids\cite{Balents2010,Savary2017,Zhou2017} have been the subject of considerable interest since the concept was first introduced in 1973 by Anderson, who suggested that geometrical frustration on the triangular lattice could lead to a resonating valence bond ground state of the antiferromagnetic Heisenberg model\cite{Anderson1973}.  Although it is now known that the Heisenberg model on the triangular lattice in fact exhibits a three-sublattice $120^\circ$ order in the ground state\cite{Huse1988,White2007}, antiferromagnetic models on the triangular lattice remain some of the most promising systems to realize a phase in which spins remain disordered even down to zero temperature.  The triangular lattice has seemed particularly promising since the work of Shimizu {\it et al.}, who found that the organic crystal $\kappa$-(BEDT-TTF)$_2$Cu$_2$(CN)$_3$, which is well-described by independent 2D layers with nearly isotropic triangular lattice structure, shows no sign of spin-ordering even down to tens of mK, indicative of a possible spin liquid ground state\cite{Shimizu2003}.   Subsequent studies of this crystal have found that the heat capacity is $T$-linear at low temperature\cite{Yamashita2008}, suggesting the presence of low-lying gapless excitations, but also that the thermal conductivity has no such $T$-linear contribution\cite{Yamashita2008b}, indicating to the contrary that there is a gap in the energy spectrum.  Another triangular lattice material, EtMe$_3$Sb[Pd(dmit)$_2$]$_2$, was until recently believed to show $T$-linear behavior in both the heat capacity and thermal conductivity\cite{Itou2008,Itou2009,Yamashita2010,Itou2010}, but new experiments show that it too may exhibit gapped thermal transport\cite{Ni2019,BourgeoisHope2019,Yamashita2019}.  The true nature of the spin liquid phases in these and other triangular lattice materials\cite{Law2017,Li2019,Zeng2019} such as YbMgGaO$_4$\cite{Li2016,Shen2017} remains unclear.

Substantial theoretical effort has gone into answering this question, primarily in studying the antiferromagnetic Heisenberg model with additional terms, such as second-neighbor interactions and ring exchanges, that frustrate the expected three-sublattice order\cite{Motrunich2005,Sheng2009,Grover2010,Block2011,Mishmash2013,Kaneko2014,
Hu2015,White2015,Gong2017,Saadatmand2017,ZhuZ2018,Sahoo2018, Riedl2019}.  The Heisenberg model and its extensions are derived from a perturbative expansion of a model of itinerant electrons, the Hubbard model\cite{MacDonald1988}; by studying the Hubbard model directly, we can capture additional effects that may be important in actual materials, at the cost of increased computational effort---compared with spin-$1/2$ models, the size of the local Hilbert space is doubled, so the system sizes that can be accessed by full-Hilbert-space numerical methods are only about half as large.

Although there is now a wide variety of theoretical evidence pointing to the existence of a non-magnetic insulating phase of the triangular lattice Hubbard model\cite{Morita2002,Motrunich2005,Kyung2006,Sahebsara2008,Tocchio2009,Yoshioka2009,Yang2010,Antipov2011,Tocchio2013, Laubach2015,Mishmash2015,Misumi2017,Shirakawa2017}, there is still little agreement on the precise nature of the phase.  Some candidates, suggested by results on both the Hubbard and extended Heisenberg models, include a $U(1)$ spin liquid with a spinon Fermi sea\cite{Motrunich2005,Sheng2009,Yang2010,Block2011,Kaneko2014,Mishmash2015}, a nodal spin liquid\cite{Tocchio2013,Mishmash2013}, a gapped chiral spin liquid\cite{Kalmeyer1987,Baskaran1989,Hu2015,Hu2016,Wietek2017}, and a $\mathds{Z}_2$ spin liquid\cite{White2015,Hu2015}.  In this work, we confirm the existence of a nonmagnetic insulating phase of the Hubbard model on the triangular lattice at half filling, provide strong evidence that it is a gapped chiral spin liquid, and comment on possible experimental signatures.

We study the triangular lattice Hubbard model on infinite cylinders with finite circumference using the density matrix renormalization group (DMRG) technique\cite{White1992,White1993,Ostlund1995,Schollwock2011}, a variational method to find the ground state of a Hamiltonian within the matrix product state (MPS) ansatz.  This method has previously been applied to an extended Hubbard model on a triangular lattice two-leg ladder, providing evidence for a $U(1)$ spin liquid phase with a spinon Fermi surface\cite{Mishmash2015}.  For systems larger than the two-leg ladder, to our knowledge there exists only one prior paper\cite{Shirakawa2017} that uses DMRG to study the triangular lattice Hubbard model. The authors of that study used the finite-system DMRG to confirm the existence of a nonmagnetic insulating phase.  In our infinite-system DMRG study, we investigate the nature of the phase by studying the entanglement spectrum and the response to adiabatic spin-flux insertion through the cylinder as accomplished by twisting boundary conditions; we also study the MPS transfer matrix spectra, allowing us to rule out the possibility that we observe a Dirac spin liquid.  We study the model on a variety of cylinders with different circumferences and boundary conditions.  With some cylinder geometries we find a chiral spin liquid phase regardless of how we twist the boundary conditions, while for the others the chiral phase exists for a range of twisted boundary conditions, and in particular for those for which the ground state is closest to obeying the symmetries of the full two-dimensional lattice.  While caution is required when extrapolating from cylinders to the 2D limit, taken together, the results for the various cylinders point to the existence of the chiral spin liquid phase in the full two-dimensional lattice as well.

The organization of the paper is as follows: in section \ref{sec:model}, we introduce the model we study and the mixed-space representation used in the simulations. In section \ref{sec:phase_diagram}, we demonstrate the existence of metallic, nonmagnetic insulating, and magnetically ordered phases of the model, and furthermore show that the intermediate phase breaks time reversal symmetry.  We present detailed results for five different cylinder geometries.  Readers wishing to see even more complete data are encouraged to read the Supplemental Material; those interested primarily in the identification of the chiral spin liquid phase can proceed to section \ref{sec:CSL}, in which we show that the intermediate phase is in fact a chiral spin liquid.  Finally, in section \ref{sec:discussion}, we discuss the results, placing them in the context of recent experiments and other theoretical studies.


\section{The Model:\label{sec:model}}
\begin{figure}
\includegraphics[width = 0.48\textwidth]{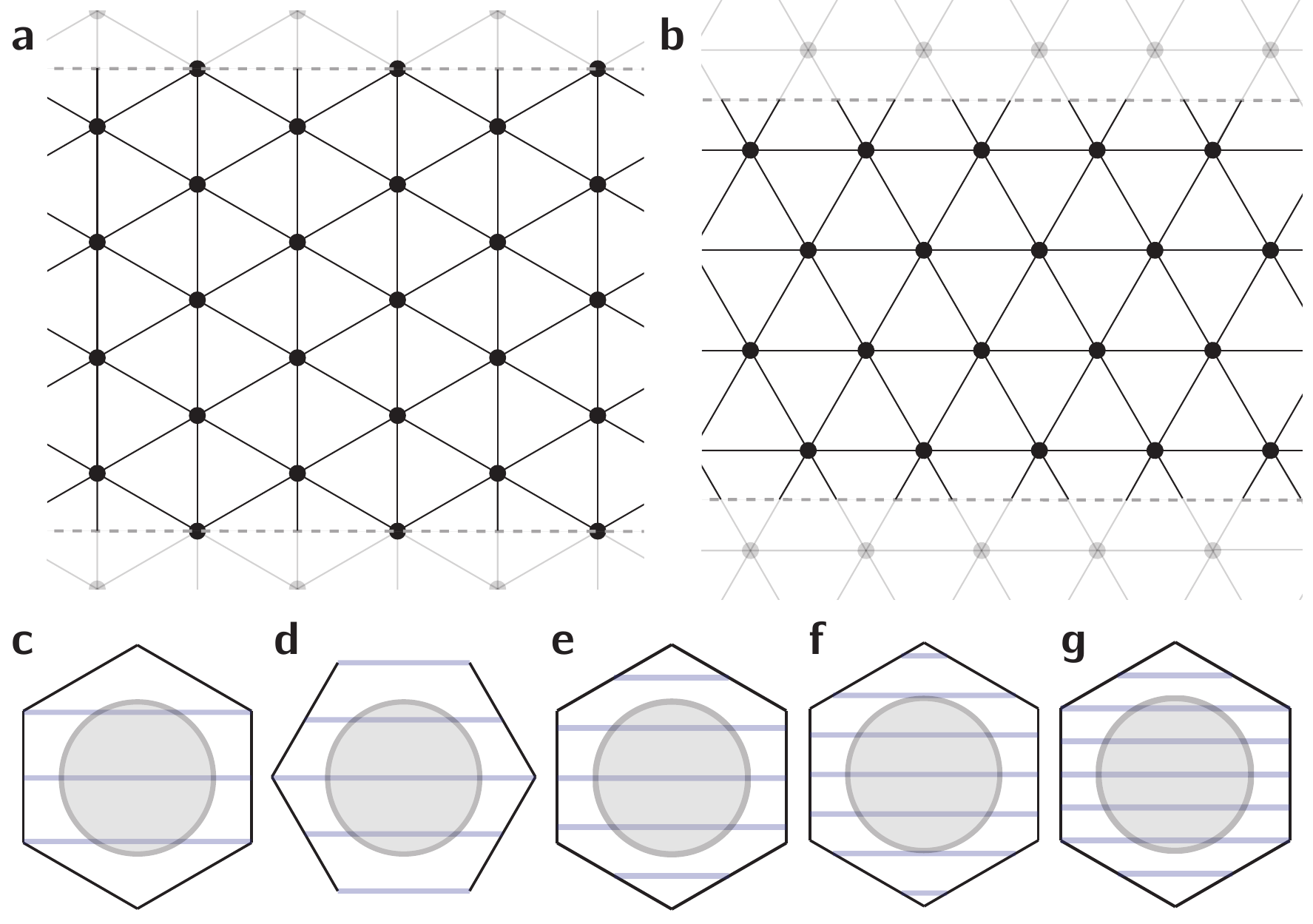}
\caption{{\bf (a)} Triangular lattice on a cylinder of circumference 4 with YC boundary conditions (YC4 cylinder); the dashed lines are identified together and run along the length of the cylinder. {\bf (b)} XC4 cylinder.  {\bf (c)}-{\bf (g)} Horizontal lines show allowed momenta in the Brillouin zone for the YC3, XC4, YC4, YC5, and YC6 cylinders, in order of increasing circumference. The shaded circle shows the Fermi surface for noninteracting electrons ($U=0$).  \label{fig:lattices}}
\end{figure}

The model we study is the standard Hubbard Hamiltonian,
\begin{equation}
H = -t\sum_{\langle i j\rangle\s} c_{i\s}^\dg c^{ }_{j\s} + \text{H.c.} + U\sum_i n^{ }_{i\su}n^{ }_{i\sd},\label{eq:Hubbard_H}
\end{equation}
where $c^{ }_{i\s}$ ($c_{i\s}^\dg$) is the fermion annihilation (creation) operator for spin $\s$ on site $i$  and $n=c^\dg c$ is the number operator; $\langle\cdot\rangle$ indicates nearest neighbor pairs on the triangular lattice (Figure \ref{fig:lattices}).  We work at half filling with net zero spin, so that $\sum_i\langle n^{ }_{i\su}\rangle = \sum_i\langle n^{ }_{i\sd}\rangle = N/2$, where $N$ is the total number of sites.  This model has a single tunable parameter, $U/t$.  In the limit $U=0$, the model is exactly solvable and at half filling forms a metal with a nearly circular Fermi surface; in the limit $U\rightarrow\infty$, double occupancy is disallowed,
so to lowest order in perturbation theory in $t/U$, the model reduces to the nearest-neighbor antiferromagnetic Heisenberg model\cite{MacDonald1988}, whose ground state exhibits a three-sublattice spin order\cite{Huse1988,White2007}.  Between these two limits of $U = 0$ and $U\rightarrow\infty$ there must be at least one phase transition, from the metallic to the Mott-insulating phase; it is in the vicinity of this metal-insulator transition that a spin liquid phase is likely to be found.  

To study this model using the DMRG method, we wrap the two-dimensional triangular lattice onto an infinitely long cylinder of finite circumference.  We primarily use the so-called YC boundary conditions\cite{Yan2011,SuppMat}, for which the triangles are oriented such that one of the sides runs along the circumference of the cylinder.  The YC4 lattice is shown in Figure \ref{fig:lattices}(a) as an example, with the dashed gray lines identified together with periodic boundaries to form a cylinder.  We also consider XC boundary conditions, for which one triangle side runs along the length of the cylinder.  We show the XC4 lattice in Figure \ref{fig:lattices}(b); an XC$n$ cylinder, which exists only for even $n$, has a physical circumference of $n\sqrt{3}/2$ lattice constants.

Denoting translation by one lattice constant around the cylinder by $T_y$, the YC$n$ cylinder has a discrete translation symmetry $T_y^n=1$; we explicitly conserve the momentum quantum numbers associated with this symmetry by rewriting the Hamiltonian in a mixed real- and momentum-space basis with single-particle operators $c_{x, k_y, \sigma}$, which both gives substantial improvements in computational efficiency and allows us to separately find the ground state in different momentum sectors.\cite{Motruk2016,Ehlers2017}  Similarly, for the XC$n$ cylinders we define the translation operator $T_y^{XC}$ that translates between two-site unit cells around the circumference, with $\left(T_y^{XC}\right)^{n/2}=1$; we can again exploit momentum conservation, but with only half as many quantum numbers.

In this paper, we particularly focus on the YC4 and YC6 cylinders, and we also present and discuss data for the YC3, XC4, and YC5 cylinders.  For the various cylinders, the finite circumferences and periodic boundary conditions restrict the accessible momenta in the Brillouin zone as shown in Figures \ref{fig:lattices} (c) through (g).


\section{Phase diagram:\label{sec:phase_diagram}} 
Our goal is to show that the Hubbard model on the full two-dimensional triangular lattice has a chiral spin liquid phase; we begin by establishing the phase diagram more generally, showing the existence of the expected metallic, nonmagnetic insulating (NMI), and magnetic phases, and we furthermore show that the NMI phase breaks time reversal symmetry.

Of course, we have access in our simulations not to the full two-dimensional model but rather to a collection of finite circumference cylinders.  To overcome this impediment, we employ three methods: (1) each phase that exists in the two-dimensional model should leave characteristic signatures when restricted to a finite circumference cylinder, and we can look for these signatures; (2) for each cylinder we can twist the boundary conditions to scan the allowed momentum cuts (Figure \ref{fig:lattices}(c)-(g)) through the full two-dimensional Brillouin zone; and (3) we can compare the results for the various cylinders and look for trends and commonalities.  The third is self-explanatory; before presenting the data, we elaborate on (1) and (2).

We first discuss how the various possible phases of the two-dimensional model should manifest on the infinite cylinders we study.  A metallic state will be gapless, as indicated by a nonzero value for the central charge $c$ of the one-dimensional conformal field theory corresponding to the restriction of the two-dimensional model to the one-dimensional allowed momentum cuts; in particular, if the Fermi surface intersects $N_F$ of the allowed momentum lines in the Brillouin zone (see Figure \ref{fig:lattices} (c) through (g)), the central charge will be $c=2N_F$\cite{CFT,SuppMat}.   We expect the 120-degree magnetically ordered phase to be fully gapped ($c=0$) and symmetric on even circumference cylinders due to the integer-spin Haldane gap\cite{Affleck1989} induced by the reduced dimension, but gapless on odd circumference cylinders; the 2D spin-order should qualitatively manifest as large peaks in the spin-structure factor at the $K^\pm$ points which diverge linearly with cylinder circumference.  If the intermediate phase is a $U(1)$ spin liquid with a spinon Fermi surface, there will be a charge gap but no spin gap, leading to cylinder central charge $c = 2 N_F - 1$ and $2 k_F$-singularities in the structure factors\cite{Sheng2008,Sheng2009,Mishmash2015,Geraedts2016}.  Finally, a gapped spin liquid will have $c=0$ and feature several ``topologically-degenerate'' low-lying states whose energy splitting decreases exponentially with circumference\cite{Wen1990}, along with other topological signatures we will return to in detail. The chiral spin liquid in particular will spontaneously break time-reversal and parity symmetry, while retaining all others; time-reversal symmetry breaking is indicated by a nonzero scalar chiral order parameter $\langle {\bf S}_i \cdot ({\bf S}_j \times {\bf S}_k)\rangle$, where $i$, $j$, and $k$ label the vertices of a triangle in the lattice\cite{Wen1989}. In the simulations, all these properties must be assessed as a function of the DMRG accuracy as captured by the bond-dimension $\chi$ of the MPS ansatz.

We next discuss how, for a given cylinder geometry, twisting of boundary conditions grants access to the full two-dimensional Brillouin zone.  In particular, instead of using periodic boundaries $c_{x,y=0,\s} = c_{x,y=L,\s}$, we set $c_{x,y=0,\s} = e^{i\theta\s/2}c_{x,y=L,\s}$, followed by the gauge transformation $c_{x,y,\s} \mapsto e^{i\theta \s y/(2 L)} c_{x,y,\s}$, where $\s$ in the exponent is $+1$ for spin up and $-1$ for spin down. Physically, this is equivalent to inserting a flux through the cylinder of $\theta/2$ for spin up electrons and $-\theta/2$ for spin down; this corresponds to flux $\theta$ for the spin degrees of freedom.  Note that because the flux insertion is opposite for spin up and spin down, this transformation does not break time reversal symmetry.  

\begin{figure}
\includegraphics[width=0.48\textwidth]{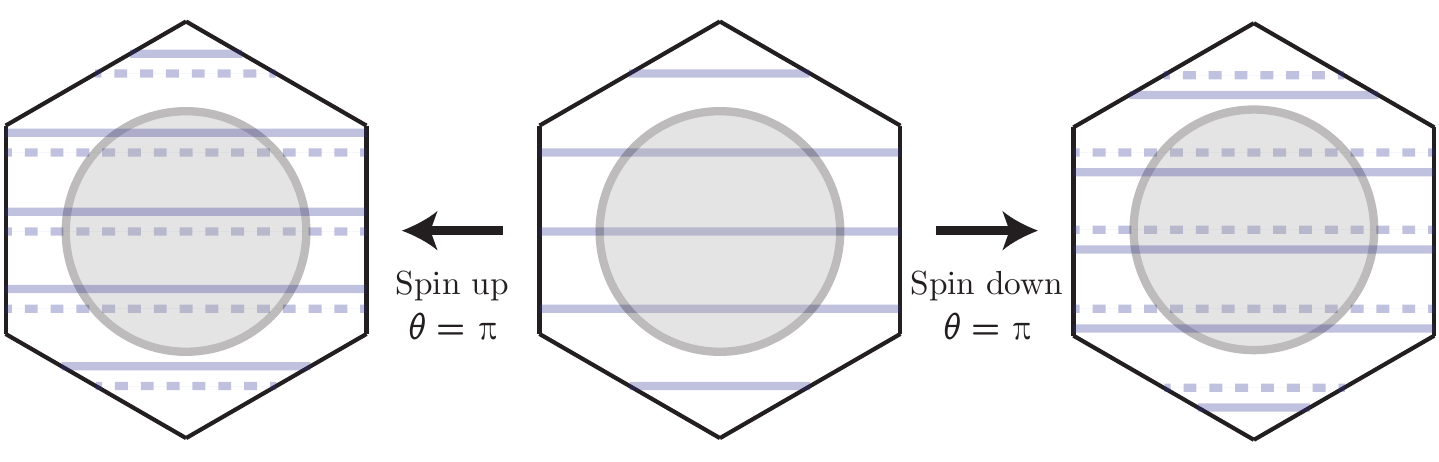}
\caption{The effect of flux insertion on the mixed-space model is to shift the allowed momentum cuts through the Brillouin zone.  They shift upwards for spin up electrons and downwards for spin down electrons, thus preserving time-reversal symmetry.  Note that for $\theta = 4\pi n$ for any integer $n$, the cuts are again in their original positions.
\label{fig:flux_insertion_cuts}}
\end{figure}

When the original Hamiltonian with periodic boundaries is written in the mixed-space picture, some coefficients will depend on the momentum $k$ around the cylinder; the only effect of the flux insertion is to transform those coefficients, with $k = (2\pi/L)n\mapsto (2\pi/L)(n+\theta\s/2)$.  This can be viewed as shifting the momentum cuts in the Brillouin zone, upwards for spin up and downwards for spin down, as illustrated in Figure \ref{fig:flux_insertion_cuts}.  Thus, by scanning $\theta$ from 0 to $4\pi$, we can access the full two-dimensional Brillouin zone, giving substantial additional evidence for the two-dimensional model despite using only a single cylinder geometry.

The only physical effect of this flux insertion is from the twisted boundary conditions, and in the two-dimensional limit where the cylinder circumference becomes infinitely large, the effect on local properties like order parameters and short-range correlations functions must go to zero.  Thus the variation in these quantities with flux insertion serves as an indication of the degree of ``two-dimensionality'' of the cylinders we study and thus of the reliability of our results in predicting the behavior of the full two-dimensional model.

Note that the flux insertion can be performed adiabatically by first computing the ground state with periodic boundary conditions and then increasing $\theta$ in small increments, at each step using the converged ground state from the previous step as the initial state for the new simulation.  Notably, this procedure allows for detection of spin pumping from a quantized spin Hall effect, which is a hallmark of the chiral spin liquid phase.  

We now present results for the various cylinder geometries we have studied.

\subsection{YC4} Out of the five different cylinders we consider, our most extensive data is for YC4, which strikes a balance between two-dimensionality (favoring larger cylinders) and ability to converge the DMRG simulations (favoring smaller ones).  

\begin{figure}
\includegraphics[width=0.48\textwidth]{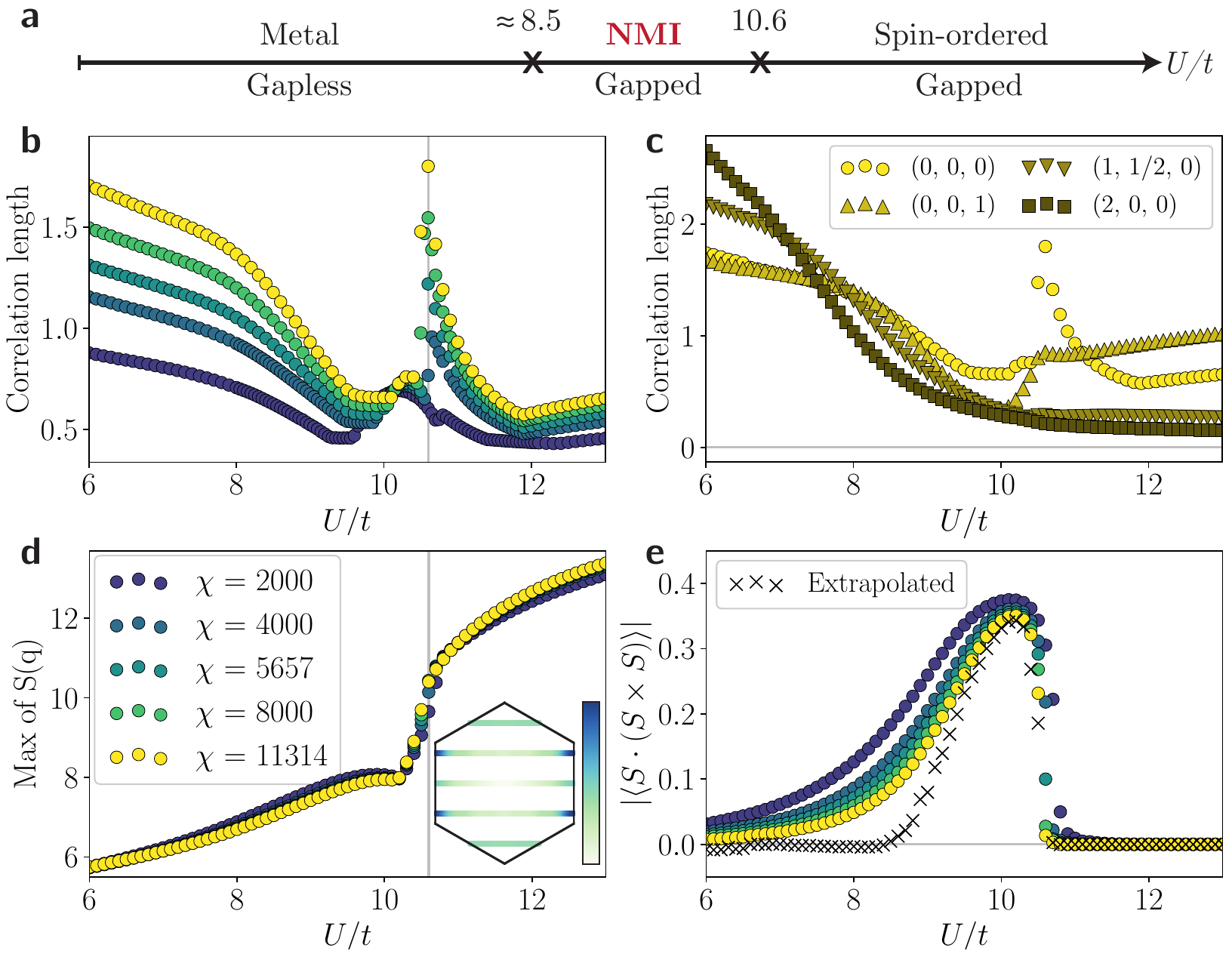}
\caption{(Color online) Results for the YC4 cylinder. Results are shown for a range of MPS bond dimensions $\chi$ as indicated in the lower left legend. {\bf (a)} A nonmagnetic insulating (NMI) phase appears between a gapless metallic phase at low $U/t$ and a magnetic phase at high $U/t$.  {\bf (b)} Correlation length in the ``charge neutral sector,'' in other words for excitations carrying no charge, spin, or momentum. The vertical line at $U/t = 10.6$ is provided as a guide to the eye.  {\bf (c)} Correlations lengths at the largest bond dimension in various charge sectors.  The sector $(Q,S,K)$ corresponds to correlations $\left\langle \mathcal{O}_1\mathcal{O}_2\right\rangle$ where $\mathcal{O}_1$ creates and $\mathcal{O}_2$ annihilates an excitation carrying charge $Q$, spin $S$, and momentum quantum number $K$.  {\bf (d)} Spin structure factor: the curve shows the maximum value of the spin structure factor in the Brillouin zone.  The inset shows the spin structure factor in the high-$U$ phase, with peaks at the closest allowed momenta to the $K^\pm$ points, where they would be expected for $120^\circ$ magnetic ordering.  Note that spin expectation values are reported here and throughout the paper with $\hbar/2 = 1$.  {\bf (e)} Chiral order parameter $\langle \mathbf{S}_i\cdot (\mathbf{S}_j\times \mathbf{S}_k)\rangle$, where $i$, $j$, and $k$ label the three vertices of a triangle in the lattice, with an additional line showing extrapolation in the DMRG truncation error\cite{Hubig2018}; see the Supplemental Material \cite{SuppMat} for details on the extrapolation. 
\label{fig:YC4}}
\end{figure}

On the YC4 cylinder with periodic boundaries we find three phases, corresponding to the expected metallic, nonmagnetic insulating (NMI), and spin-ordered phases of the full two-dimensional model; the phase diagram is summarized in Figure \ref{fig:YC4}, along with our results for several physical quantities: the correlation length, spin structure factor, and scalar chiral order parameter.

The transition from the NMI phase to the spin-ordered phase at $U/t\approx 10.6$ is indicated by a peak in the correlation length, the appearance of large peaks near the $K^\pm$ points of the Brillouin zone in the spin structure factor, and the vanishing of the chiral order parameter.  The spin structure factor in particular allows us to identify the high-$U$ side of this transition as the one-dimensional descendant of the two-dimensional magnetically ordered phase, and the intermediate-$U$ side as nonmagnetic.

\begin{figure}
\includegraphics[width=0.48\textwidth]{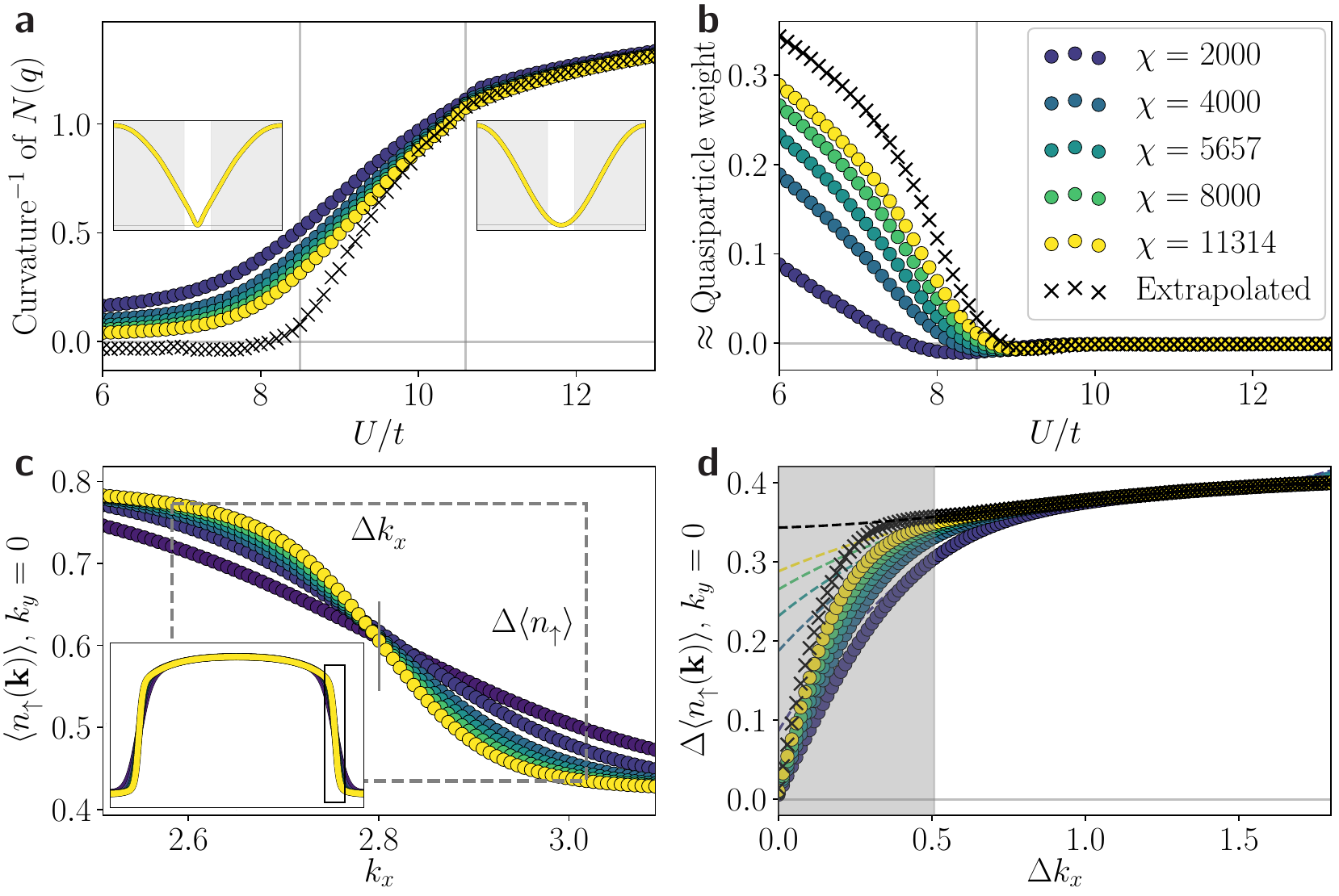}
\caption{(Color online) Measurements pertaining to the metal-insulator transition for the YC4 cylinder. 
{\bf (a)} Inverse curvature of the density-density structure factor at $k=0$, a proxy for the charge gap, with vertical lines at $U/t=8.5$ and 10.6 indicating the estimated locations of phase transitions.  Insets show that the structure factor is linear around $k=0$ for small $U/t$ and quadratic at large $U/t$, indicating gapless and gapped phases respectively; curvature is measured using a small interval around $k=0$, the non-shaded region in the insets.  {\bf (b)} Approximate fermionic quasiparticle weight, measured by discontinuity in occupation at the Fermi surface.  The vertical line is at $U/t = 8.5$.  As described in the text, this is an indirect calculation rather than a direct measurement; the calculation is illustrated in the next two panels for one value of $U/t$, with further examples available in the Supplemental Material\cite{SuppMat}. {\bf (c)} Occupation of spin up electrons as a function of $k_x$, with $k_y=0$, for $U/t = 6$, both for the full Brillouin zone (inset) and zoomed in on the edge of the Fermi surface, identified as the point with steepest slope.  The gap in occupation $\Delta \langle n\rangle$ can be measured as a function of the size $\Delta k_x$ of an interval symmetric around the Fermi surface.  The result is shown in the next panel.  {\bf (d)} Change in occupation $\Delta \langle n\rangle$ across an interval $\Delta k_x$ around the Fermi surface. Small $\Delta k$ results are not reliable because of Luttinger liquid effects from the one-dimensionality of the cylinder and because occupations are calculated by a finite range of real-space correlations.  We therefore eliminate the points in the shaded region and fit the remainder with a cubic polynomial, then extrapolate to $\Delta k_x=0$ as shown.  These extrapolated results are the data shown in panel (b).   
\label{fig:YC4_MIT}}
\end{figure}

Because the metal is gapless, the metal to NMI transition ($U/t \approx 8$) is less clear from the direct physical measurements shown in Figure \ref{fig:YC4}, though it is visible from the chiral order parameter: although a nonzero value of the order parameter indicates time-reversal symmetry breaking in both the metallic and NMI phases for finite bond dimension, an extrapolation in the DMRG truncation error\cite{Hubig2018} shows that the symmetry is actually unbroken in the low-$U$ phase\cite{SuppMat}.  To make this transition clearer, and to show that it corresponds specifically to the opening of a charge gap, we consider two additional quantities that we measure indirectly from the wavefunction via more involved calculations: the small-$k$ curvature of the density-density structure factor and the fermion quasiparticle weight; these are shown in Figure \ref{fig:YC4_MIT}.

The density-density structure factor $N(\mathbf{k})$ shows the presence or absence of a charge gap: if the system is gapless, for small momentum $N(\mathbf{k})\propto |k|$, while if it is gapped $N(\mathbf{k})\propto k^2$.\cite{Capello2005,DeFranco2018}  Indeed we find that at low $U$ the structure factor is visibly linear near $k=0$, shown in Figure \ref{fig:YC4_MIT}(a), left inset, while at high $U$ it is clearly quadratic (right inset).  To observe the metal-insulator transition, we measure the curvature of the structure factor at $k=0$ and then extrapolate in the bond dimension.  This reveals a transition from linear to quadratic at $U/t \approx 8.5$, consistent with the extrapolated chiral order parameter.

In Figure \ref{fig:YC4_MIT}(b), we show an estimate of the fermion quasiparticle weight, measured via the change in electron occupation at the Fermi surface.  This calculation requires accounting for three different effects: (1) high resolution in momentum-space occupation requires very long range correlation function calculations in real space; (2) because a cylinder is a one-dimensional system, in a metallic phase it will behave as a Luttinger liquid and not actually have a discontinuity at the Fermi surface, though there is still a singularity; and (3) finite bond dimension removes the singularity at the Fermi surface.  The latter is easily dealt with by extrapolation; in the Supplemental Material\cite{SuppMat} we show that the singularity reappears in the infinite bond dimension limit.  The first effect is accounted for by measuring the change in occupation across a finite interval in $k$ symmetric around the Fermi surface.  As long as the interval is sufficiently wide, the corresponding wavelength is short enough to not be affected by the finite range of the computed real-space correlations.  The behavior above this cutoff in $\Delta k$ can then be extrapolated to $\Delta k=0$.  This procedure is illustrated for $U/t = 6$ in Figures \ref{fig:YC4_MIT}(c) and (d), and for other values of $U/t$ in the Supplemental Material\cite{SuppMat}.

The second effect cannot be accounted for rigorously using just the YC4 cylinder, since there is no well-defined quasiparticle weight in a quasi-one-dimensional system; in particular, all $\Delta k \lesssim 2\pi/L=\pi/2$ will be affected by the one-dimensionality of the cylinder, and this includes approximately all $\Delta k$ around the Fermi surface out to the edge of the Brillouin zone.  However, as can be seen in Figure \ref{fig:YC4_MIT}(d), the change in occupation does not show a qualitative change between $\Delta k = \pi/2$ and the cutoff $\Delta k \gtrsim 0.5$ used in our extrapolation, so the procedure described above and shown in Figures \ref{fig:YC4_MIT}(c) and (d) should still give an approximately correct estimate of the quasiparticle weight.  

The resulting estimate of the quasiparticle weight appears to vanish around $U/t=9$, though there is some dependence on the precise form of extrapolation used.  Since in computing the estimate we use only larger $\Delta k$ values, where there is a larger occupation gap, the actual quasiparticle weight should vanish at a slightly lower value of $U/t$, again being consistent with the chiral order parameter and structure factor curvature.

\begin{figure}
\includegraphics[width=0.48\textwidth]{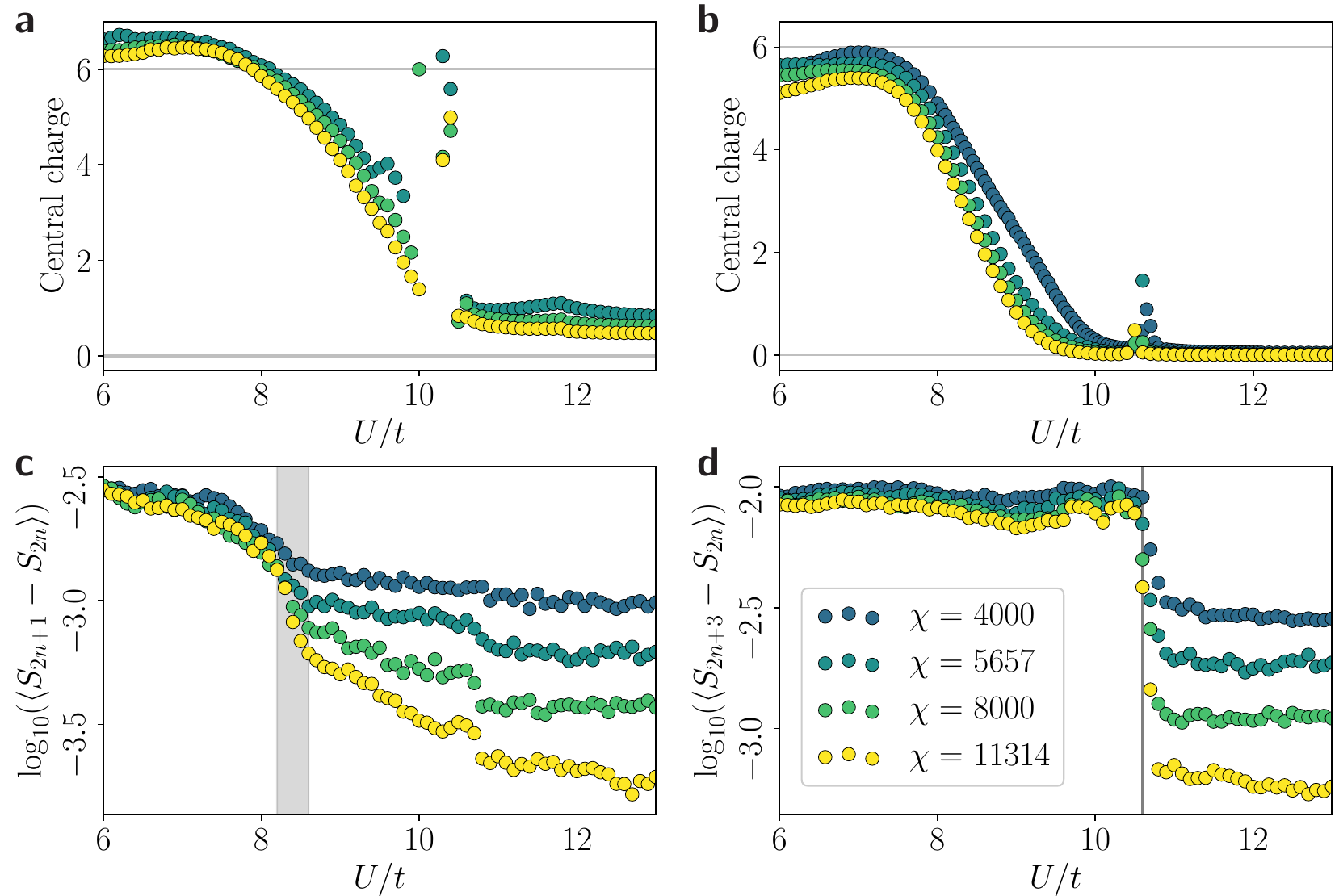}
\caption{(Color online) Entanglement results for the YC4 cylinder.  
{\bf (a)} Central charge of the effective one-dimensional state as calculated by the scaling of entanglement with correlation length, equation \eqref{eq:fin_ent_scaling}; this is the most accurate method for gapless systems. {\bf (b)} Central charge as calculated by the scaling of entanglement with bond dimension, equation \eqref{eq:fin_ent_scaling_v2}; this is the most accurate method for gapped systems. {\bf (c)} Average separation between pairs of levels from the entanglement spectrum; the shaded region is $U/t=8.2$ to $8.6$, where the spectrum approaches exact two-fold degeneracy with increasing bond dimension. {\bf (d)} Average separation among groups of four levels from the spectrum; the vertical line at $U/t=10.6$ marks the onset of four-fold degeneracy.
\label{fig:YC4_entanglement}}
\end{figure}

Further information about the three phases and the transitions between them comes from studying the entanglement in the system, and in particular from finite entanglement scaling\cite{Tagliacozzo2008,Pollmann2009,Pirvu2012}. If we cut the infinite cylinder into two semi-infinite halves, we can calculate the entanglement entropy $S$ between them from the eigenvalues $\lambda_i^2$ of the reduced density matrix of either side of the cut,
\begin{equation}
S \equiv -\sum_i \lambda_i^2\log(\lambda_i^2).\label{eq:S_def} 
\end{equation}
In the true ground state this is an infinite sum; however, when running DMRG simulations the MPS bond dimension $\chi$ upper-bounds the number of non-zero $\lambda_i$ in equation \eqref{eq:S_def} and thereby bounds $S \leq \log(\chi)$.
In a gapless state the true $S$ is infinite, as is the correlation length $\xi$, but finite entanglement scaling predicts that the two quantities will scale with $\chi$ such that \cite{Calabrese2004}
\begin{equation}
S \approx (c/6)\log(\xi),\label{eq:fin_ent_scaling}
\end{equation}
which can be used to estimate the central charge $c$ of the conformal field theory corresponding to the gapless metallic phase. We show the central charge computed using equation \eqref{eq:fin_ent_scaling} in Figure \ref{fig:YC4_entanglement}(a).

In a gapped state $S$ is finite \cite{Hastings2007, Arad2013}, so the DMRG estimate of $S$ should converge as $\chi$ is increased; however, $\xi$ will also converge, and the two quantities may converge at different rates so that the relative scaling between them becomes less reliable.  In such a case, the central charge can be more accurately computed by direct scaling of entanglement with bond dimension,\cite{Tagliacozzo2008,Pollmann2009,Pirvu2012}
\begin{equation}
S\approx \left(1 + \sqrt{12/c}\right)^{-1}\log\left(\chi\right).\label{eq:fin_ent_scaling_v2}
\end{equation}
We show the central charge computed using equation \eqref{eq:fin_ent_scaling_v2} in Figure \ref{fig:YC4_entanglement}(b).

Until $U/t\approx 8$, the central charge is constant with respect to $U/t$ and is near to the value $c=6$ that we would expect for a metallic state\cite{CFT,SuppMat}.  For $U/t \gtrsim 9$, it is clear from Figure \ref{fig:YC4_entanglement}(b) that $c=0$, indicating that the phases are gapped.  For intermediate values of $8 \lesssim U/t \lesssim 9$, the central charge is still far from converged with bond dimension, but it is plausible that it will extrapolate to zero; see the SM for details\cite{SuppMat}.  Note that the apparently unsystematic behavior in Figure \ref{fig:YC4_entanglement}(a) near the previously identified transition at $U/t\approx 10.6$ is due to a slight shift in the location of the peak in the correlation length with bond dimension.

\begin{figure}
\includegraphics[width=0.48\textwidth]{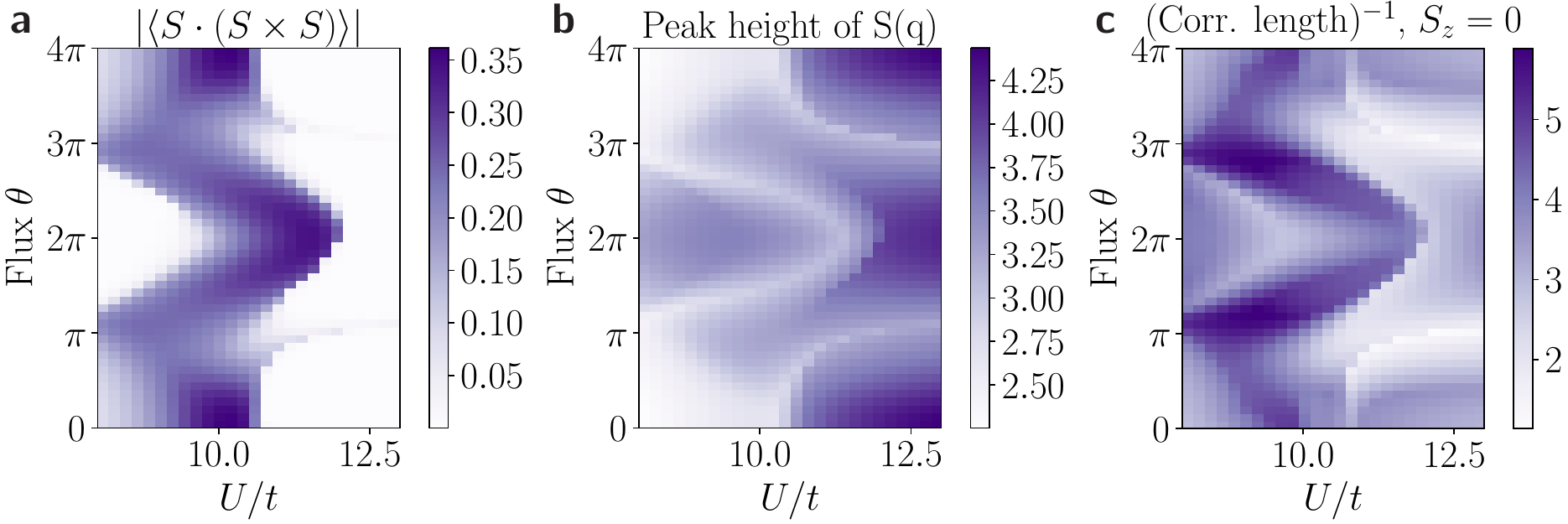}
\caption{(Color online) YC4 cylinder with flux insertion $\theta$, for $\chi = 4000$. {\bf (a)} Absolute value of chiral order parameter.  The chiral phase exists for all twisted boundary conditions, but the phase boundaries shift with $\theta$. {\bf (b)} Maximum of $\langle S_z S_z\rangle$ structure factor on allowed momentum cuts in the Brillouin zone. {\bf (c)} Inverse correlation length for excitations with $S_z=0$, which serves as a proxy for the spin-singlet gap.  
\label{fig:YC4_flux_insertion_data}}
\end{figure}

We can identify the locations of both phase transitions with more precision by studying the entanglement spectrum, which is the list of values $\{-\log(\lambda_i)\}$, for the same $\{\lambda_i\}$ appearing in equation \eqref{eq:S_def}; the full low-lying entanglement spectrum is shown as a function of $U/t$ in the Supplemental Material\cite{SuppMat}.  In particular, by pairing the levels and then finding the average separation between the levels in each pair, we observe that the entire spectrum acquires an exact two-fold degeneracy for $U/t \gtrsim 8.4$; this is shown in Figure \ref{fig:YC4_entanglement}(c).  Although at finite bond dimension there remains for any $U/t$ a finite separation between the levels in each pair, for $U/t < 8.2$ the separation does not visibly depend on bond dimension, whereas for $U/t \geq 8.6$, the separation goes to 0 with increasing bond dimension; the exact location of the transition is not clear from our data, but appears to be in the range of $8.2 \lesssim U/t \lesssim 8.6$, which is consistent with the location of the metal-insulator transition as found above.  We can similarly group the entanglement spectrum into sets of four levels and consider the average separation of the highest and lowest levels in each group, revealing the onset of four-fold degeneracy at $U/t\approx 10.6$, at the spin-ordering transition.  This four-fold degeneracy corresponds to the different projective representations of the symmetry group carried by the entanglement spectrum\cite{Pollmann2010}.

Taken together, the data in Figures \ref{fig:YC4}, \ref{fig:YC4_MIT}, and \ref{fig:YC4_entanglement} demonstrate that the YC4 cylinder with periodic boundary conditions exhibits three distinct phases, corresponding to metallic, time-reversal symmetry-breaking nonmagnetic insulating, and magnetically ordered phases in two dimensions.  The nature of the transitions is a more challenging question.  The metal-insulator transition appears to be continuous. No quantity we measure, including correlation length, spin order, chiral order, estimate of the quasiparticle weight, and the entanglement spectrum, shows any kind of discontinuity; even in a weakly first-order transition, we would expect to see some such signature in our data.  However, despite the very large bond dimensions we use, we are unable to pinpoint the location of the transition, so further properties of the metal-insulator transition such as critical exponents are not feasible to calculate.  The magnetic ordering transition shows much more abrupt changes in various quantities, especially the chiral order parameter, though none are clearly discontinuous.  There may be a very small discontinuity in the entanglement spectrum (Figure S23 in the Supplemental Material\cite{SuppMat}), which could indicate a weakly first order transition, but much higher bond dimension would be required to make a definitive statement.  To summarize, the metal-insulator transition appears to be continuous, while the spin-ordering transition is either continuous or very weakly first order.

We now turn to the results of flux insertion.  We perform the flux insertion adiabatically, twisting the boundary conditions in intervals of $\theta = \pi/12$.  Due to the much larger parameter space spanned by both $U/t$ and $\theta$, we restrict our computations to a single bond dimension, $\chi=4000$.  Based on the data shown in Figure \ref{fig:YC4}, we believe this bond dimension is sufficient to capture the qualitative behavior of the system; furthermore, we increase the bond dimension to 8000 for several values of $U/t$ and confirm that there is no qualitative change.

In Figure \ref{fig:YC4_flux_insertion_data} we show several quantities computed as a function of both $U/t$ and $\theta$, namely the chiral order parameter, the maximum value of the $\langle S_z S_z\rangle$ structure factor on the allowed momentum cuts, and the inverse of the correlation length for operators carrying quantum numbers $(Q,S_z)=(0,0)$ as computed from the MPS transfer matrix spectrum.  In the infinite bond-dimension limit, the latter quantity is a proxy for the gap to excitations with $S_z=0$;\cite{Zauner2015,SuppMat} we present data for only a single bond dimension, but nevertheless a comparison of this inverse correlation length across parameter space can indicate which phases are likely to have a spin-singlet gap.  All three quantities would be independent of $\theta$ in the limit of a very wide cylinder; here we see substantial variation, but at each $\theta$ the qualitative behavior as $U/t$ is varied remains essentially the same.  


Most notably, the chiral order parameter is nonzero in a region of roughly constant width; furthermore, if for each $\theta$ we find the maximum value of the chiral order parameter versus $U/t$, these maxima vary with $\theta$ by only about 1/3 of the maximum at $\theta=0$.  The comparison between the three figures also reveals behavior for all $\theta$ that is in good agreement with what we found with periodic boundary conditions.  In particular, the degree of short-range magnetic ordering rapidly increases at the right edge of the chiral phase, and furthermore the chiral phase appears to be strongly gapped, consistent with the analysis of central charge.

\subsection{YC6\label{sec:YC6}}

We next present data for the YC6 cylinder, which is the largest, and thus presumably the least impacted by finite-size effects, of those we study; this has the potential drawback that the MPS bond dimension required to achieve a given level of precision scales exponentially with $L$, so the simulations are less converged than for smaller cylinders, but we find that the qualitative behavior of the system is nevertheless clear.  

\begin{figure}
\includegraphics[width=0.48\textwidth]{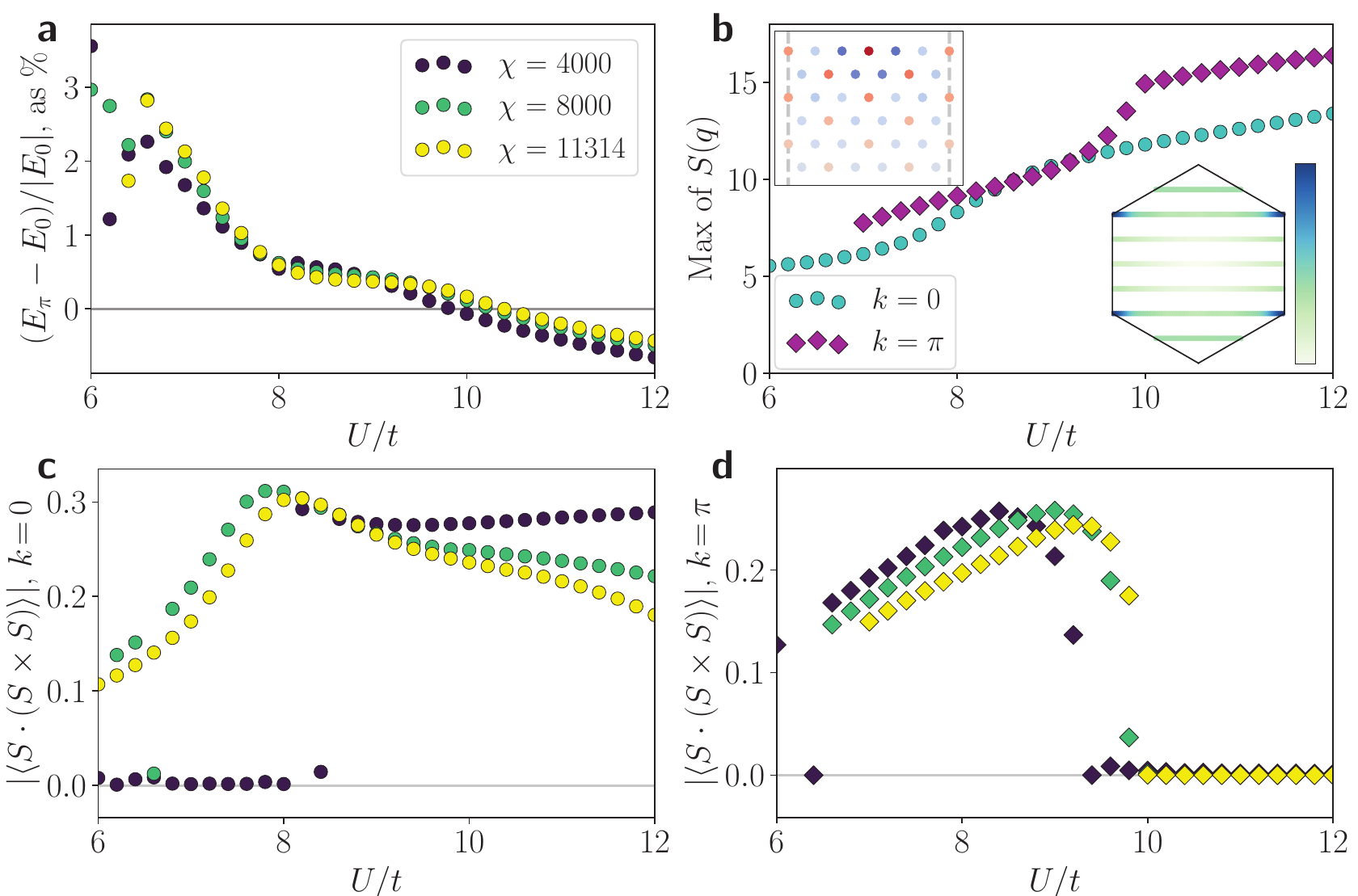}
\caption{(Color online) Results for the YC6 cylinder.  {\bf (a)} Relative energy (percent difference) between ground states in the symmetry sectors with $k=\pi$ and $k=0$ around each ring.  {\bf (b)} Maximum value of the spin structure factor for the two momentum sectors.  Insets show (lower right) the high-$U$ spin structure factor for the $k=\pi$ sector, with peaks at the $K^\pm$ points as expected for $120^\circ$ magnetic ordering, and (upper left) the corresponding real-space $\langle \mathbf{S}\cdot \mathbf{S}\rangle$ correlations to a chosen point (center on the top). {\bf(c)} Chiral order parameter for the $k=0$ ground state.  {\bf(d)} Chiral order parameter for the $k=\pi$ ground state.\label{fig:YC6}}
\end{figure}

The YC6 cylinder is notable not just because it is the widest of those we study but also because, as we show now, it has topologically degenerate ground states in two different momentum sectors.  Because we employ a mixed real- and momentum-space basis, we can initialize the DMRG with states in different sectors of momentum around the cylinder per unit length\cite{Zaletel2017}, $k$, and thus separately find the ground state in each sector. On the YC4 cylinder, the ground state always lies in the $k=0$ sector, but for the YC6 cylinder we observe low-lying states in two different momentum sectors, $k=0$ and $k=\pi$.  The relative energy difference between the ground states in the two sectors is shown in Figure \ref{fig:YC6}(a).  There are three apparent regimes of behavior: at low $U$, the $k=0$ sector is clearly the ground state; at intermediate $U$, the two sectors become close in energy, and the difference is decreasing with bond dimension; at high $U$, the $k=\pi$ sector becomes the ground state, though again the relative difference in energy decreases with bond dimension.

\begin{figure}
\includegraphics[width=0.48\textwidth]{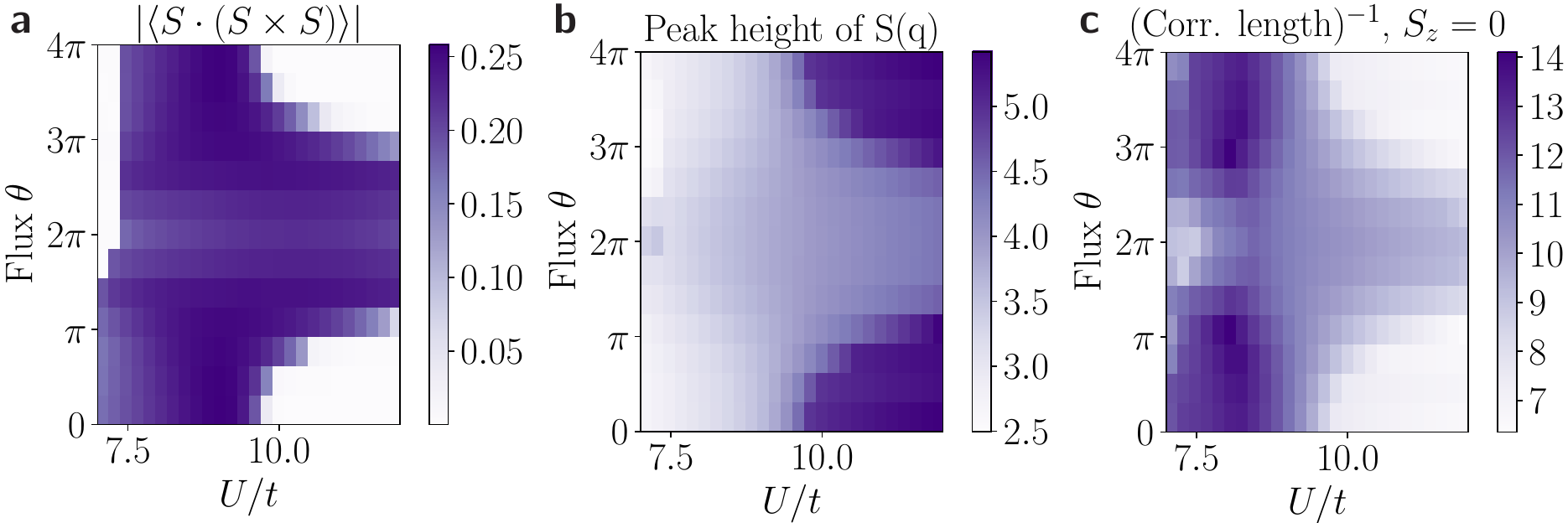}
\caption{(Color online) YC6 cylinder with flux insertion $\theta$, for $\chi = 8000$. {\bf (a)} Absolute value of chiral order parameter.  {\bf (b)} Maximum of $\langle S_z S_z\rangle$ structure factor on allowed momentum cuts in the Brillouin zone. {\bf (c)} Inverse correlation length for excitations with $S_z=0$.  
\label{fig:YC6_flux_insertion_data}}
\end{figure}

The low-$U$ phase is expected to be metallic, with central charge $c=10$.\cite{CFT, SuppMat} Finite entanglement scaling indeed suggests that the phase is gapless\cite{SuppMat}, though an accurate measurement of the central charge would require a bond dimension currently inaccessible to us, on the order of 50,000.  (Extremely high entanglement in the low-$U$ region leads to very large DMRG truncation error, on the order of $10^{-4}$, even with $\chi\sim10,000$.)  The high-$U$ phase should be the one-dimensional descendant of the two-dimensional $120^\circ$ N\'{e}el ordered phase, and indeed at approximately the same value of $U/t$ where the $k=\pi$ sector becomes the ground state, there is a rapid increase in peak height of the spin structure factor in the $k=\pi$ sector, as shown in Figure \ref{fig:YC6}(b).  In this phase, we observe the expected peaks in the structure factor at the corners of the Brillouin zone (lower right inset) and short range spin-ordering in the real-space spin-spin correlations (upper left inset).

The intermediate phase, for $U/t\approx 8$ to $U/t\approx 10$, is the region where the relative energy difference between the two momentum sectors is small and approximately constant; the spin structure factors are also approximately equal. We identify the transition to the right by the onset of the afore-mentioned spin ordering.  To the left, the transition can be observed by the $k=0$ sector becoming the sole ground state and from the transition in that sector to a metallic phase; as we show in the SM\cite{SuppMat}, the latter can be seen qualitatively from the entanglement spectrum and finite entanglement scaling---the low-$U$ phase appears gapless while the intermediate phase is likely gapped.

As with the YC4 cylinder, spontaneous breaking of time-reversal symmetry leads to a nonzero value of the chiral order parameter in the metallic and intermediate phases, as shown for the two momentum sectors in Figure \ref{fig:YC6} (c) and (d), though in the metal we would expect the symmetry to be restored at larger bond dimensions.  In the $k=\pi$ sector, which is the true ground state for high $U$, the chiral order parameter rapidly vanishes at the spin-ordering transition.  In the $k=0$ sector, the chirality does not seem to drop abruptly to zero; however, as can be seen in Figure \ref{fig:YC6}(c), the chirality does rapidly decrease with increasing bond dimension for $U \gtrsim 10$.

We can again acquire more information about the full two dimensional model by performing adiabatic flux insertion to scan the allowed momentum cuts through the full Brillouin zone; we perform the flux insertion using the $k=\pi$ ground state as the initial state with $\theta =0$, and we perform all computations with $\chi=8000$.  Although the bond dimension is twice that used for YC4 flux insertion, the results are much less converged.  Nevertheless, some qualitative features can be captured at least qualitatively, as shown in Figure \ref{fig:YC6_flux_insertion_data}.  In particular, there is a chiral phase for all $\theta$, which has weak local magnetic order and a sizable spin singlet gap (indicated by the inverse correlation length).  The chiral region extending to higher $U$ around $\theta = 2\pi$ is likely an artifact of the finite bond dimension: all local properties at $2\pi$ flux are essentially identical to those of the $k=0$ ground state with periodic boundaries, and as noted above, the chiral order parameter is far from converged at $\chi=8000$ above $U\gtrsim 10$.

\subsection{YC5}

\begin{figure}
\includegraphics[width=0.48\textwidth]{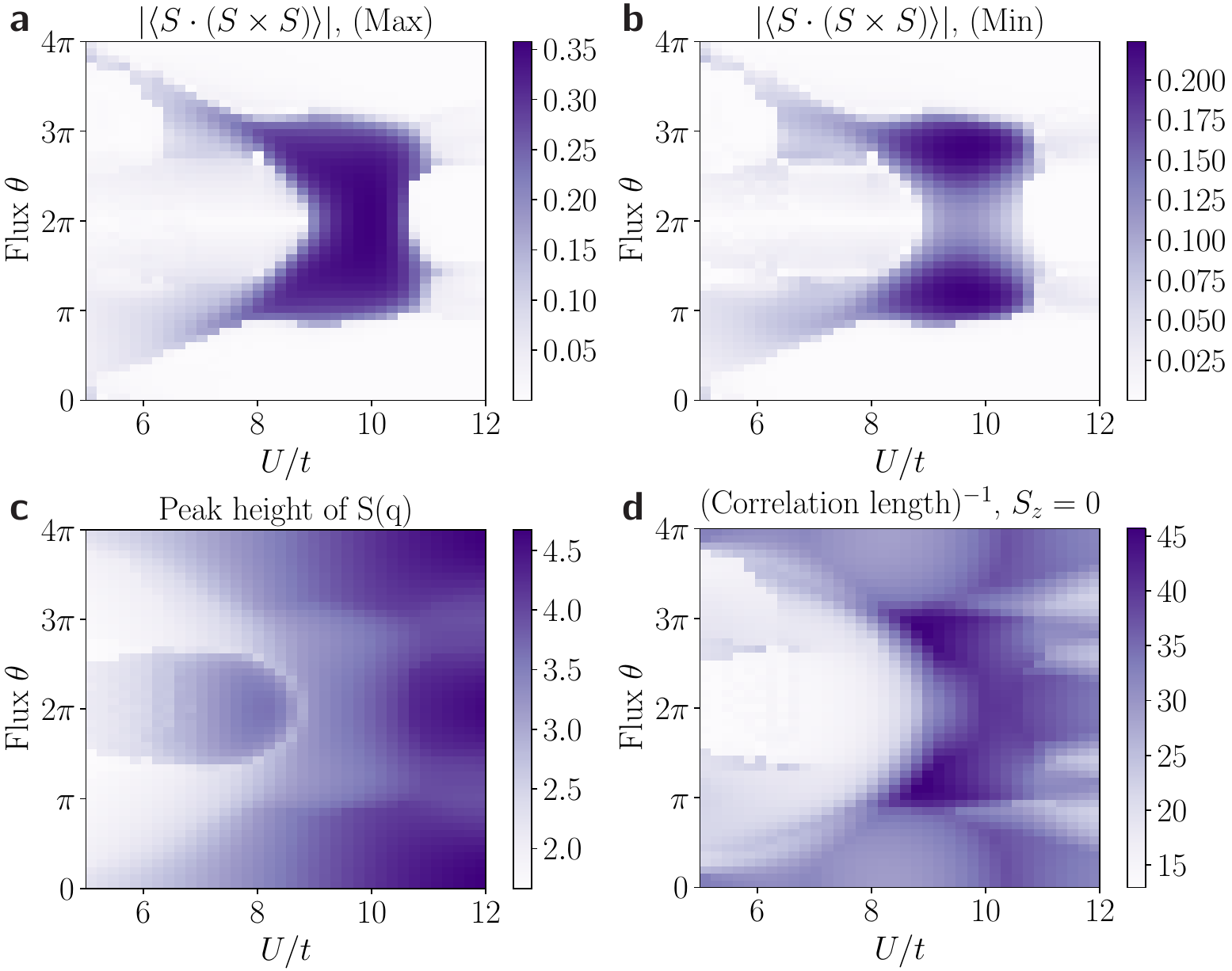}
\caption{(Color online) YC5 cylinder with flux insertion $\theta$, for $\chi = 4000$. {\bf (a)} Absolute value of chiral order parameter; at finite bond dimension the translation symmetry is broken along the cylinder, and here we show the larger of the chiral order parameters between the two distinct rings.  The chiral phase exists at intermediate $U$ when $\theta$ is approximately in the range $\pi$ to $3\pi$. {\bf (b)} Smaller of the two chiral order parameters. {\bf (c)} Maximum of $\langle S_z S_z\rangle$ structure factor in the Brillouin zone. {\bf (d)} Inverse correlation length for excitations with $S_z=0$.  
\label{fig:YC5_flux_insertion_data}}
\end{figure}

The YC4 and YC6 phase diagrams discussed above are qualitatively similar; both show a chiral intermediate phase in the vicinity of $U/t = 10$, which is present regardless of the twisting of the boundary conditions.  The same is not true for the YC5 cylinder---with periodic boundary conditions, $\theta = 0$, there is no spontaneous time-reversal symmetry breaking for any $U$.  However, when we perform flux insertion we find that the chiral intermediate phase does still exist, for $\pi \lesssim \theta \lesssim 3\pi$ and $8 \lesssim U/t \lesssim 10$.  This is shown in Figure \ref{fig:YC5_flux_insertion_data}.  

\begin{figure}
\includegraphics[width=0.48\textwidth]{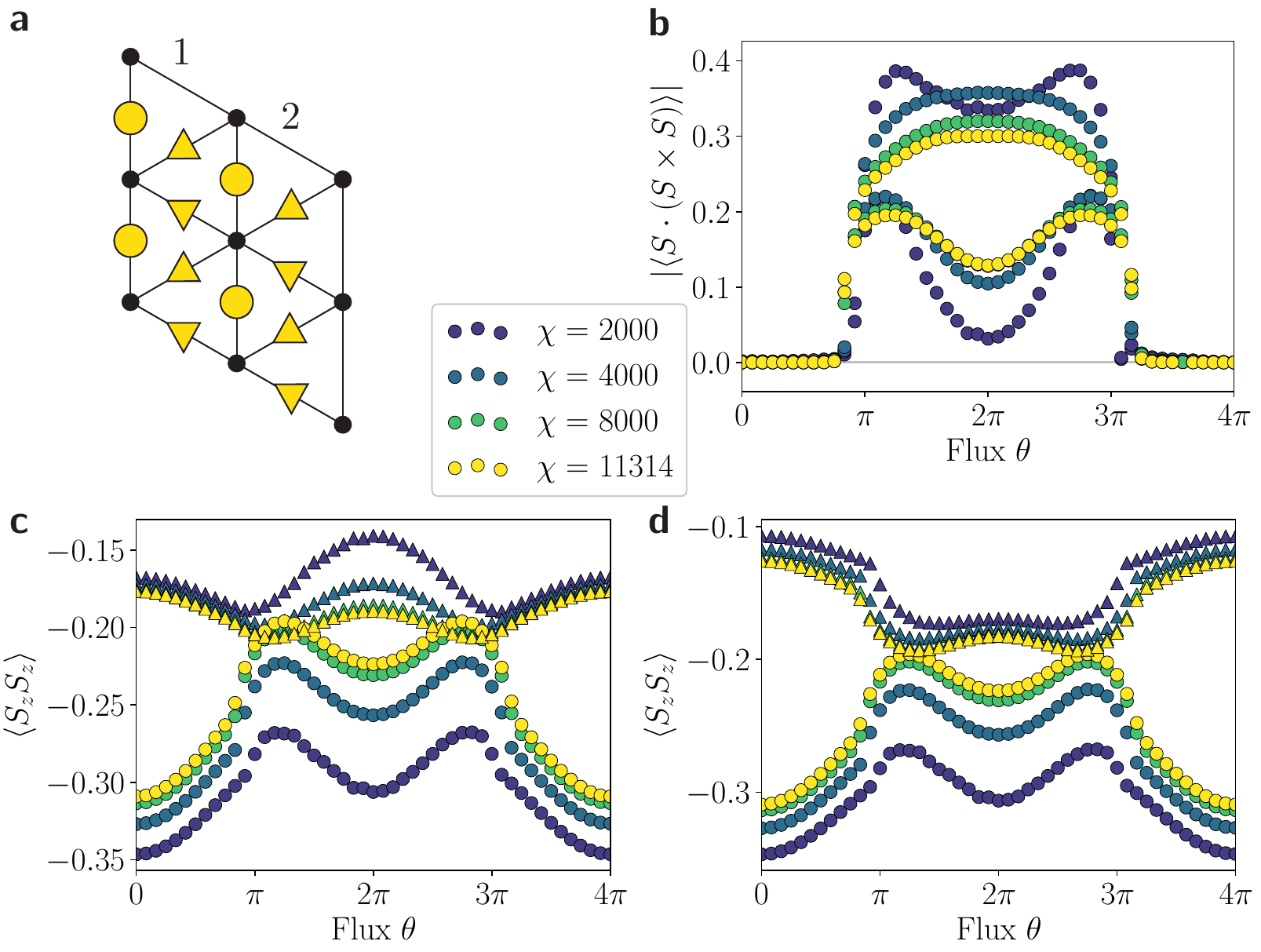}
\caption{(Color online) Flux insertion for YC5 cylinder with $U/t = 10$, for a range of bond dimensions.  {\bf(a)} As noted in the text, a two-ring unit cell allows for translation symmetry-breaking.  Here we label the distinct rings of the cylinder in the unit cell and the three bonds in each ring that may have different $\langle S_z S_z\rangle$ correlations.  {\bf(b)} Chiral order parameter on ring 1 (upper curves) and ring 2 (lower curves) of the unit cell.  It appears that lower curve is converged for the largest $\chi$, while the upper one is not, but it does not appear that the two will become equal even in the infinite $\chi$ limit.  {\bf(c)} $\langle S_z S_z\rangle$ for nearest neighbor bonds on ring 1 as shown in panel (a); the symbol for each data point indicates it corresponds to the bond labeled by that symbol in (a).  We do not show the strength of the down-triangle bond because for each $\chi$ and $\theta$ it is equal to that of the up-triangle bond to better than one part in $10^8$.  In the 2D model, all three bonds are equivalent; on the YC cylinder the vertical bonds are inequivalent to the two diagonal ones. {\bf(d)} $\langle S_z S_z\rangle$ for ring 2.
\label{fig:YC5_U10}}
\end{figure}

To understand this data, it is important to note that, unlike for YC4 and YC6, we have used a two-ring unit cell; this allows us to initialize the DMRG simulation with a product state that is half-filled both for spin up and spin down. The two ring unit cell allows the ground state to break translation symmetry along the cylinder, which indeed occurs; this is expected even in a spin liquid phase\cite{Zaletel2015}. Figure \ref{fig:YC5_flux_insertion_data}(a) and (b) show the chiral order parameter on the two rings of the unit cell.  As shown in Figure \ref{fig:YC5_U10}(b), the degree of symmetry-breaking decreases as the MPS bond dimension used in running DMRG is increased, though it appears that even at infinite bond dimension the symmetry will remain broken.  

The chiral phase observed for YC5 seems to be the same as that found in YC4 and YC6 even if it does not extend through all boundary condition twists $\theta$.  This is partially confirmed by considering the peak height of the $\langle S_zS_z\rangle$ structure factor and the inverse correlation length for excitations with $S_z=0$, shown in Figures \ref{fig:YC5_flux_insertion_data}(c) and (d), respectively.  As with YC4 and YC6, the chiral phase has a degree of short-range spin ordering which is intermediate between that of the metal and of the high-$U$ phase and has the largest spin singlet gap of any region of the phase diagram.  We also show below, in section \ref{sec:CSL}, that this chiral phase shows the same signatures of the topological chiral spin liquid as do the YC4 and YC6 phases.

As evidence for the existence of the chiral phase in the full two-dimensional model, the YC5 results are somewhat ambiguous.  Neither $\theta=0$, for which there is no chiral phase, nor $\theta=2\pi$, for which the chiral phase exists, is a priori ``better'' or more representative of the two-dimensional model.  However, further insight can be gleaned by understanding the effect of the twisted boundaries on the spin degrees of freedom that are the relevant ones for a spin liquid phase.  Indeed, we believe that the $\theta = 2\pi$ boundary conditions turn out to be the more representative ones.

In particular, we can look at the strength of $\langle S_z S_z\rangle$ correlators on bonds between adjacent sites; the results are shown for four bond dimensions up to $\chi = 11314$ for $U/t = 10$ in Figure \ref{fig:YC5_U10}(c) and (d).  Evidently, for flux near $\theta = 0$, there is huge anisotropy, with spin correlations much stronger on bonds around the cylinder circumference than for diagonal ones.  As flux increases from zero, the anisotropy steadily decreases and shows only a change in slope upon entering the chiral phase; the anisotropy is smallest precisely where the chiral order parameter is largest.  Assuming that the true intermediate phase of the two-dimensional model does not break the model's $C_3$ rotation symmetry, the $\theta$ for which the YC5 cylinder exhibits a chiral phase are precisely those in which the symmetry of the spin correlations is most two-dimensional.  We also test this explanation by explicitly adding anisotropy to the model to weaken the bonds around the cylinder circumference; indeed, with the hopping strength on these bonds reduced by $10\%$, a chiral phase appears even at zero flux.\cite{SuppMat}  

\subsection{YC3}

\begin{figure}
\includegraphics[width=0.48\textwidth]{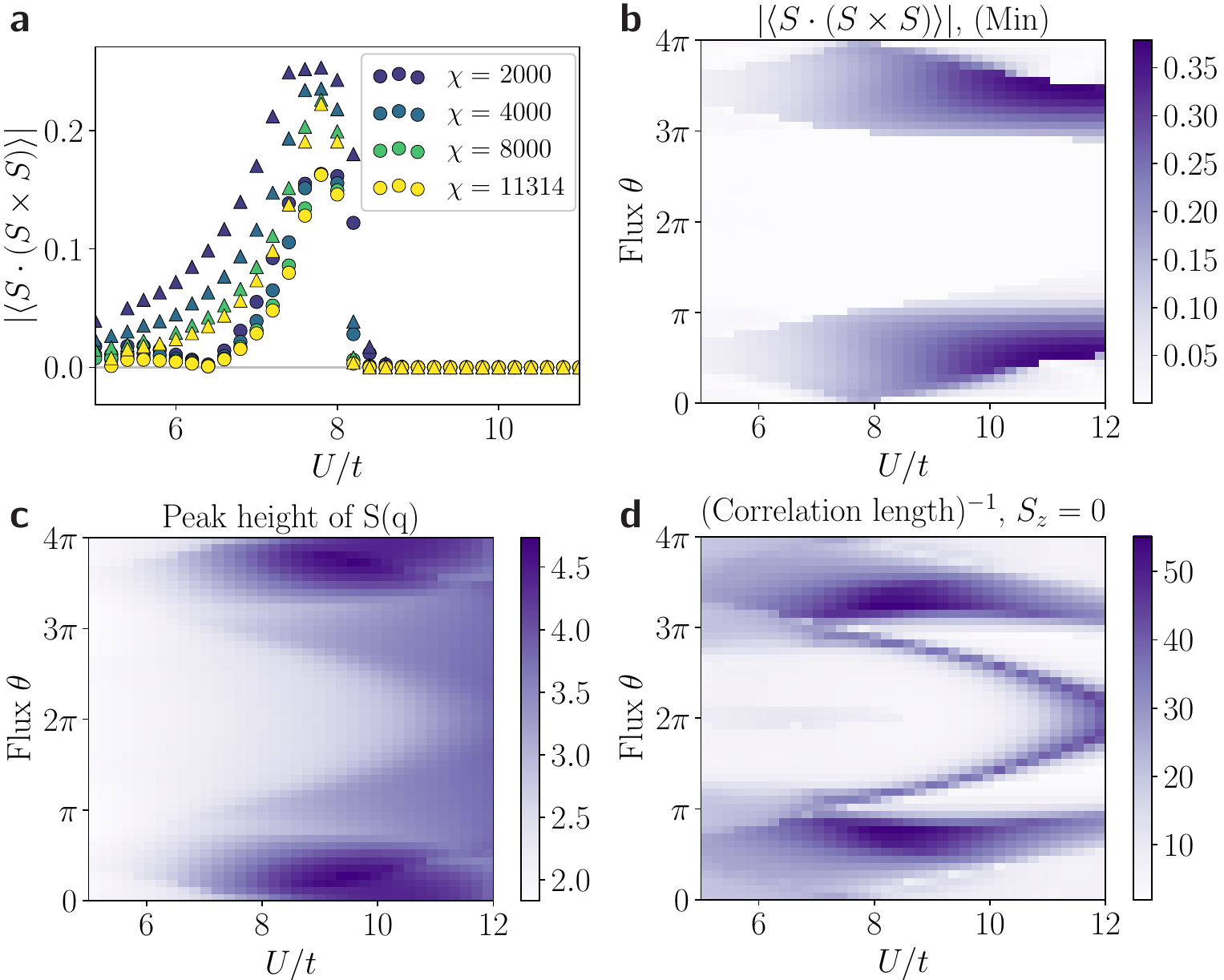}
\caption{(Color online) Results for the YC3 cylinder.  {\bf(a)} Chiral order parameter on each of two ring
s (circle and triangle symbols, respectively), plotted versus $U/t$ for a range of bond dimensions.  The behavior is qualitatively similar to that of YC4 and YC6.  {\bf(b)} Chiral order parameter versus $U/t$ and flux insertion $\theta$, for $\chi = 4000$.  Here we show just the smaller of the chiral order parameters on the two rings, but the qualitative behavior is essentially identical at this bond dimension.  {\bf (c)} Peak height of the $\langle S_z S_z\rangle$ structure factor in the Brillouin zone. {\bf (d)} Inverse correlation length for excitations with $S_z=0$.
\label{fig:YC3}}
\end{figure}

The YC3 cylinder is the smallest, and thus presumably least representative of the two-dimensional model, of all those we have studied; we nevertheless include our data for completeness.  With periodic boundaries, $\theta=0$, we find much the same behavior as for YC4 and YC6, with an intermediate chiral phase between a metallic phase and a short-range magnetically ordered one.  As partial evidence, we show the chiral order parameter versus $U/t$ in Figure \ref{fig:YC3}(a), with additional data available in the SM\cite{SuppMat}.  Note that as with YC5, we use a larger unit cell (in this case four rings) and find that for finite bond dimension the model has a only a two-ring translation symmetry; in the figure, the two curves for each bond dimension correspond to the chiral order parameter on the two distinct rings.

With flux insertion the behavior is quite different, and, as we show in Figure \ref{fig:YC3}(b), the chirality vanishes for $\pi\lesssim \theta \lesssim 3\pi$, essentially the opposite of the behavior observed for YC5.  In Figure \ref{fig:YC3}(c) and (d) we also show the peak height of the spin structure factor and the inverse correlation length for excitations with $S_z=0$.  The relationship between these quantities and the chirality is quite different from what we observe for all three cylinder geometries discussed above, so it is not clear that the chiral phase observed here corresponds to the one found for the larger cylinders.

\subsection{XC4}

\begin{figure}
\includegraphics[width=0.48\textwidth]{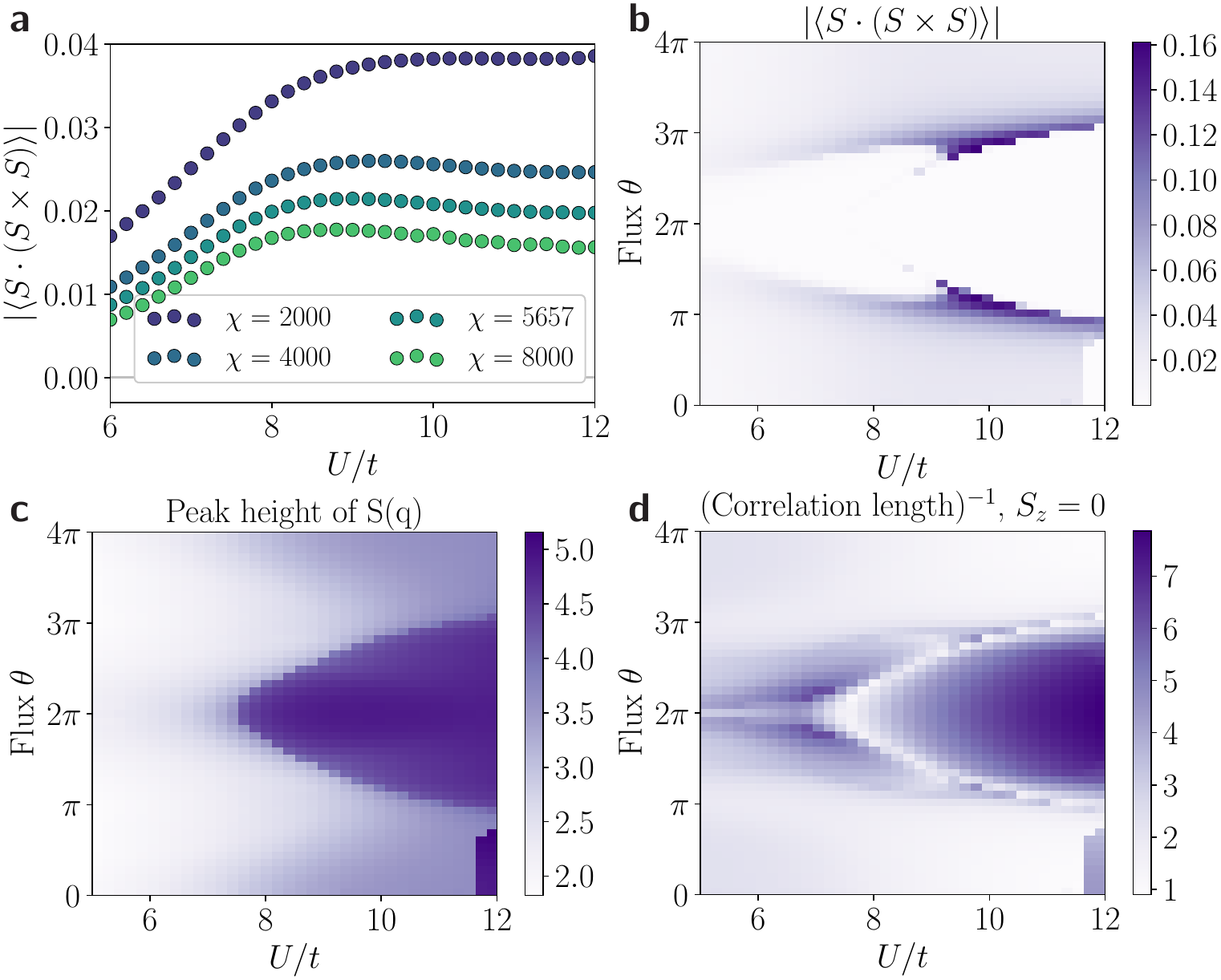}
\caption{(Color online) Results for XC4 cylinder. {\bf (a)} Chiral order parameter versus $U/t$ for a range of bond dimensions, with periodic boundary conditions.  This likely extrapolates to zero. {\bf (b)} Chiral order parameter with flux insertion. {\bf (c)} Maximum of $\langle S_z S_z\rangle$ structure factor on allowed momentum cuts in the Brillouin zone. {\bf (d)} Inverse correlation length for excitations with $S_z=0$.
\label{fig:XC4}}
\end{figure}

Finally, we place the model on the XC4 cylinder, which is the second smallest cylinder after YC3.\footnote{Ideally we would also consider the XC6 cylinder, but we are unable to reach large enough bond dimension to converge the DMRG; at our largest accessible bond dimensions, there remains a strong symmetry-breaking effect from the orientation of the DMRG snake.} With periodic boundaries, we find very weak time reversal symmetry-breaking for all $U/t$, which decreases with bond dimension; this is shown in Figure \ref{fig:XC4}(a).  The extrapolation to infinite bond dimension is not entirely clear, but it is likely that the true ground state preserves the symmetry.

With flux insertion, we find that a chiral phase again appears, as shown in Figure \ref{fig:XC4}(b).  We also show the peak height of the $\langle S_z S_z\rangle$ structure factor and the inverse correlation length for excitations with $S_z=0$, in Figures \ref{fig:XC4}(c) and (d), respectively.  As with YC3, there is no clear relation between the three quantities that we found for YC4-6.  However, like with YC5, the chirality appears near where the nearest neighbor spin-spin correlations are most isotropic.  In the high-$U$, mid-flux region (with large spin singlet gap in Figure \ref{fig:XC4}(d)), the diagonal bonds are much stronger than the horizontal ones, whereas in the rest of the phase diagram the opposite is true; the chirality is strongest precisely on the border between these two regions.  Furthermore, the anisotropy is much larger in the region with exactly zero chirality than in the region where the chirality likely extrapolates to zero but is nonzero at finite bond dimension.


\section{Identification as a chiral spin liquid:\label{sec:CSL}} 
We have demonstrated, for both the YC4 and YC6 cylinders, the existence of an intermediate phase which is nonmagnetic and which breaks time-reversal symmetry; we have also demonstrated that the phase is gapped for YC4 and likely gapped for YC6.  We have furthermore observed this same phase for the YC5 cylinder for a range of twisted boundary conditions, and we have observed some similar behavior for the YC3 and XC4 cylinders.  We now show that the chiral phase observed on the YC4-6 cylinders can in fact be identified as a chiral spin liquid (CSL)\cite{Kalmeyer1987,Wen1989}.

\begin{figure*}[t]
\centering
\includegraphics[width = \textwidth]{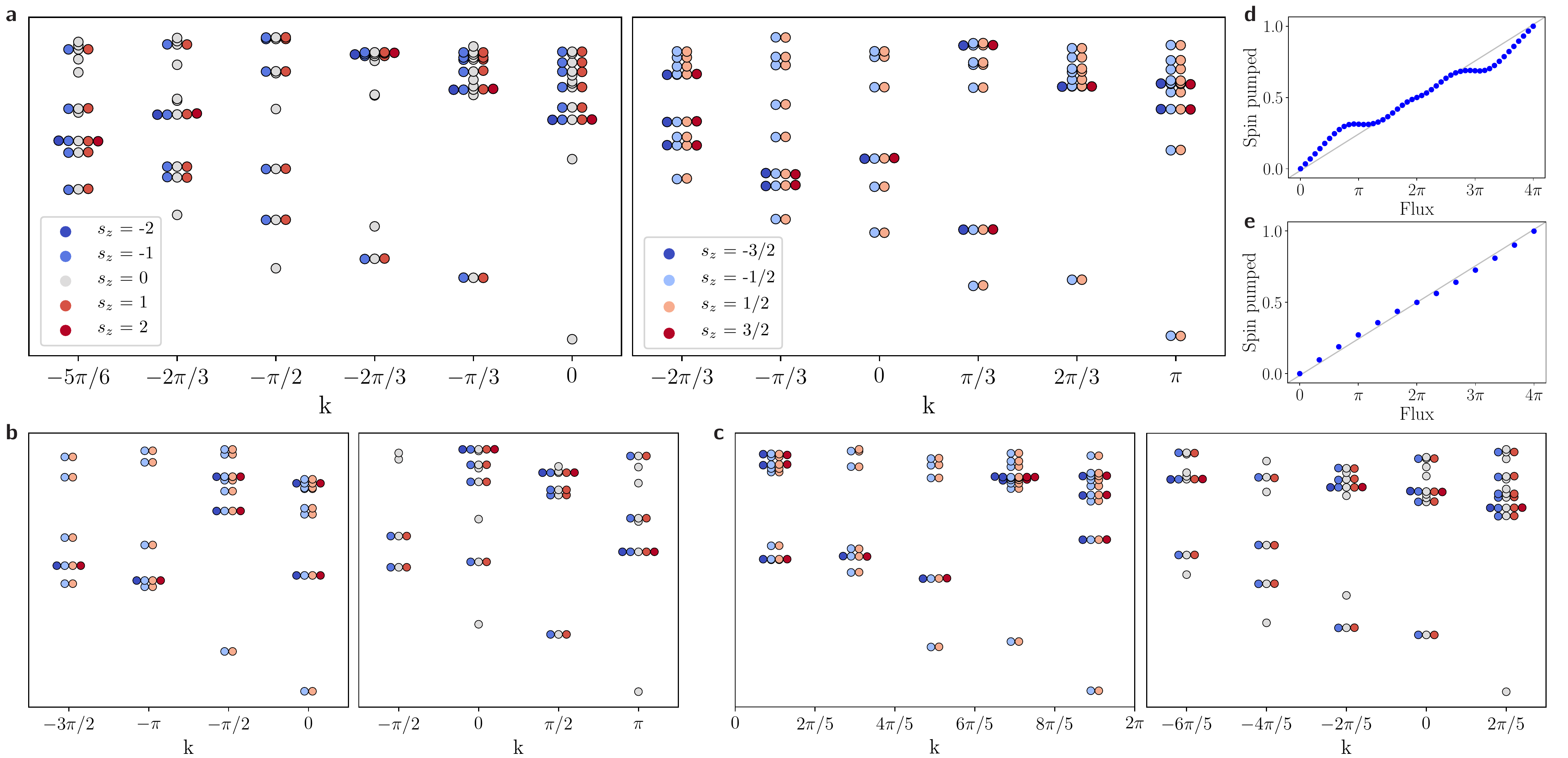}
\caption{(Color online) {\bf (a)} Momentum- and spin-resolved entanglement spectrum for the YC6 cylinder in the intermediate phase, for the ground state in the $k=0$ (left) and $k=\pi$ (right) sectors; these correspond to the trivial and semion sectors of a chiral spin liquid (CSL) respectively.  Insertion of $2\pi$ flux interchanges the two topological sectors, though as discussed in the text there is a subtlety due to working with a fermion model.  {\bf (b)} Momentum- and spin-resolved entanglement spectrum for the YC4 cylinder, with periodic boundaries at $U/t=10.2$ (left) and with $2\pi$ flux inserted at $U/t = 11.6$ (right), corresponding to the highest chirality in each of the two topological sectors.  {\bf (c)} Entanglement spectrum for YC5 with $2\pi$ flux, $U/t=10$, between two-ring unit cells (left) and between the rings in the unit cell (right), again corresponding to the two topological sectors. {\bf (d)} Spin pumping as a function of flux insertion in the intermediate phase for YC4 ($U/t=10$) and {\bf (e)} for YC6 ($U/t=9$).  
\label{fig:CSL}}
\end{figure*}


A CSL is a topological phase with four degenerate ground states on the infinite cylinder\cite{Wen1989a}.  Each minimally entangled ground state\cite{Zhang2012} spontaneously breaks time-reversal ($T$) and parity ($P$) symmetries, as indicated by a nonzero value of the chiral scalar order parameter; the two possible chiralities account for a two-fold degeneracy in the ground state manifold, which could be lifted by a $P, T$-breaking perturbation such as a magnetic field.  

The remaining degeneracy is topological and is robust to such perturbations; the two topologically degenerate sectors, called the trivial and semion sectors, are distinguished by the respective absence or presence of a pair of semionic spinons, fractional excitations that carry spin-$1/2$ but no charge, separated to the ends of the cylinder at $\pm\infty$.\cite{Wen1989a,Oshikawa2006} In a pure spin system, insertion of $2\pi$ flux creates a pair of spinons and separates them to the ends of the cylinder, thus exchanging the two ground states and also pumping a net spin of exactly $1/2$ across any cut through the cylinder; this latter property indicates that the CSL has a spin Chern number of $1/2$ and a corresponding quantized spin Hall conductance\cite{Gong2014,Grushin2015}. 

In contrast, insertion of $2\pi$ spin-flux in the Hubbard model imposes antiperiodic boundary conditions on the cylinder, since $e^{2 \pi i S^z} = -1$.  The Hamiltonian is thus modified by $2\pi$ flux insertion, so that the question of whether the two ground state sectors are exchanged under flux insertion, as they are in a spin-model CSL, is ill-defined; instead, $2\pi$ flux insertion converts between one sector of the original Hamiltonian (with periodic boundaries) and the opposite sector of the Hamiltonian with antiperiodic boundaries, which should still lead to the same quantized spin pumping as for a spin model.

Each ground state of a CSL has a chiral edge mode with a universal low-lying spectrum; when the state is placed on an infinite cylinder, this edge spectrum appears in the entanglement spectrum for a cut between rings of the cylinder.\cite{Kitaev2006,Li2008,Qi2012}  The edge modes are described by a chiral $SU(2)_1$ Wess-Zumino-Witten (WZW) conformal field theory\cite{Wen1991}; labeling them by spin and momentum quantum numbers\cite{Zaletel2013, Cincio2013, SuppMat}, for a given spin the number of levels at successive discrete momenta around the cylinder follows the counting $(1,1,2,3,5,\cdots)$.\cite{Moore1997}  The spectrum is degenerate under $s_z\rightarrow -s_z$, where $s_z$ is the spin quantum number of the entanglement level; the spin quantum numbers are integers in the trivial sector and half-integers in the semion sector, leading to two-fold degeneracy of the spectrum in the latter case.  

We observe all of these signatures of the CSL phase.  On the YC6 cylinder, we have already identified above two nearly degenerate low-lying states, in the $k=0$ and $k=\pi$ momentum sectors; within each sector, by initializing the DMRG with different product states, we are able to converge to both chiralities (see SM\cite{SuppMat}), thus finding all four degenerate ground states.  The chiral order parameter in each sector, indicative of time-reversal and parity symmetry-breaking, has already been shown above in Figures \ref{fig:YC6} (c) and (d) and Figure \ref{fig:YC6_flux_insertion_data}(a); note that these figures show the absolute value of the order parameter, which is independent of the chirality to which the DMRG spontaneously converges.  

The spin- and momentum-resolved entanglement spectra for the ground states in the two sectors are shown in Figure \ref{fig:CSL}(a), where we have excluded levels corresponding to charge fluctuations between rings of the cylinder in order to highlight the spin degrees of freedom.  Both spectra show the expected WZW level counting in the low-lying states for those momenta where the low-lying states can be distinguished from the continuum, and the spin quantum numbers of the entanglement levels are integer for the $k=0$ ground state and half-integer for $k=\pi$, allowing us to identify the low-lying states in the two momentum sectors with the trivial and semion topological sectors respectively.  

Alternatively, $2\pi$ flux insertion should convert between the two topological sectors.  We already noted in section \ref{sec:YC6} above that indeed the local properties like spin-spin correlations and the chiral order parameter look nearly identical between the $k=\pi$ sector with $2\pi$ flux and the $k=0$ sector with periodic boundaries, which is consistent with this picture.  (In principle these should also be equal to the local properties of the $k=\pi$ sector with periodic boundaries, but that may not be true at finite bond dimension, and will be weakly violated even in the true ground state due to the finite circumference of the cylinder.)  In the SM, we show the equivalent of Figure \ref{fig:CSL}(a) with the $k=0$ entanglement spectrum replaced by the $k=\pi$ spectrum with $2\pi$ flux, and evidently it is nearly identical.

To see the equivalent of Figure \ref{fig:CSL}(a) for the YC4 cylinder, because we find only one ground state sector, with $k=0$, we must use the latter method.  In Figure \ref{fig:CSL}(b), we show the spin- and momentum-resolved entanglement spectrum for YC4 in the $k=0$ sector, with periodic boundaries at $U/t=10.2$ and with $\theta = 2\pi$ at $U/t=11.6$; as shown in Figure \ref{fig:YC4_flux_insertion_data}(a), these values of $U/t$ are each at the peak of the chiral order for their respective amounts of flux insertion, $\theta$.  For the YC5 cylinder, as with YC4 we find a ground state only in the $k=0$ sector, although with two rings per unit cell, this includes both $k=0$ and $k=\pi$ per ring.  In this case, however, we cannot observe both topological sectors by looking at $\theta = 0$ and $2\pi$ since the chiral phase exists only for $\pi \lesssim \theta \lesssim 3\pi$.  Instead, we make use of the fact that, for any cylinder with an odd number of spin-1/2 per ring, translation along the cylinder converts between topological sectors\cite{Zaletel2015}, so that we can just consider a single wavefunction and examine its entanglement spectrum both between two-ring unit cells and between the two rings in the unit cell; the result is shown in Figure \ref{fig:CSL}(c). 

With flux insertion, we also observe the quantized spin Hall effect, as shown for the YC4 and YC6 cylinders at $U/t=9$ and 10, respectively, in Figures \ref{fig:CSL} (d) and (e), with a pumping of exactly spin $1/2$ per $2\pi$ flux insertion.  For YC6, for which the chiral order is roughly constant at $U/t=10$, the flux insertion proceeds at a constant rate.  For YC4, the shifting boundary of the chiral phase with flux insertion causes some deviation, but the qualitative behavior is the same.  


\section{Discussion:\label{sec:discussion}} By employing the DMRG method to study the triangular lattice Hubbard model on infinite cylinders in a mixed real- and momentum-space basis, we have observed that the model exhibits three phases: a metallic phase, a nonmagnetic insulating phase, and a magnetically ordered phase.  While the nature of the intermediate phase depends on the precise boundary conditions used, with flux insertion through the cylinder we find that for each cylinder geometry there is a region with spontaneous time-reversal symmetry breaking, as indicated by a nonzero chiral order parameter.  In particular, this chiral intermediate phase exists for all values of flux insertion for the YC4 and YC6 cylinders and for a large range of flux for the YC5 cylinder; the YC5 chiral intermediate phase appears precisely for those amounts of flux insertion for which spin-spin correlations are most consistent with the symmetries of the two-dimensional lattice.

Furthermore, we have shown for the YC4, YC5, and YC6 cylinders that the chiral phase shows the characteristic entanglement spectrum of a CSL with two topologically degenerate ground state sectors, and for YC4 and YC6 we have demonstrated a fractionally quantized spin Hall effect.  The phase additionally appears to be gapped.  Along with the nonzero chiral order parameter, this evidence strongly suggests that the nonmagnetic insulating phase is, in fact, a chiral spin liquid.  This is, to our knowledge, the first clear demonstration of a chiral spin liqid in a time-reversal symmetric model of itinerant fermions.

The apparent gapped nature of the spin liquid in our simulations is consistent with the thermal conductivity measurements on $\kappa$-(BEDT-TTF)$_2$Cu$_2$(CN)$_3$ reported in reference \onlinecite{Yamashita2008b}; some recent experiments\cite{Ni2019,BourgeoisHope2019} also suggest gapped thermal conductivity in EtMe$_3$Sb[Pd(dmit)$_2$]$_2$, although this is disputed\cite{Yamashita2019}.  On the other hand, our conclusions about the nature of the spin liquid do not agree with those of past studies of this model using the DMRG method: the study on the two-leg ladder found a gapless spin liquid phase\cite{Mishmash2015}, while the DMRG study on a finite XC6 cylinder found an intermediate phase that appeared gapped but with a rapidly decaying chiral-chiral correlation function\cite{Shirakawa2017}.  The two-leg ladder study used a modified Hamiltonian with some longer-range interactions, so the disagreement on the nature of the spin liquid is not surprising.  The discrepancy with the XC6 finite cylinder study is more difficult to explain.  One possibility is that, as with the XC4 and YC5 cylinders in our study, the XC6 cylinder will exhibit a chiral phase after flux insertion; we are not able to reach high enough bond dimension to converge the XC6 cylinder, and thus are unfortunately not able to test this possibility.

It is also useful to briefly consider other candidates for the intermediate phase.  In particular, it is worth investigating the possibility of the intermediate phase being a Dirac spin liquid (DSL), both because there has recently been evidence in support of a DSL in frustrated spin models\cite{He2017,Hu2019} and because the CSL can be derived by gapping out the Dirac cones in a DSL, so that one might imagine a DSL in two dimensions becoming a CSL due to finite cylinder circumference or finite bond dimension.  The first scenario is difficult to rule out, given that the CSL would still be the true ground state up to some cylinder circumference which could be much larger than what is accessible, but there is also no particular evidence from our data to support this scenario.  The second scenario we do rule out, by analyzing the low-lying excitation spectrum using the MPS transfer matrix spectra; this analysis is described in detail in the SM.\cite{SuppMat}

If the CSL is indeed the ground state in the full two-dimensional triangular lattice Hubbard model, in real materials well described by this model we would expect regions of both possible chiralities to coexist, with a finite temperature phase transition to long-range chiral order at a temperature of the same order of magnitude as the chiral domain wall tension, possibly reduced due to entropy from the gapless edge modes located at the domain walls.  We measure this domain wall tension for the YC4 cylinder by finding an optimized composite wave function that transitions from the ground state with one chirality to the ground state with the other, and we find a domain wall tension of approximately 0.0065$t$ per lattice constant; using estimates for $t$ for real materials\cite{Shimizu2003}, this is about 4$K\times k_B^{ }$.\cite{SuppMat}  The corresponding phase transition may be related to the observed feature in the heat capacity, thermal conductivity, and magnetic relaxation rate at about 6$K$ in $\kappa$-(BEDT-TTF)$_2$Cu$_2$(CN)$_3$\cite{Shimizu2006,Yamashita2008,Yamashita2008b}.

At very low temperatures in a single-domain sample, we would observe no longitudinal thermal transport, in agreement with experimental data\cite{Yamashita2008b,Ni2019,BourgeoisHope2019}, and a quantized thermal Hall conductance, $K_{xy} =  \frac{\pi^2 {k_B}^2 T}{3 h}$; note that the latter is twice the value of the Majorana-like plateau recently reported in $\alpha$-RuCl$_3$~\cite{kasahara:2018majorana,Yokoi2020}.  In the presence of time-reversal symmetry-breaking disorder, there would be regions of both possible chiralities with gapless edge modes between them. It would be interesting to further investigate the resulting behavior of the specific heat and the thermal conductivity in this scenario.

An applied magnetic field could in principle break the degeneracy between the two chiralities, but this effect is extremely small at experimentally accessible field strengths.  If the magnetic flux through a triangle in the lattice is $\phi$, perturbation theory in $t/U$ gives a term $\left[24(t^3/U^2)\sin(\phi)\left(\mathbf{S}\cdot (\mathbf{S}\times\mathbf{S})\right)/\hbar^3\right]$ in the effective spin Hamiltonian; using our measured value for the chiral order parameter and estimated parameters for $\kappa$-(BEDT-TTF)$_2$Cu$_2$(CN)$_3$\cite{Komatsu1996,Shimizu2003}, in a 10 T field the energy splitting between ground states for the two chiralities is about 1 $\mu$eV per lattice site, so at 1 K the favored chirality would be expected to be just 1\% more prevalent.  So it would not be surprising for experiments to see no significant effect for applied magnetic fields up to 10 T.\cite{Yamashita2008}

In addition to studying the chiral spin liquid phase, we also considered the transitions to the neighboring metallic and spin-ordered phases.  For the YC4 cylinder in particular, for which our data is the most extensive, every quantity we computed, including correlation length, spin order, chiral order, and an estimate for the quasiparticle weight, shows no discontinuity at the metal-insulator transition, suggesting that it is second order. However, we are not aware of any field theory description of a direct metal to chiral spin liquid transition, and furthermore a recent experiment tuning across the metal-insulator transition by doping found the transition to be first-order\cite{Saito2019}, so further study is very much warranted. Some possibilities include a weakly first order transition in the full two-dimensional model or the presence of a very small intermediate phase such as the aforementioned Dirac spin liquid; alternatively a theory of this transition may simply be waiting to be discovered.  Future numerical work focusing specifically on the metal-insulator transition will hopefully be able to resolve the exact nature of the transition, including finding the critical exponents if it is indeed continuous.

More generally, further theoretical work must address the question of whether the chiral phase we find on the cylinders we have studied indeed extrapolates to the full two-dimensional model.  Our results strongly support this conclusion: on the YC4 and YC6 cylinders the chiral phase exists for a large range of $U/t$ independent of the flux insertion that scans the allowed momentum cuts through the full two-dimensional Brillouin zone, and furthermore on the YC5 cylinder the same phase appears when the twisted boundary conditions lead to spin correlations that approximately obey the symmetries of the full two-dimensional lattice.  In other words, the chiral spin liquid is always present in the model as a competing phase, and it seems to be favored in those situations that best represent the two-dimensional system.  The existence of the chiral spin liquid in two dimensions could be further confirmed either by using larger circumferences, which would be computationally expensive, or by fully 2D methods such as projected entangled pair states (PEPS)\cite{Verstraete2004,Corboz2010}.


\begin{acknowledgments}
We would like to thank Bryan Clark, Ryan Mishmash, Roger Mong, Frank Pollmann, Tomonori Shirakawa, Shigetoshi Sota, Senthil Todadri, Hiroshi Ueda, Ruben Verresen, and Seiji Yunoki for helpful conversations.  We have also used the TenPy tensor network library\cite{Kjall2013}, which includes contributions from MPZ, Mong, Pollmann, and others.  Numerical computations were primarily performed using the Lawrencium cluster at Lawrence Berkeley National Laboratory.  This work was funded by the U.S. Department of Energy, Office of Science, Office of Basic Energy Sciences, Materials Sciences and Engineering Division under Contract No.\ DE-AC02-05-CH11231 through the Scientific Discovery through Advanced Computing (SciDAC) program (KC23DAC Topological and Correlated Matter via Tensor Networks and Quantum Monte Carlo) (AS, MZ, JEM) and the Theory Institute for Materials and Energy Spectroscopies (TIMES) (JM). JM acknowledges support through DFG research fellowship MO 3278/1-1.  Research at Perimeter Institute is supported in part by the Government of Canada through the Department of Innovation, Science and Economic Development Canada and by the Province of Ontario through the Ministry of Economic Development, Job Creation and Trade.
\end{acknowledgments}

\bibliography{spin_liquid_triangle_bib}

%

%

\bigskip

\onecolumngrid
\newpage



\section*{Supplementary material}

# Chiral spin liquid phase of the triangular lattice Hubbard model: a density matrix renormalization group study

Aaron Szasz,[1,2,3,*] Johannes Motruk,[1,2] Michael P. Zaletel,[4,1] and Joel E. Moore[1,2]

[1]*Department of Physics, University of California, Berkeley, California 94720, USA*

[2]*Materials Sciences Division, Lawrence Berkeley National Laboratory, Berkeley, California 94720, USA*

[3]*Perimeter Institute for Theoretical Physics, Waterloo, Ontario N2L 2Y5, Canada*

[4]*Department of Physics, Princeton University, Princeton, New Jersey 08540, USA*

(Dated: February 27, 2020)

## I. COMPACTIFICATION TO A CYLINDER

As described in the main text, we use the density matrix renormalization group (DMRG) method to study the Hubbard model on the triangular lattice[1,2]. The DMRG is a method for finding the ground state of a one-dimensional model, so it cannot be used to study the full two-dimensional system directly. Instead, we take 1D strips of the lattice with some finite width. In particular, we identify the two edges of the strip with each other, using periodic boundary conditions; this eliminates edge effects, giving the best approximation to the 2D model that we can achieve with a strip of finite width.

To pick the 1D strip that defines the cylinder, we follow these steps:

1. Pick two points of the lattice, and declare them to be equivalent.

2. The line between the two points is the width of the strip or equivalently the circumference of the cylinder.

3. The line passing through the identified point (ie both points, since they are the same) and perpendicular to the circumference is the glued edge of the cylinder.

It is important to note that choosing any cylinder of this type automatically guarantees periodicity of the Hamiltonian along the cylinder, so we once again have a translation-invariant system. To see this, let the lattice vectors $\mathbf{a}$ and $\mathbf{b}$ be as shown in Figure S1, and $a$ a lattice spacing. Then, noting that $\mathbf{a}^2 = \mathbf{b}^2 = a^2$ and $\mathbf{a} \cdot \mathbf{b} = a^2/2$, if the edges are perpendicularly separated by $n_a \mathbf{a} + n_b \mathbf{b}$ for some integers $n_a$ and $n_b$ (as must be true given the above procedure), then one can check that the vector $(2n_b + n_a)\mathbf{a} - (2n_a + n_b)\mathbf{b}$ is perpendicular, and both coefficients are integers. This is a vector that points along the length of the cylinder, and it is an integer linear combination of the lattice vectors, so the Hamiltonian is invariant under this translation. (In some cases, the actual period may be smaller than this, eg if $n_b = 0$ and $n_a$ is even.)

### A. Allowed cylinders and the consequences of choosing one

We now have a general procedure for generating cylinders to which the 2D triangular lattice Hubbard Hamiltonian can be restricted in a natural way, namely by picking pairs of points on the lattice to identify with each other. If

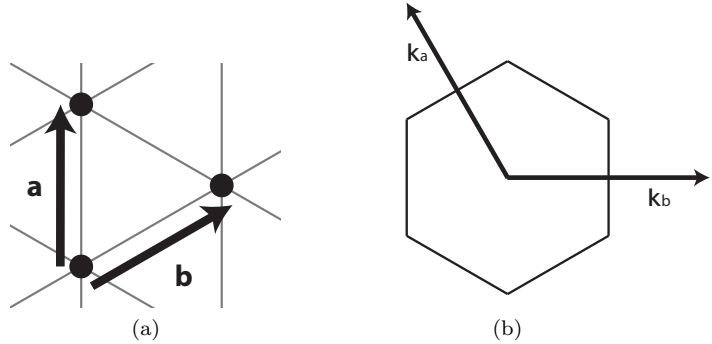

(a)

(b)

FIG. S1. (a) Lattice vectors $\mathbf{a}$ and $\mathbf{b}$ on the triangular lattice. (b) Corresponding reciprocal lattice vectors and the first Brillouin zone.

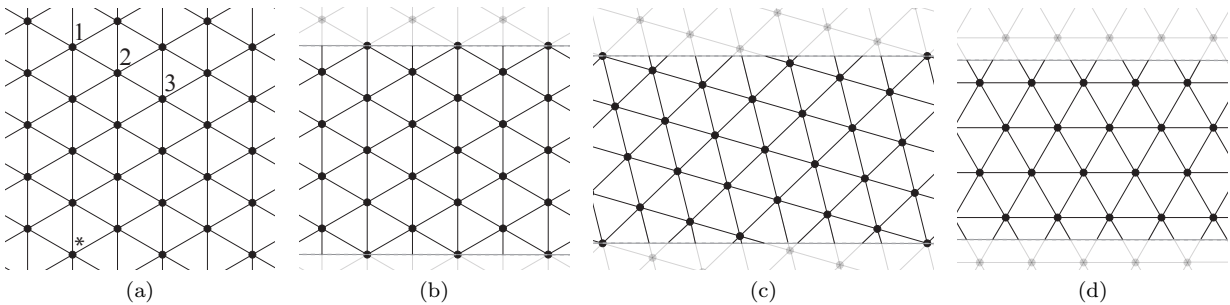

FIG. S2. (a) Periodic boundary conditions on the cylinder are defined by identifying the point ⋆ with the points labeled in the figure as 1, 2 and 3, which correspond to $(n_a, n_b) = (4,0)$, $(3,1)$, and $(2,2)$ respectively. (b) Cylinder from identifying point ⋆ with point 1, called YC4 boundary conditions. This is the same cylinder one would get by adding sites at the hexagon centers of a zigzag nanotube. (c) Cylinder from identifying point ⋆ with point 2. Note that this case has a 26-site unit cell, making it computationally intractable. (d) Cylinder from identifying point ⋆ with point 3, called XC4 boundary conditions. This corresponds to adding sites at the hexagon centers of an armchair nanotube.

we fix the cylinder circumference (in Manhattan distance, ie the minimum number of lattice vectors to go between equivalent points; this is not the physical circumference in general) to be a particular integer, $L$, there are exactly $\lfloor (L+1)/2 \rfloor$ unique cylinders of this type that can be constructed, which are given by fixing one point in the 2D lattice and identifying it with each of the $\lfloor (L+1)/2 \rfloor$ points separated by $n_a \mathbf{a} + n_b \mathbf{b}$ such that $n_a + n_b = L$ and $n_a \in \{\lfloor L/2 \rfloor, \lfloor L/2 \rfloor + 1, \cdots, L\}$. The three points for $n = 4$ are shown in Figure S2a. All other lattice points that are equidistant (in Manhattan distance) from the fixed point give physically equivalent cylinders by rotating or reflecting the 2D lattice. The resulting one-dimensional strips (with a cylinder formed by identifying the edges) are shown in Figures S2b, S2c, and S2d; the first and third cylinders are called YC4 and XC4, indicating that a lattice vector runs, respectively, along the $y$ or the $x$ direction[3]. In general, the YC$L$ cylinder is one with $(n_a, n_b) = (L, 0)$ and is defined for any $L$, while the XC$L$ cylinder can be constructed only when $L$ is even and corresponds to $(n_a, n_b) = (L/2, L/2)$.

The choice of boundary conditions has important consequences, both for the physics of the model and for the computational efficiency of the DMRG. The first implication of the choice of boundary conditions is that the allowed momenta in the Brillouin zone are restricted. The allowed inequivalent momenta in the full 2D model are those in the first Brillouin zone, which is shown for this model in Figure S3a. However, if we define a cylinder by identifying, with periodic boundary conditions, two points that are separated by $n_a \mathbf{a} + n_b \mathbf{b}$, then an eigenstate at momentum $\mathbf{k} = c_a \mathbf{k}_a + c_b \mathbf{k}_b$ ($c_a$ and $c_b$ can be arbitrary real numbers) must satisfy $\psi_{\mathbf{k}}(\mathbf{x}) = \psi_{\mathbf{k}}(\mathbf{x} + n_a \mathbf{a} + n_b \mathbf{b})$, or equivalently (due to Bloch's theorem)

$$1 = e^{i(n_a \mathbf{a} + n_b \mathbf{b})\cdot(c_a \mathbf{k}_a + c_b \mathbf{k}_b)} = e^{2\pi i(n_a c_a + n_b c_b)} \tag{1}$$

which requires that $n_a c_a + n_b c_b$ be an integer. Each integer corresponds to a particular line through the Brillouin zone. For example, in the case of the YC4 cylinder, where $n_a = 4$ and $n_b = 0$, there is no restriction on $c_b$ but $c_a$ must be an integer multiple of $1/4$. This leads to the cuts through Brillouin zone shown in Figure S3b. The corresponding cuts for the other two possible choices of boundary conditions are shown in Figures S3c and S3d.

A related consequence of the choice of boundary conditions is that certain types of multi-sublattice orders may or may not be allowed. This is extremely important for the triangular lattice Hubbard model, which in the limit $U \to \infty$ reduces to the nearest neighbor Heisenberg model and thus should have a three-sublattice 120° Néel order. Notably, this order is *not allowed* on the YC4 cylinder, since the four sites around the circumference cannot be assigned to three distinct sublattices in a consistent way.

Another physical consequence of the choice of boundary condition is that the final cylinder circumference can vary in size. In the case of YC4 boundaries, the cylinder has circumference $4a$, while for XC4 boundaries it is just $2\sqrt{3}\,a$. This is also reflected in total length of the allowed cuts through the Brillouin zone; these have lengths $4 \times (4\pi/a\sqrt{3})$ and $2\sqrt{3} \times (4\pi/a\sqrt{3})$ respectively. This means that with a given number of sites $L$ in the unit cell, the YC$L$ cylinder may be "more two-dimensional" than the corresponding XC$L$ cylinder, though this effect is presumably less important than the question of which multi-sublattice orders are or are not allowed.

Finally, an appropriate choice of boundary conditions can dramatically speed up numerical computations by introducing additional conserved quantities. The YC$L$ cylinders have discrete $L$-fold translation symmetry around the cylinder, leading to $L$ conserved momenta. These correspond to the cuts through the Brillouin zone. The XC$L$ cylinders (well-defined for even $L$) similarly have $L/2$-fold discrete translation symmetry, giving $L/2$ conserved quantities.

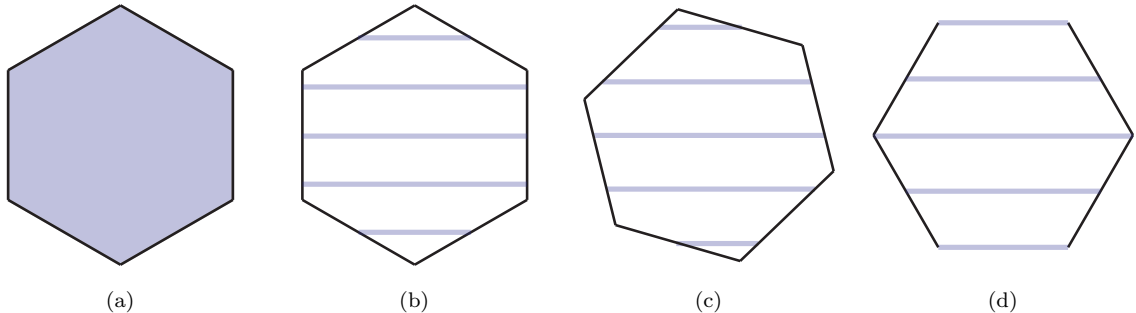

FIG. S3. (a) Allowed momenta in the first Brillouin zone of the full 2D triangular lattice. (b) Allowed momenta for YC4 boundary conditions. (c) Allowed momenta for the $(n_a, n_b) = (3, 1)$ boundary conditions. (d) Allowed momenta for the XC4 boundary conditions. Note that in (b)-(d), if the hexagons are tiled, then the allowed cuts form 4, 1, and 2 distinct lines respectively, corresponding to different numbers of conserved momenta in the Hamiltonian.

The distinct conserved momenta correspond to distinct allowed cuts through the Brillouin zone (Figure S3): if the BZ is tiled, then the allowed cuts actually form 4, 1, and 2 distinct lines for the three respective boundary conditions.

As explained in the main text, we primarily use the YC boundary conditions. There are two main reasons: (1) by choosing different cylinder circumferences, we can try to stabilize/destabilize different phases and in particular we can frustrate the expected high-$U$ magnetic order to make a spin liquid phase more robust and easier to observe; and (2) with YC$L$ boundary conditions we can use a mixed real- and momentum-space basis with $L$ conserved momenta, which both gives a dramatic improvement in computational efficiency and allows us to separately find the ground state in different momentum sectors.

## II. EXPECTED CENTRAL CHARGE IN THE METALLIC PHASE

As reported in the main text, we numerically observe for the YC4 cylinder a central charge $c \approx 6$. This is the expected result for the metallic phase, based on an exact tight-binding solution for the non-interacting limit of $U = 0$ on the full 2D lattice. In that limit, the Hamiltonian becomes:

$$H = -2t \sum_{kq\sigma} n_{kq\sigma} \big( \cos(2\pi k) + \cos(2\pi q) + \cos(2\pi(k-q)) \big)$$ (2)

where the momentum in the Brillouin zone is given by $\mathbf{k} = k\mathbf{k}_a + q\mathbf{k}_b$ for the reciprocal lattice vectors $\mathbf{k}_a$ and $\mathbf{k}_b$ as shown in Figure S1(b). In the ground state, all states with negative energy will be occupied and all with positive energy will be empty, defining a nearly circular Fermi surface with approximate radius $4\pi/(3\sqrt{3}\,a)$ [for comparison, the side length of the hexagonal Brillouin zone is $4\pi/(3a)$].

When the system is restricted to a finite cylinder, we can then count how many of the allowed momentum cuts cross the Fermi surface. This is shown visually in Figure S4 for the YC4 and YC6 cylinders; the number of cuts crossing the Fermi surface is 3 for YC4 and 5 for YC6.

Each distinct cut through the Fermi surface corresponds to two species of free fermion, one for spin up and one for spin down, and each free fermion contributes a central charge of 1[4]. Thus we conclude that the expected central charges at $U = 0$ and therefore throughout the metallic phase are 6 and 10 for the YC4 and YC6 cylinders, respectively.

## III. LABELING THE ENTANGLEMENT SPECTRUM BY QUANTUM NUMBERS

Recall that the entanglement spectrum is the set of values $\{\log(\lambda_i)\}$ where the $\{\lambda_i\}$ are the coefficients of the Schmidt decomposition

$$|\psi\rangle = \sum_i \lambda_i |\psi_i^L\rangle |\psi_i^R\rangle$$ (3)

for a cut between any two rings of the cylinder[5].

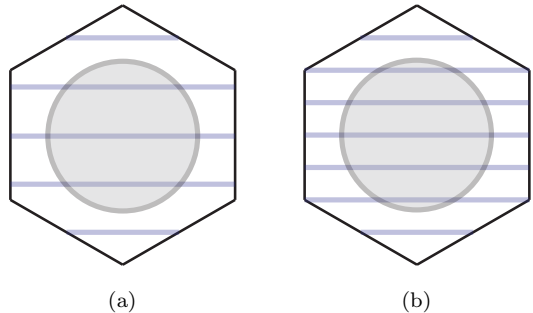

FIG. S4. **(a)** The shaded circle denotes single-particle eigenstates that are filled in the $U = 0$ limit of the model on the full two-dimensional lattice. The blue lines are the allowed momentum cuts for the YC4 cylinder; evidently, three of them cross the Fermi surface. **(b)** Same for the YC6 cylinder, with 5 lines crossing the Fermi surface.

We use a matrix product state with all legs labeled by conserved charges, so that when we perform the Schmidt decomposition as in equation (3), each Schmidt state $|\psi_i^L\rangle$ is an eigenstate of three operators: total momentum around the cylinder, spin up occupation number, and spin down occupation number. We then label the $\lambda_i$ by the corresponding integer eigenvalues.

However, for iDMRG the left and right Schmidt states extend to infinity, and it is not obvious how these integer charge labels correspond to "physical" values of the charge because, for example, the total spin up occupation is infinite in each of the two halves. Thus our charge labels actually give the total charge relative to some point on the cylinder (arbitrarily chosen as a result of details of the DMRG algorithm) which may be far from the cut we consider in the Schmidt decomposition.

We can fix this ambiguity by subtracting a constant offset from all charge labels so that the net charge on each of the two semi-infinite halves, defined by

$$\sum_i \lambda_i^2 Q_{\lambda_i} \tag{4}$$

where $Q_{\lambda_i}$ is the charge label of $\lambda_i$, is 0. A more rigorous treatment of this subtraction is given in the Supplementary Material of reference 6, section II(B).

After making this correction, each Schmidt value $\lambda_i$ is labeled by a set of "physical charges" including the momentum and total spin $(\left(n_\uparrow - n_\downarrow)/2\right)$. The latter may be a half-integer if, as in the semion sector of a chiral spin liquid (CSL), there are fractionalized quasiparticles.

## IV. MATRIX PRODUCT STATE TRANSFER MATRIX AND THE EXCITATION SPECTRUM

Excitation gaps in the physical system can be estimated from the eigenvalues of the matrix product state (MPS) transfer matrix spectrum. Here we elaborate, showing how flux insertion can be used to partially map out the two-dimensional excitation spectrum versus $k_x$ and $k_y$. This is similar to the method used in references 7 and 8. We use this technique to show that our data do not clearly indicate a susceptibility towards a Dirac spin liquid state; see sections V H, VI D, and VII B below.

### A. One-dimensional version

Consider a one-dimensional system, with its ground state given by an MPS. For clarity, we suppose a one-site unit cell and a right-normalized MPS with the transfer matrix as shown in Figure S5a. If all legs of the MPS tensors are labeled by conserved charges, then transfer matrix eigenvalues can be labeled as well. As an example, suppose the conserved charge is $S_z$. Then the transfer matrix can be decomposed into independent sectors with a conserved charge on each leg, as in Figure S5a. The total charge coming in from the right is $s_2 - s_4$, while the total charge going out on the left is $s_1 - s_3$. An eigenvector of this sector of the transfer matrix will have outgoing charge to the left of $s_2 - s_4$ so that the product is nonzero, and the outgoing charge of the product will be the left charge of the transfer matrix, or $s_1 - s_3$; since the product is a scalar multiple of the original vector, we can conclude that $s_1 - s_3 = s_2 - s_4$ in any sector of the diagonalized transfer matrix. (In other words, the transfer matrix being diagonalizable means

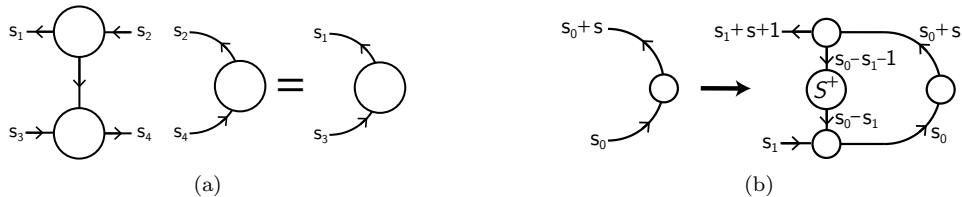

(a)                                    (b)

FIG. S5. **(a)** The MPS transfer matrix, here shown for a right-normalized MPS with a one site unit cell, can be decomposed into blocks labeled by a conserved charge on each leg. $s_1 - s_3 = s_2 - s_4$ for every nonzero block if the transfer matrix is diagonalizable. **(b)** The charge of the transfer matrix corresponds to the physical charge of an excitation, here demonstrated via a spin-1 excitation produced by the $S^+$ operator.

that all blocks with $s_1 - s_3 \neq s_2 - s_4$ must be identically zero.) In this way, each eigenvalue of the transfer matrix can also be labeled by this conserved charge, $s_1 - s_3$.

We can then relate these conserved charges to physical excitations in the system. Assume a transfer matrix eigenvector with total outgoing conserved charge $s$ and apply an operator $S^+$ inside the transfer matrix as shown in Figure S5b; the outgoing conserved charge of the combined object is $s+1$, and it will transform according to the $s+1$ block of the transfer matrix. Thus if the 1 eigenvalue corresponding to the normalization of the MPS is in the $s = 0$ sector, the $s = 1$ sector of the transfer matrix will describe the physical evolution with translation in $x$ of excitations with $S_z = 1$.

In fact, the eigenvalues of the transfer matrix spectrum give very specific information about the excitation spectrum in each charge sector. Letting the eigenvalues be denoted by $\lambda = e^{-\epsilon + i\phi}$, it is believed to be true that for each such eigenvalue there is an excitation of the system with energy $E = v\epsilon$ and momentum $k_x = \phi/a$; $v$ is a scale factor that is the same for every excitation.[9] At finite bond dimension it will be impossible to map out the full excitation spectrum since $k_x$ is continuous, but the transfer matrix spectrum will include the minima of the excitation spectra versus $k_x$. Thus by looking at the low-lying eigenvalues in each charge sector, it is possible to observe the smallest excitation energies for different types of excitations and the corresponding $k_x$. For example, the low-lying $s = 0$ eigenvalues give the spin singlet gap; it is similarly possible to compute the triplet gap from the $s = 1$ sector.

## B. Hubbard model and the two-dimensional excitation spectrum

Now suppose that the system is the square lattice Hubbard model (lattice constant $a$) on a cylinder with a circumference of $L$ sites, in particular in mixed real- and momentum-space, with conserved quantities $(n_\uparrow, n_\downarrow, k_y)$, so that the conserved spin is $(n_\uparrow - n_\downarrow)/2$ and the conserved charge is $n_\uparrow + n_\downarrow$. Here we label the $k_y$ eigenvalues by integers from 0 through $L-1$ for a cylinder of circumference $L$, and they should be multiplied by $2\pi/a$ to get the true momentum.

Then suppose that in some charge sector, for example $k_y = 1$, $n_\uparrow = 0$, $n_\downarrow = -1$, the smallest eigenvalue is at $k_x = \pi/a$. This means that in that charge sector, the lowest energy excitation has $\Delta k_x = \pi/a$, and that it has a $\Delta k_y$ corresponding to a transition between two of the allowed momentum cuts, at $k = n$ and $k = n+1$. This would seem to indicate that $\Delta k_y = 2\pi/a$, but there is an important, and very powerful, exception.

As described in the main text, when we perform flux insertion that twists the boundaries oppositely for spin up and spin down electrons, we shift the momentum cuts through the Brillouin zone in opposite directions. Compared with the charge neutral sector, the excitation in sector $(1, -1, 1)$ is created by some combination of the operators $c_{\uparrow, k+1}^\dagger c_{\downarrow, k}$ and the physical shift in $k_y$ between the annihilated electron and the created one is actually $(2\pi/a)(1 + \theta/(2\pi))$.

Using this, we can, for each $\theta$, find the $\Delta k_x$ for the lowest energy excitation in each charge sector and its corresponding $\Delta k_y$, then scan $\theta$ from 0 to $4\pi$ to map out how both the minimum excitation energy and its corresponding $\Delta k_x$ vary with $\Delta k_y$. (Note that the particular way that momentum cuts shift will depend on the charge sector.) This provides a one-dimensional cut through the minimum of the excitation spectrum vs $k_x$ and $k_y$.[10]

To demonstrate the technique, consider the noninteracting model shown in Figure S6a, with hopping $-t$ on solid bonds and $+t$ on dashed bonds; this is just the $U = 0$ Hubbard model on a distorted triangular lattice with staggered $\pi$ flux. The exact solution is given by

$$E(n_a, n_b) = \pm \left( 4\cos^2(2\pi n_b) + \left| e^{i2\pi n_b} + e^{-i2\pi n_a} + e^{-i2\pi(n_a + n_b)} - 1 \right|^2 \right) \tag{5}$$

$$= \pm \left( 4\cos^2(2\pi n_b) + 2(2 - \cos(2\pi n_a) + \cos(2\pi n_a + 4\pi n_b)) \right) \tag{6}$$

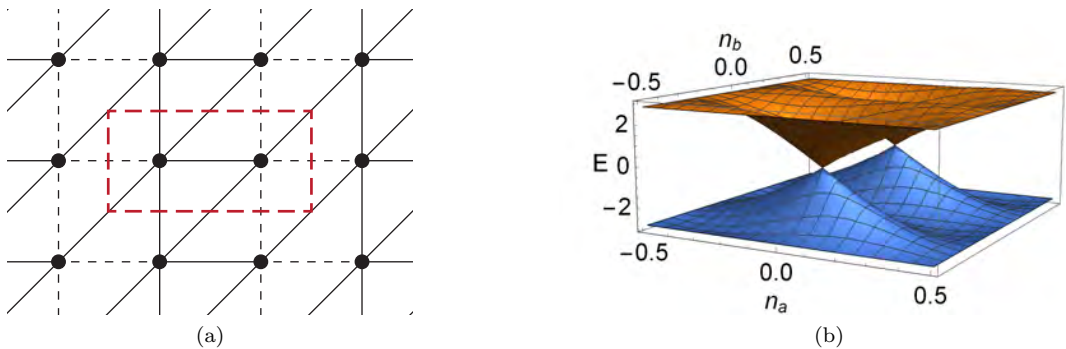

FIG. S6. (Color online) **(a)** Sample tight-binding model for demonstration of excitation spectra from MPS transfer matrix spectrum. Hopping is $-t$ on solid bonds, $+t$ on dashed bonds, and interaction is zero. The red rectangle shows the unit cell. **(b)** Dispersion for this model, plotted as a function of $(n_a, n_b)$ where $\mathbf{k} = n_a \mathbf{k}_x + n_b \mathbf{k}_y$. The reciprocal lattice vectors are $\mathbf{k}_x = (\pi/a)\hat{x}$ and $\mathbf{k}_y = (2\pi/a)\hat{y}$ where $a$ is the lattice constant.

where $n_a$ and $n_b$ give the momentum as $\mathbf{k} = n_a \mathbf{k}_x + n_b \mathbf{k}_y$. This dispersion is shown in Figure S6b and evidently has Dirac cones at $(n_a, n_b) = (0, \pm 1/4)$. In Figure S7 we show in the left two columns the following quantities for a circumference 4 cylinder: the allowed momentum cuts through the Brillouin zone at $\theta = 0$; the allowed cuts at $\theta = 2\pi$; the low-lying transfer matrix spectrum versus $\theta$ in the spin 0 and spin 1 sectors, with eigenvalues colored by momentum quantum number; the low-lying transfer matrix spectrum for all theta versus $n_a$ in the spin 0 and spin 1 sectors; and the low-lying spectrum for all theta versus $n_b$ in the spin 0 and spin 1 sectors. In the right two columns we show the same for $L = 5$.

The $S_z = 0$ spectra in Figure S7 are discrete with respect to both momenta $n_a$ and $n_b$, a fact which we briefly explain. These excitations correspond to applying either a $c_\uparrow^\dagger$ followed by $c_\uparrow$ or $c_\downarrow^\dagger$ followed by $c_\downarrow$. For $n_b$, at zero flux each creation and annihilation operator must lie on one of the allowed momentum cuts, and the difference in the momenta of $c^\dagger$ and $c$ is discretized to one of the $L$ allowed transitions between cuts. As flux is inserted, the momentum cuts shift, but they shift together for $c_\sigma^\dagger$ and $c_\sigma$ so that the momentum transfer remains discretized. (This does not happen for $S_z = 1$ because the cuts for $c_\sigma^\dagger$ and $c_{-\sigma}$ shift oppositely.) This discretization in $n_b$ will be present for any spinful fermion model in the $S_z = 0$ sector. The discretization in $n_a$, on the other hand, is specific to this model. In particular, for this model the local minima of the excitation spectrum with respect to $n_a$ at fixed $n_b$ remain at the same $n_a$ when the initial and final momenta in the $\mathbf{k}_y$ direction (separated by the momentum of the excitation, $n_b$) are varied. Recalling from above that the transfer matrix spectrum eigenvalues generally give the local minima in the excitation spectra, this leads to discretization in $n_a$. In more complicated models, the $S_z = 0$ will not be exactly discretized, though it may still have a tendency to cluster around lines of discretized momentum.

To confirm that the transfer matrix really gives the two-dimensional excitation spectrum as claimed, we point out several features of the spectra. (1) For spin 0 excitations, where the momenta of $c^\dagger$ and $c$ move together, $\Delta k_y$ is quantized, and the Dirac cones appear in the $k = 0$ sector whenever an allowed momentum cut hits one. For $L = 4$, they also appear simultaneously for $k = 2$ because the separation between the cones is twice the separation between allowed cuts. (2) For spin 1 excitations, the Dirac cones appear for $L = 5$ in the $\Delta k = 2$ sector at $\theta = \pi$ corresponding to a transition from the lower cone with $k_y = 4$ for spin down to the upper one with $k_y = 1$ for spin up and at $\theta = 3\pi$ in the $\Delta k = 1$ sector corresponding to a transition from the upper cone with $k_y = 2$ for spin down to the lower one with $k_y = 3$ for spin up. (3) Dirac cones appear at $\Delta k_y = 0$ for L=5 in the spin 0 sector but not in the spin 1 sector, since cuts for spin up and spin down can never both pass through the same Dirac cone at the same time given their spacing.

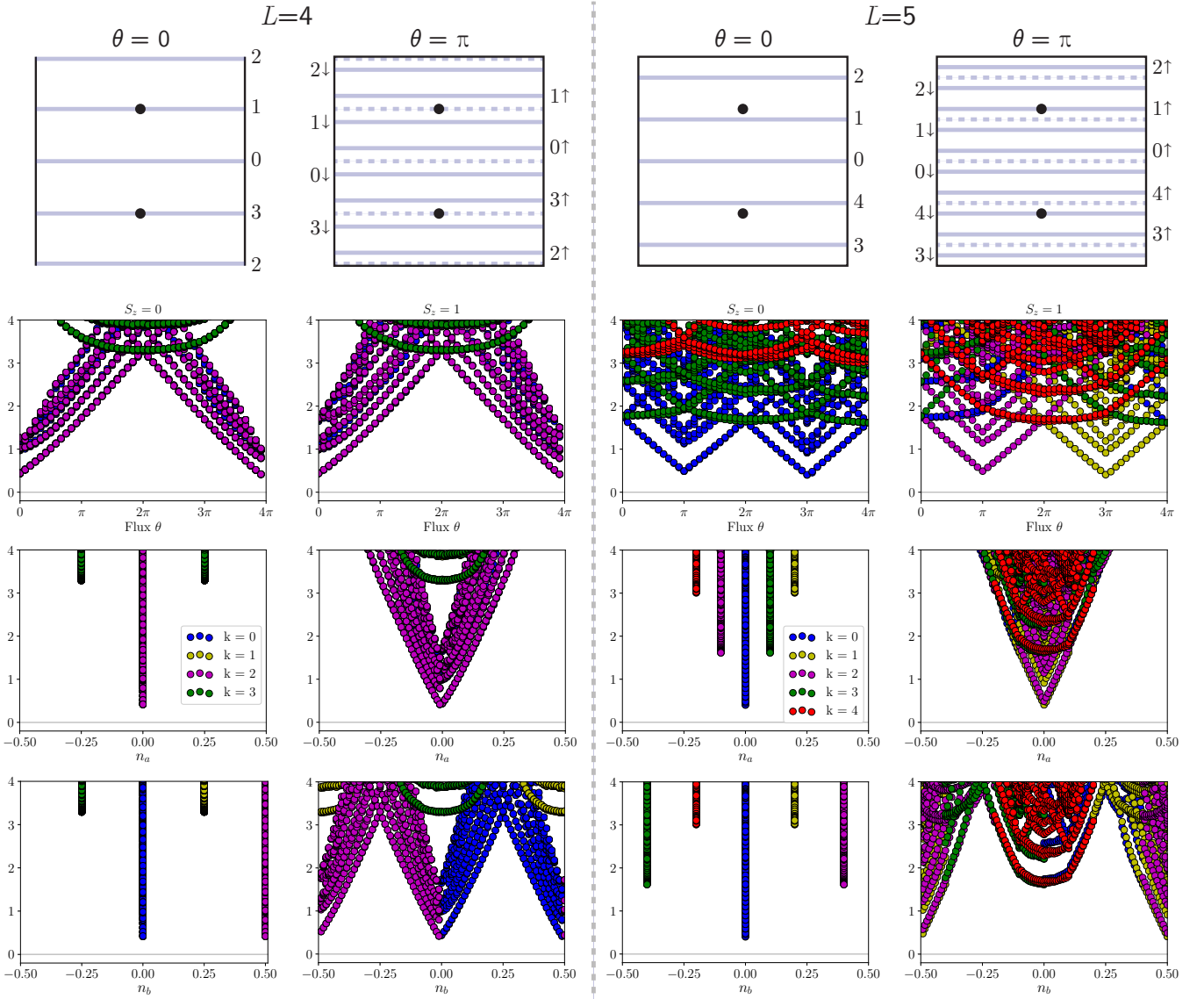

FIG. S7. (Color online) Transfer matrix spectra for the ground state of the model shown in Figure S6a, computed using DMRG with a bond dimension of 4000. **Left two columns, top row:** allowed momentum cuts for a square lattice on a circumference 4 cylinder at (left) zero flux and (right) $\pi$ flux. Black dots indicate the locations of the Dirac cones. **Second row:** low-lying transfer matrix spectrum versus flux $\theta$ in the $S_z = 0$ (left) and $S_z = 1$ (right) sectors. Colors indicate the momentum quantum number of each eigenvalue, as given by the legend in the third row. Notes: (1) we omit the eigenvalue of the transfer matrix that equals 1, which just gives the normalization of the state; (2) the bottoms of the Dirac cones are at noticeably nonzero energy due to finite bond dimension; and (3) some data points may be hidden, e.g. $k = 0$ behind $k = 2$. **Third row:** transfer matrix spectrum for spin 0 and spin 1 versus $n_a$. **Fourth row:** transfer matrix spectrum for spin 0 and spin 1 versus $n_b$. Note that part of the $S_z = 0$ spectrum is near the right edge. **Right two columns:** same data for a cylinder with circumference $L = 5$.

### C. On the triangular lattice

Finally, we describe how this approach can be applied to the triangular lattice. To actually perform our DMRG simulations, we convert the triangular lattice into a distorted square lattice, as shown for YC and XC boundary conditions in Figure S8. Following the method above will give the excitation spectrum as a function of $k_x$ and $k_y$ in the effective square/rectangular Brillouin zone.

We then write the overall momentum as $\mathbf{k} = n_a \mathbf{k}_x + n_b \mathbf{k}_y$ as above, and simply keep $n_a$ and $n_b$ constant while replacing $\mathbf{k}_x$ and $\mathbf{k}_b$ by the corresponding reciprocal lattice vectors of the triangular lattice, as shown in Figure S8.

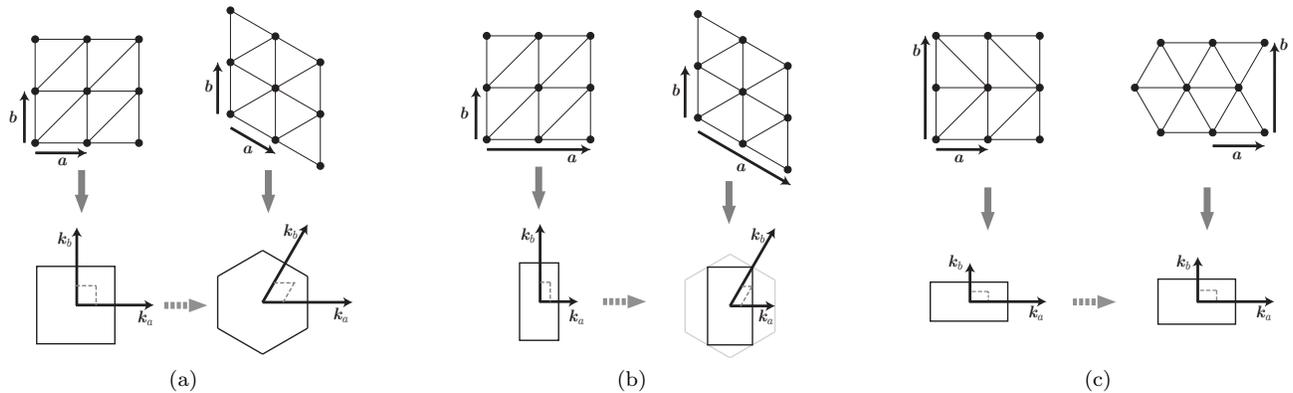

FIG. S8. (Color online) Conversion from momentum space on the square lattice representation of the triangular lattice back to momentum space for the original lattice. **(a)** The original triangular lattice is shown in the top right, with a one ring unit cell. For running DMRG with YC boundary conditions, we convert to the square lattice shown at top left. Using the MPS transfer matrix from the computed ground state, we find the excitation spectrum as a function of **k** in the Brillouin zone for the square lattice, bottom left; the momentum is given in terms of $n_a$ and $n_b$, fractions of the reciprocal lattice vectors $\mathbf{k}_a$ and $\mathbf{k}_b$. The corresponding momenta in the Brillouin zone for the original triangular lattice, bottom right, can be computed by using the corrected $\mathbf{k}_a$ and $\mathbf{k}_b$ while keeping $n_a$ and $n_b$ the same; this is demonstrated by the point $n_a = n_b = 1/4$ at the intersection of the dashed gray lines. **(b)** Same but for the YC boundary conditions with a two-ring unit cell. At bottom right, we show how the reduced Brillouin zone with two rings fits into the one-ring Brillouin zone. **(c)** Same but for the XC boundary conditions.

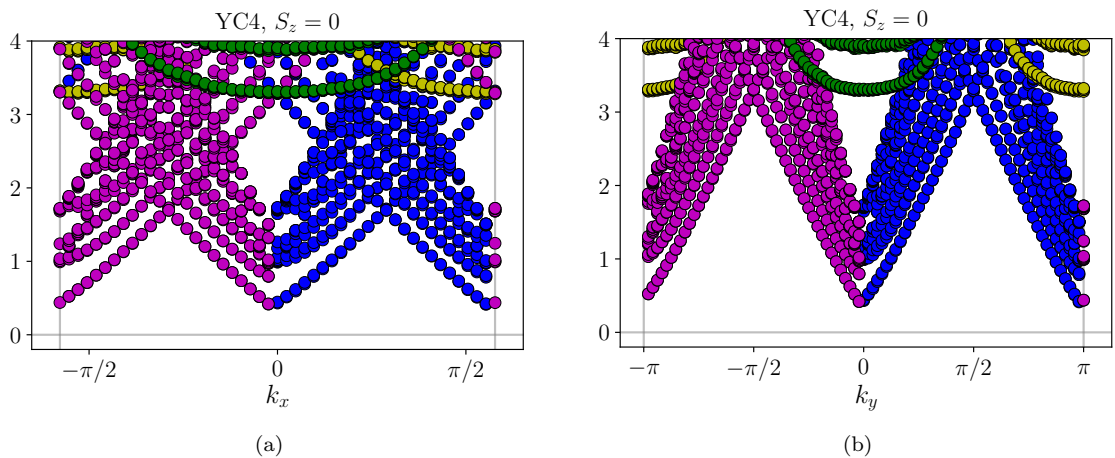

FIG. S9. (Color online) **(a)** Low-lying part of the transfer matrix spectrum for the noninteracting model of Figure S6a, vs $k_x$ in the Brillouin zone of the original triangular lattice. (The lattice constant is set to 1.) Note that transitions between the Dirac cones appear at $\Delta k_x = 0, \pm \pi/\sqrt{3}$ as expected. **(b)** Same, versus $k_y$, with transitions between Dirac cones appearing as expected at $\Delta k_x = 0, \pm \pi$.

As an example, we can place the staggered-$\pi$ flux hopping model above onto the original triangular lattice; in terms of the new $k_x$ and $k_y$, the $L = 4$, spin 1 spectrum versus $k_x$ and $k_y$ from Figure S7 will be as shown in Figure S9.

## V. YC4 ADDITIONAL DATA AND ANALYSIS

### A. Correlation lengths in different charge sectors

As described above, the MPS transfer matrix can be decomposed into blocks in different conserved charge sectors, and this allows the computation of a correlation length $\xi$ in each sector. In Figure 3(b) of the main text, we show the correlation length as a function of $U/t$ for a range of bond dimensions in the $(Q, S_z, K) = (0, 0, 0)$ sector, in other words the correlation length for operators carrying no charge, spin, or momentum around the cylinder. In Figure

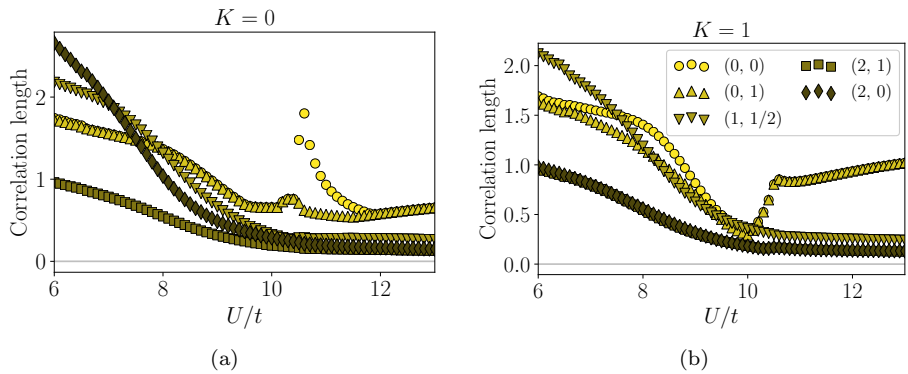

FIG. S10. (Color online) Correlation lengths computed from the MPS transfer matrix ($\chi = 11314$) for operators with **(a)** $K = 0$ and **(b)** $K = 1$. Each line corresponds to a charge sector $(Q, S_z)$, see legend in panel (b).

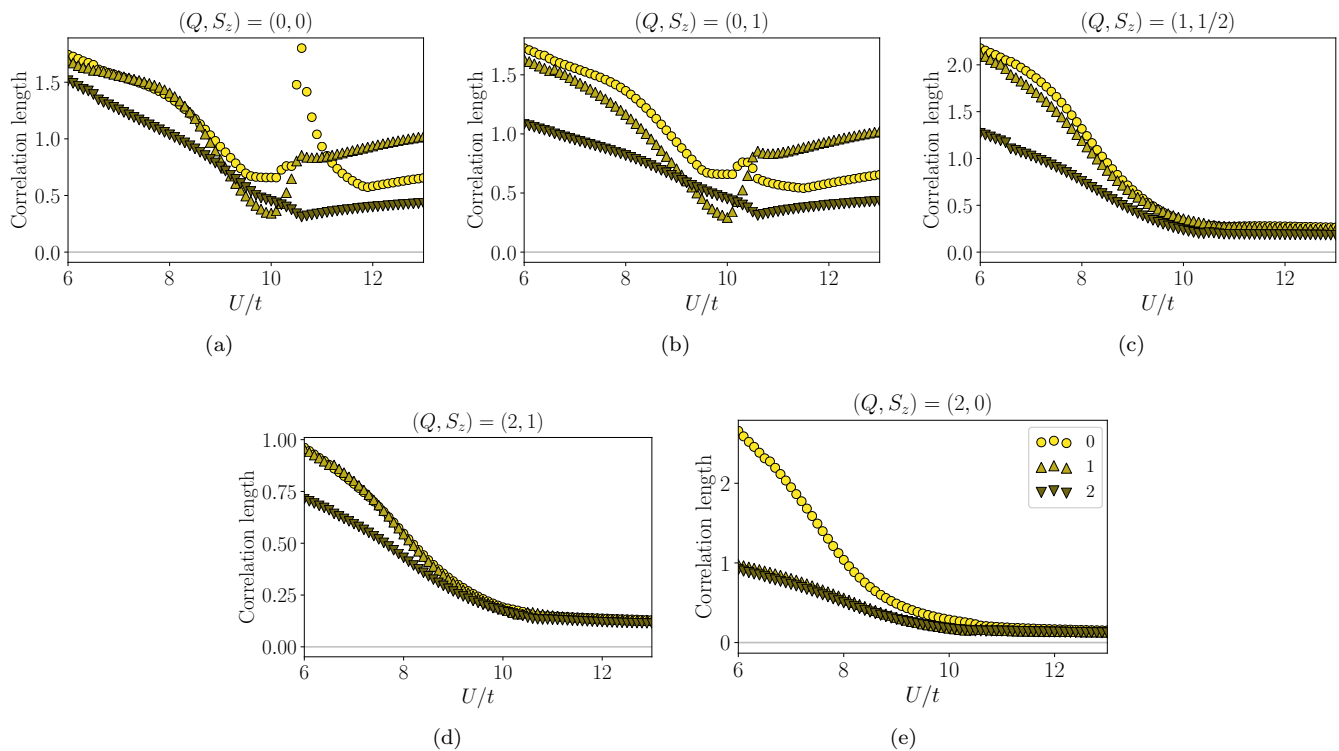

FIG. S11. (Color online) Correlation lengths for operators with $(Q, S_z) =$ **(a)** (0,0), **(b)** (0,1), **(c)** (1,1/2), **(d)** (2,0), and **(e)** (2,1). Each line corresponds to a charge sector $K$, see legend in panel (e).

3(c), we show for the highest bond dimension the correlation length in all four charge sectors that have the largest $\xi$ for some $U/t$. Here we show the comparison between correlation lengths in different sectors in slightly more detail.

In Figures S10 and S11 we show the correlation lengths for the $(Q, S_z) = (0,0)$, $(1,1/2)$, $(0,1)$, $(2,0)$, and $(2,1)$ sectors with $K = 0$ and 1; in the first figure the correlation lengths are separated by $K$ and in the second by $(Q, S_z)$. There are several notable features: (1) The large peak at $U/t \approx 10.6$ appears only in the $(0,0,0)$ sector. (2) The longest correlation length is in the $K = 1$ sector for large $U$, and in the $K = 0$ sector elsewhere. (3) Below $U/t \approx 8$, the longest correlation lengths correspond to charge fluctuations. Where the $(Q, S_z) = (1, 1/2)$ fluctuations have the longest correlation length, this is consistent with a metallic state. Above this, charge fluctuations are gapped and the spin correlations are the largest. (4) At lower $U$ the $(Q, S_z) = (2, 0)$ fluctuations have the largest correlation length, potentially indicating a susceptibility to superconductivity. However, this is far from the chiral phase in which we are interested, so we have not investigated this possibility.

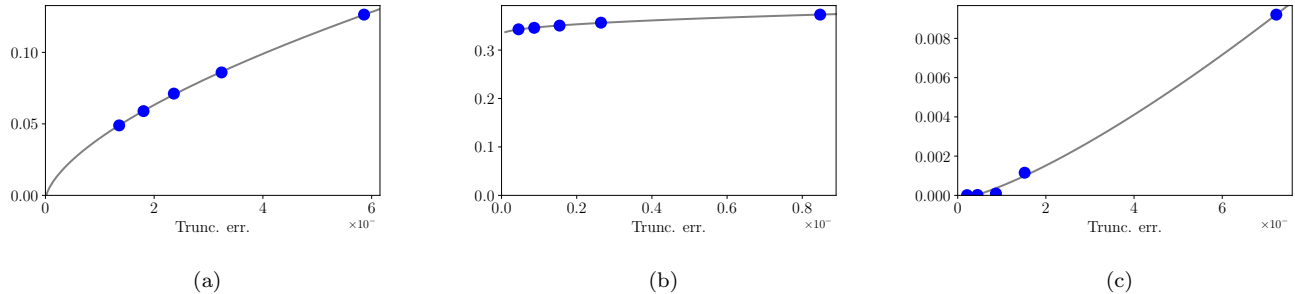

(a)

(b)

(c)

FIG. S12. Chiral order parameter versus DMRG truncation error, for **(a)** $U/t = 8$ in the low-$U$ phase, **(b)** $U/t = 10$ in the intermediate phase, and **(c)** $U/t = 11$ in the high-$U$ phase. Gray lines show best fit curves of the form $C + A \times p^B$, allowing for extrapolation to the limit of no truncation error/infinite bond dimension.

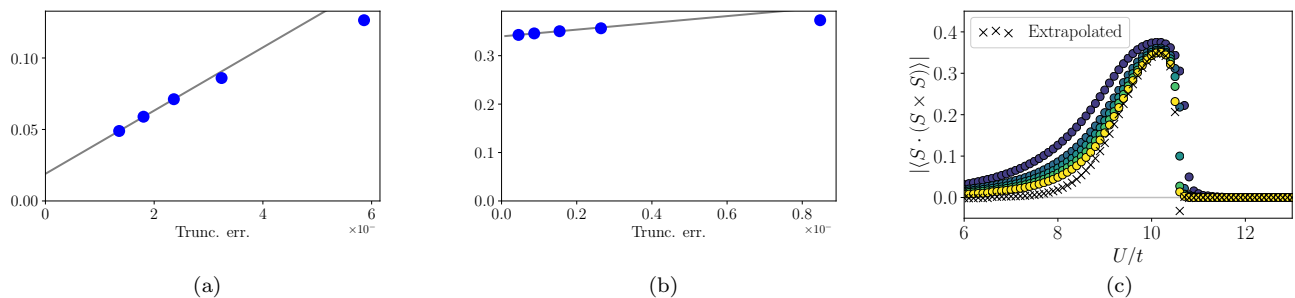

(a)

(b)

(c)

FIG. S13. (Color online) In **(a)** and **(b)**, we show the same data as in Figures S12a and S12b respectively, but with linear best fit lines computed using the points with the three lowest truncation errors in each case. The result is essentially unchanged in the intermediate phase. In **(c)**, we show the extrapolation as a function of $U/t$ using this linear fit method. The actual data is the same as in Figure 3(e) of the main text.

### B. Chiral order parameter and extrapolation

In Figure 3(e) of the main text, we show the chiral order parameter $\langle \mathbf{S}_i \cdot (\mathbf{S}_j \times \mathbf{S}_k) \rangle$, where $i$, $j$, and $k$ label three lattice sites at the vertices of a triangle, as a function of $U/t$ at different bond dimensions. We additionally show an extrapolation in the DMRG truncation error; here we explain the details of the extrapolation method.

At each value of $U/t$, we have values of the order parameter for five different bond dimensions, namely 2000, 4000, $5657 \approx 4000\sqrt{2}$, 8000, and $11314 \approx 8000\sqrt{2}$, and corresponding DMRG truncation errors, $p$. The error in the energy of a state should scale linearly with the truncation error, $E = E_{\lim} + A \times p$,[11] but the error in other observables may scale in a more complicated manner. For the chiral order parameter, we assume a scaling of the form

$$\langle \mathbf{S}_i \cdot (\mathbf{S}_j \times \mathbf{S}_k) \rangle = \langle \mathbf{S}_i \cdot (\mathbf{S}_j \times \mathbf{S}_k) \rangle_{\lim} + A \times p^B \tag{7}$$

used in reference 11, Figure 8. The data and best fit curves for several specific values of $U/t$ are shown in Figure S12; in particular, we show $U/t = 8$ at the upper end of the metallic phase, $U/t = 10$ where the chiral order parameter is near its peak, and $U/t = 11$ in the high-$U$ phase. (The `optimize.curve_fit` function from Python's `scipy` library fails to find the best fit of this form for $U/t \gtrsim 11.5$, beyond which we use instead a linear extrapolation from the few highest bond dimensions.)

We also show in Figure S13 the best fit results if we do a simple linear extrapolation from the three highest bond dimensions; we show the best fit line with the data for $U/t = 8$ and $U/t = 10$, as well as the equivalent of Figure 3(e) from the main text with the extrapolation line determined using this linear fit. (The best linear fit at $U/t = 11$ is simply a flat line at 0.) This method makes it seem that some time-reversal symmetry breaking may survive in the low-$U$ phase, but comparing the fitted curves for $U/t = 8$ using the two methods, it appears that the nonlinear fit is significantly better, and that one predicts the expected vanishing of the chiral order parameter in the low-$U$ phase.

Despite the disagreement at low $U$, in intermediate phase the two extrapolation methods give essentially similar

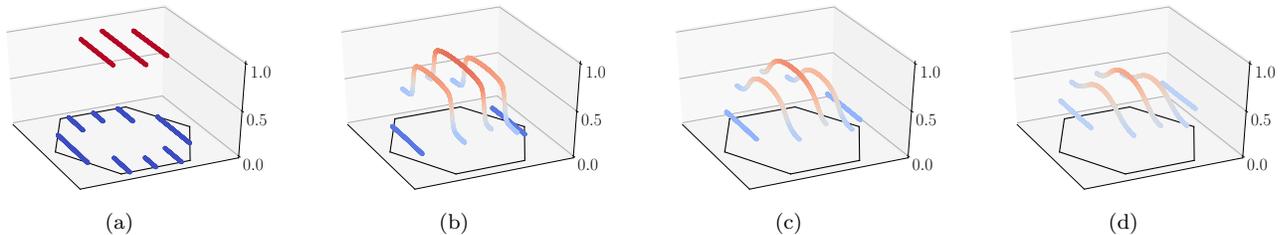

FIG. S14. (Color online) Spin up occupation in the Brillouin zone, $\langle n_{k\uparrow} \rangle$, for **(a)** $U/t = 0$ (exact result), **(b)** $U/t = 6$ in the metallic phase, **(c)** $U/t = 9$ in the spin liquid phase, and **(d)** $U/t = 12$ in the high-$U$ phase.

results, as seen in Figures S12b and S13b; the chiral order clearly remains nonzero in the limit of infinite bond dimension/zero truncation error.

## C. Metal-insulator transition

### 1. Fermi surface singularity and quasiparticle weight

One sign of a metallic state is the presence of a singularity at the Fermi surface in the occupation $\langle n_k \rangle$. In a fully two-dimensional system, where the metal is a Fermi liquid, this singularity takes the form of a discontinuity, whose size is equal to the quasiparticle weight. For a cylinder with finite width, which is actually one-dimensional, the metal is instead a Luttinger liquid and consequently has no discontinuity, but there nevertheless remains a singularity exactly at the Fermi surface, with infinite gradient.

We do not expect to observe a singularity at any finite MPS bond dimension, but by measuring $\langle n_k \rangle$ as a function of $U/t$ and bond dimension we can observe the approximate location where the singularity would appear. We compute the correlators $\langle c_{0,k_y,\uparrow}^\dagger c_{x,k_y,\uparrow} \rangle$ for $x$ in the range $-50$ to $50$, then compute the occupation for spin up by

$$\langle n_{k_x,k_y,\uparrow} \rangle = \sum_{x=-50}^{50} e^{ik_x x} \langle c_{0,k_y,\uparrow}^\dagger c_{x,k_y,\uparrow} \rangle. \tag{8}$$

The range of 50 is about an order of magnitude larger than the correlation length, and the results are converged in the sense that when $\langle n_{k\uparrow} \rangle$ is plotted, the curves from using 40 vs 50 points are essentially indistinguishable and oscillations visible when using 10 or 20 points have disappeared. However, the finite range of correlations does still impose a limit on the momentum resolution of the occupation, and small scale features such as the singularity at the Fermi surface will not be completely resolved.

In Figure S14 we show the spin up occupation in the Brillouin zone (computed with bond dimension $\chi = 4000$) for $U/t = 6$ in the metallic phase, $U/t = 9$ in the spin liquid phase, and $U/t = 12$ in the high-$U$ phase, as well as the exact tight-binding result for $U = 0$ as a comparison. The behavior is clearly qualitatively different at high $U$ compared with low $U$.

A more thorough understanding comes from considering how the occupation changes at each $U/t$ with increasing bond dimension. It is most helpful to consider the $k_y = 0$ slice of the Brillouin zone in particular. The occupations $\langle n_{(k_x,0)\uparrow} \rangle$ for $U/t = 6$, 9, and 12 are shown in Figure S15, both in the full Brillouin zone and in a small range around the Fermi surface, which is identified as the point with steepest gradient of the occupation. Evidently, at small $U$ the maximum gradient dramatically increases with bond dimension, while at high $U$ it remains approximately constant; the singularity at the Fermi surface exists in the physical system if the slope at the Fermi surface extrapolates to infinity with increasing bond dimension.

This can be used to find the location of the metal-insulator transition. To do so, we show the maximum gradient of the occupation vs $U/t$ for several bond dimensions in Figure S16a. The gradient appears to extrapolate to infinity at small $U$, consistent with the existence of a singularity, while the gradient clearly converges with bond dimension for $U/t \gtrsim 10.6$, demonstrating that there is no singularity. The exact location of the transition remains unclear, however, since for $8 \lesssim U/t \lesssim 10$, it is not clear whether the gradient will diverge or not.

Another possibility for observing the transition is to add a factor of $|x|$ in the Fourier transform in equation (8), which converts the singularity at the Fermi surface into a peak. We can then plot the peak height, which is shown

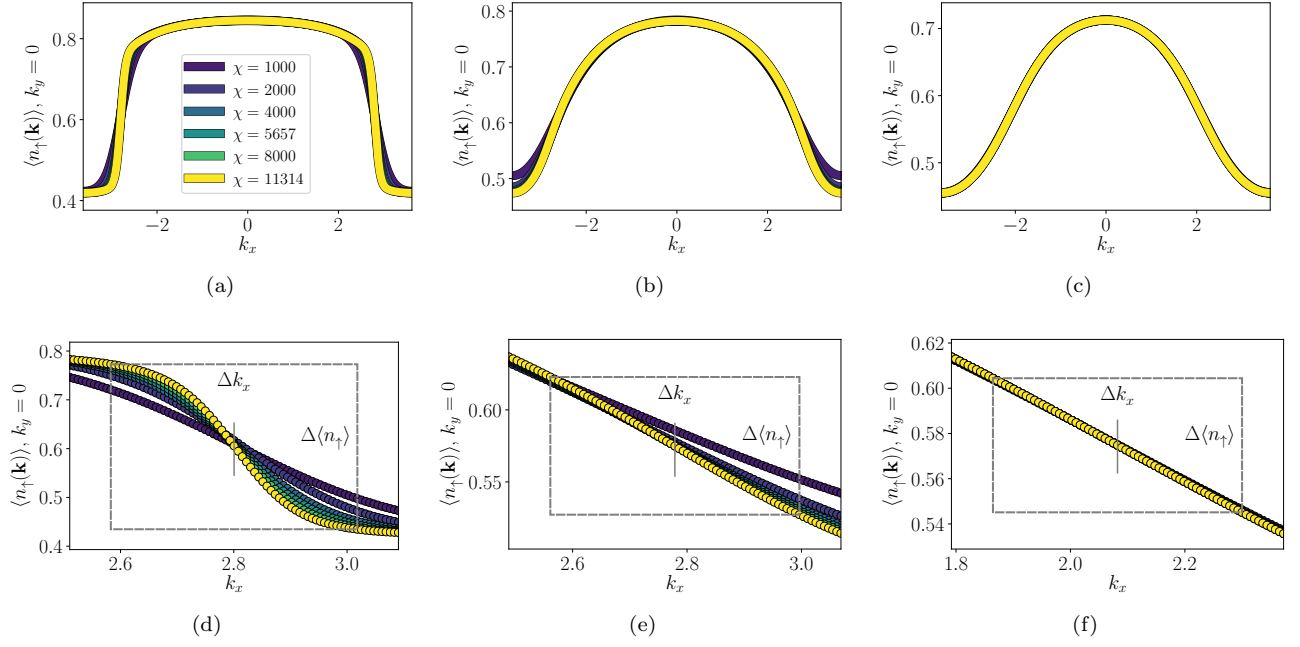

FIG. S15. (Color online) Spin up occupation, $\langle n_{k\uparrow} \rangle$, on the $k_y = 0$ slice through the Brillouin zone, computed for a range of bond dimensions as given in the legend in the first panel. **(a)-(c)** Occupation for the full range of $k_x$, for $U/t = 6$, 9, and 12, respectively. **(d)-(f)** Close-up view of the vicinity of the Fermi surface. For estimating the quasiparticle weight, it is useful to consider how the occupation gap $\Delta\langle n \rangle$ depends on the size of an interval $\Delta k_x$ around the Fermi surface, as shown with the dashed box in each panel.

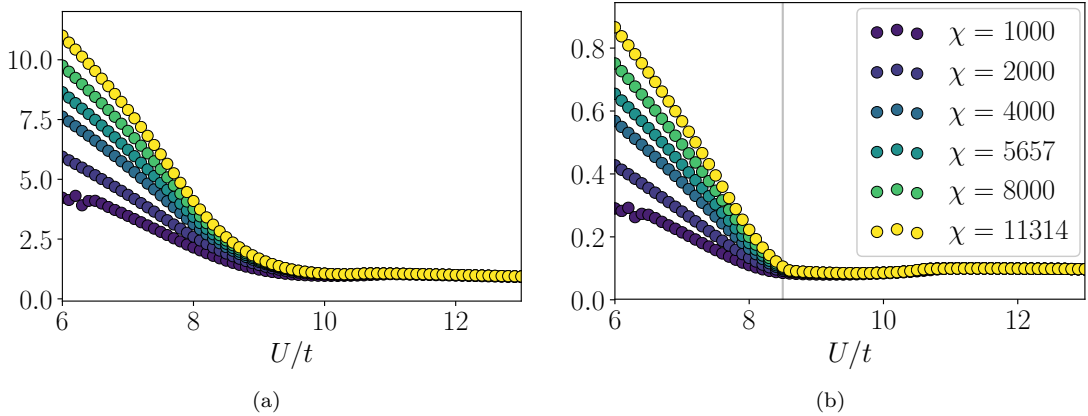

FIG. S16. (Color online) **(a)** Maximum gradient of the occupation, as a function of $U/t$ and bond dimension, showing a transition between $U/t \approx 8$ and $U/t = 10$. **(b)** Height of the peak at the Fermi surface found by including a factor of $|x|$ in the Fourier transform in equation 8. The vertical line at $U/t = 8.5$ appears to be the approximate location of the transition.

in Figure S16b. This allows for a somewhat more precise determination of the transition location, at $U/t \approx 8.5$, the location of the vertical line in the figure.

A related indication of the metal-insulator transition comes from Luttinger's theorem: as long as the system remains metallic, the location of the Fermi surface cannot change as interaction strength is increased. If we identify the Fermi surface on the $k_y = 0$ cut as the point with steepest gradient, we can plot the resulting $k_x$ versus $U/t$. As shown in Figure S17, this point remains fixed until $U/t \approx 8.6$, after which the metallic phase ends, Luttinger's theorem no longer applies, and the point with steepest gradient begins to shift.

Finally, we can estimate the quasiparticle weight from the discontinuity at the Fermi surface. As discussed in the main text, this cannot be computed directly from our data because three different effects all prevent an actual

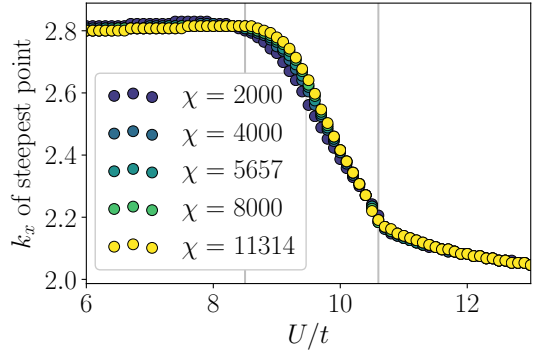

FIG. S17. (Color online) Location ($k_x$) of the point on the $k_y = 0$ line with the steepest change in $\langle n_{k\uparrow} \rangle$, plotted vs $U/t$ for various MPS bond dimensions; vertical lines are at $U/t = 8.6$ and 10.6. The location of the Fermi surface is fixed by Luttinger's theorem inside the metallic phase, so the metal-insulator transition must occur at or before $U/t = 8.6$.

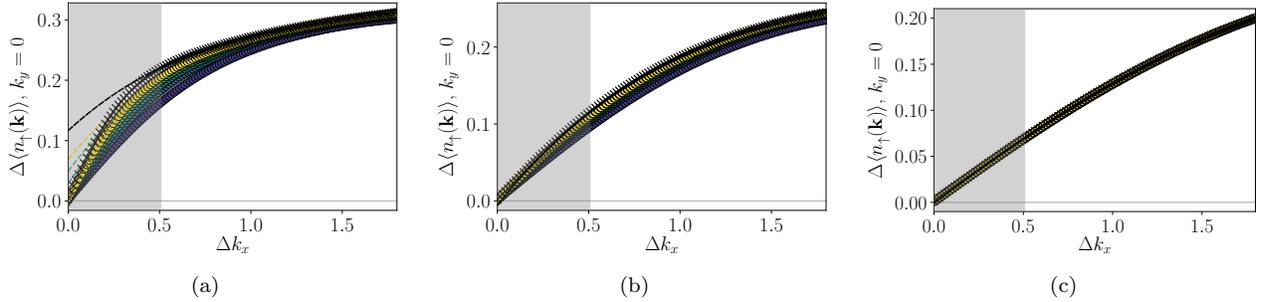

FIG. S18. (Color online) Estimating the quasiparticle weight from the gap in occupation $\Delta n$ across a finite interval $\Delta k_x$ around the Fermi surface; the data at very small $\Delta k_x$, in the shaded region, is not reliable due to finite cylinder circumference and finite range of correlation function calculations, but we fit a cubic polynomial to the remaining data and extrapolate to $\Delta k_x = 0$ to get the quasiparticle weight. This is shown in the main text for $U/t = 6$ and here for $U/t =$ **(a)** 8, **(b)** 9, and **(c)** 12. The cubit fit is shown by a dashed line for each bond dimension and for the extrapolated data. That the extrapolation curves are not visible for $U/t = 9$ and 12 is because there is no $\Delta k$ beyond which the $\Delta \langle n \rangle$ vs $\Delta k$ curve qualitatively changes, since that qualitative change and the visible "corner" in the function is the result of nonzero quasiparticle weight.

discontinuity: one-dimensionality of the cylinder, finite range of real-space correlations used in the calculation, and finite bond dimension. As demonstrated above, bond dimension extrapolation restores at least the Luttinger liquid singularity, so that effect is easily dealt with. For the other two, we can consider the gap in occupation across a finite interval around the Fermi surface for a range of interval sizes. As long as the interval is not too small, the data should not be affected by either of the first two issues, and we can then extrapolate to $\Delta k_x = 0$. This is shown for $U/t = 6$ in Figure 4(d) of the main text, and here in Figure S18 we also show the equivalent results for $U/t = 8$ just below the the metal-insulator transition, $U/t = 9$ just above it, and $U/t = 12$ in the high-$U$ phase. The precise method used for the extrapolation has little to no effect either at very low or very high $U$, but does have some effect near the transition. Our goal was to capture any curvature in the data for large enough $\Delta k_x$ to be reliable, while throwing away the spurious behavior at very small $\Delta k_x$. We selected a cutoff value that, for all $U/t$, is above the visible "corner" in the $\Delta \langle n \rangle$ vs $\Delta k_x$ curve, then found the best-fit cubic polynomial for the data from that cutoff until the $\Delta k_x$ interval reached the edge of the Brillouin zone. In computing the fit, we have weighted the data so that the smallest $\Delta k_x$ above the cutoff are given full weight and then the weights decrease linearly to 0 at the edge of the Brillouin zone; this is important because the true curve is not cubic, and with equal weights the curve does not match even the included points just before the cutoff, while with this weighting scheme the full curve visibly matches quite well. These extrapolation curves are shown as dashed lines in Figure S18. Figure 4(b) in the main text shows our final estimate of the quasiparticle weight as a function of $U/t$.

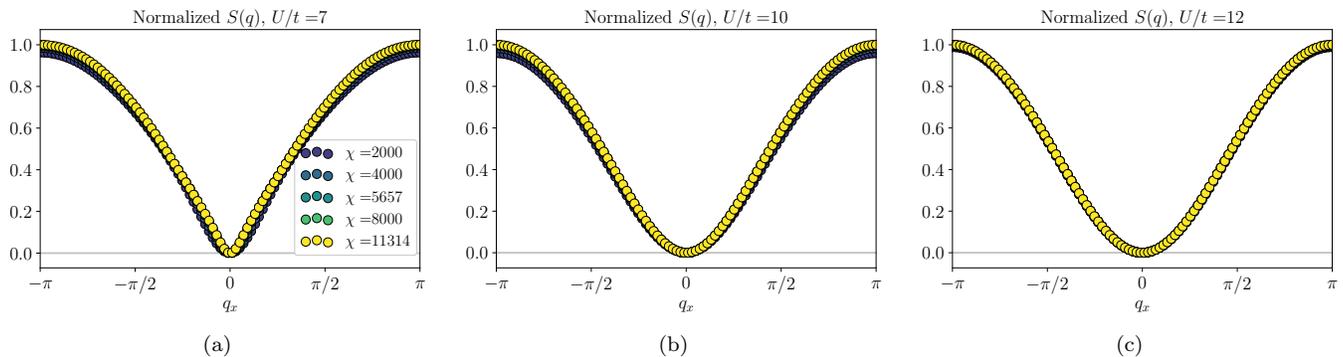

FIG. S19. (Color online) **(a)** $S(q_x)$ at $U/t = 7$, normalized so the maximum value for the highest bond dimension is 1. The dependence on $q_x$ is approximately linear near the origin. **(b)** $U/t = 10$. The dependence on $q_x$ is quadratic. **(c)** $U/t = 10$. The dependence on $q_x$ is again quadratic.

### 2. Density structure factor

In the main text, we also observe the metal-insulator transition using the density structure factor. Here, we provide some details of that calculation. The structure factor is defined by

$$S(\mathbf{q}) = \langle \delta n(\mathbf{q}) \delta n(-\mathbf{q}) \rangle = \frac{1}{L^2} \sum_{\mathbf{x}, \mathbf{y}} e^{-i\mathbf{q} \cdot (\mathbf{x} - \mathbf{y})} \langle \delta n(\mathbf{x}) \delta n(\mathbf{y}) \rangle, \tag{9}$$

where $\delta n(\mathbf{x}) = n(\mathbf{x}) - \langle n(\mathbf{x}) \rangle = n(\mathbf{x}) - 1$. Near $\mathbf{q} = 0$, the structure factor should satisfy $S(\mathbf{q}) \propto q^2$ if the state has a charge gap and $S(\mathbf{q}) \propto |q|$ if it does not.[12,13] Both behaviors can be captured by the functional form

$$S(\mathbf{q}) \propto \sqrt{q^2 + m^2} - m, \tag{10}$$

for an "effective mass" $m$ that goes to 0 when the state is gapless. Up to the proportionality constant in equation (10), $m$ is the inverse curvature of the function at $k = 0$.

To distinguish between the gapped and gapless behavior, it is sufficient to consider a one-dimensional cut through $S(\mathbf{q})$, namely $S(q_x, q_y = 0)$, and this has the benefit of being quite efficient to compute in our mixed-space MPS. Our local sites are labeled by $(x, k_y)$, and in terms of the operators $\hat{n}_{x,k_y} = c^\dagger_{x,k_y} c_{x,k_y}$ the $q_y = 0$ structure factor is

$$S(q_x) \propto \sum_{xkk'} e^{iq_x x} \left( \langle \hat{n}_{0k} \hat{n}_{xk'} \rangle - \langle \hat{n}_{0k} \rangle \langle \hat{n}_{xk'} \rangle \right). \tag{11}$$

We compute the $\langle \hat{n} \hat{n} \rangle$ correlations out to a distance of 200 rings of the cylinder; even at the largest bond dimension (11314) and the smallest $U$ (with the slowest decay of the correlations), we find that numerical error is larger than the actual correlations beyond 125 rings, so this range is more than enough for accurate results.

In Figure S19 we show $S(q_x, q_y = 0)$ for a range of bond dimensions, with the structure factor for $U/t = 7$, 10, and 12, in the metallic, intermediate, and ordered phases, respectively; for each $U$, $S(q_x)$ is normalized so that the maximum is 1 for the highest bond dimension. Qualitatively, it is clear that the dispersion at $U/t = 7$ looks linear whereas at the two higher values of $U$ it appears quadratic, which confirms that the metal-insulator transition is between 7 and 10. To get a clearer picture of the transition, we fit the part of the structure factor with $|q_x| \leq \pi/6$ to a curve of the form (10) and extrapolate the mass $m$ as a power law in the DMRG truncation error. In Figure 4(a) of the main text, we show the mass computed for each $U$ and $\chi$, as well as an extrapolation to infinite bond dimension. The metal-insulator transition appears to occur at $U/t \approx 8.5$, roughly consistent with the rest of our analysis.

### D. Finite entanglement scaling

In the main text, we show in Figure 5(a) and (b) the central charge, a characteristic property of the conformal field theory describing a gapless one dimensional system, computed as a function of $U/t$ using the scaling relations[14]

$$S \approx (c/6) \log(\xi). \tag{12}$$

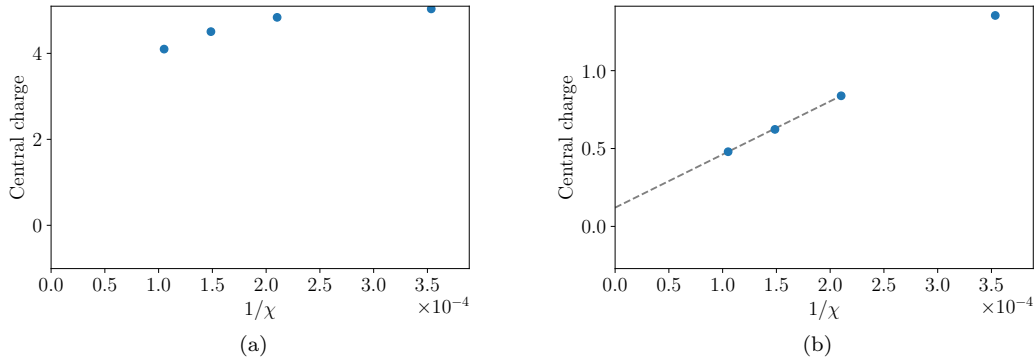

FIG. S20. Central charge vs $1/\chi$ for **(a)** $U/t = 9$ and **(b)** $U/t = 13$. Both seem to extrapolate to finite values, but this is misleading, as discussed in the text.

and[15–17]

$$S \approx \left(1 + \sqrt{12/c}\right)^{-1} \log(\chi),$$ (13)

respectively.

Both approaches clearly show a finite central charge at low $U$, with a value of approximately $c = 6$, which is exactly what we expect for a metallic state on this cylinder as discussed in section II above. At high $U$ the latter approach clearly shows $c = 0$ while the former requires some extrapolation of the finite bond dimension results. In the intermediate phase the second approach gives $c = 0$ at least for $U/t \gtrsim 9$, but the behavior between about $U/t = 8$ and 9 is unclear. To provide some further clarification, we discuss here the precise way that we compute the central charge and discuss the extrapolation to the true, infinite bond dimension ground state.

We first explain further how we calculate the central charge. At each $U/t$ and each bond dimension, we can calculate the total entanglement entropy $S$ for a cut between two rings of the cylinder, and also the correlation length (Figure 3(b) of the main text). As both become large, they should scale according to equation (12), but the relation will be inaccurate when both quantities are small because non-universal offsets will be comparatively large. The coefficient is thus best approximated by the derivative, $c/6 \approx d\log(\xi)/dS$, and we calculate discrete approximations to this derivative from the values of $S$ and $\xi$ at successive bond dimensions; the lines in Figure 5(a) and (b) of the main text are labeled by the larger of the two bond dimensions used in calculating the discretized derivative. So for example the yellow (most accurate) line in the figure is computed using the ground state wave functions for bond dimensions 8000 and 11314.

In Figure S20, we show the central charge estimates using equation (12) at $U/t = 9$ and at $U/t = 13$ versus $1/\chi$, where the $\chi$ used is the geometric mean of the two bond dimensions used to calculate the derivative. In the high-$U$ phase we show a linear extrapolation to infinite bond dimension; although it appears to show a small but nonzero central charge, that is not really a reliable result. For example, the use of the geometric mean of the two bond dimensions used in computing the derivative is an arbitrary choice, particularly because the error may be determined mostly by the smaller bond dimension, and using that bond dimension for the horizontal axis would shift the graph to the right. At $U/t = 9$, it is essentially impossible to extrapolate the central charge at all as the shape of the curve is completely unclear.

In fact, in both phases, and particularly in the high-$U$ phase, the entanglement appears to be converging with increasing bond dimension, as shown in Figure S21, which is indicative of a gapped phase that should have $c = 0$. The apparent nonzero central charge comes from the fact that both the entanglement and the correlation length are still growing very slightly at the accessible bond dimensions.

Another option, then, is to compute the central charge directly from the scaling of entanglement with bond dimension, using equation (13). In the gapless low-$U$ phase, this is less accurate than the computation of $c$ from scaling with $\xi$, but at higher $U$ it is indeed more converged. Slices at $U/t = 9$ and 10 in the intermediate phase and $U/t = 13$ in the high-$U$ phase are shown in Figure S22. It is still not completely clear that system is gapped at $U/t = 9$, but that is likely a finite bond dimension effect, as the later point in the same phase, at $U/t = 10$, clearly shows $c = 0$.

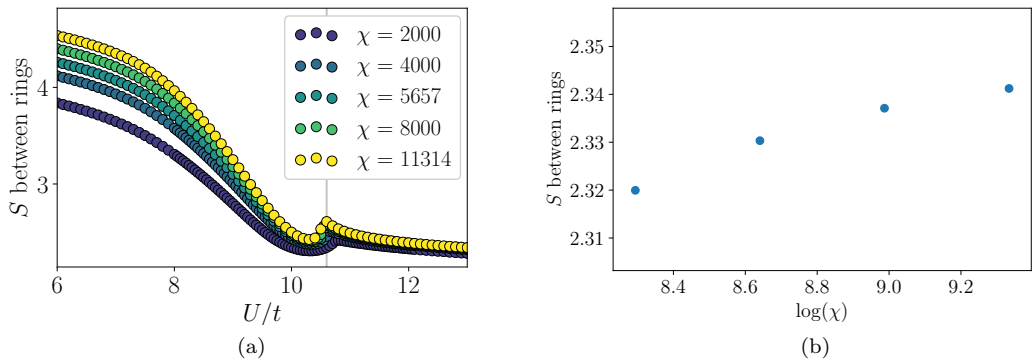

FIG. S21. (Color online) **(a)** Entanglement between rings of the cylinder as a function of $U/t$ for different bond dimensions. It is nearly converged in the intermediate and high-$U$ phases, indicating that they are gapped. **(b)** A close-up slice at $U/t = 13$. Note that the vertical scale is only about 2% of the value of $S$.

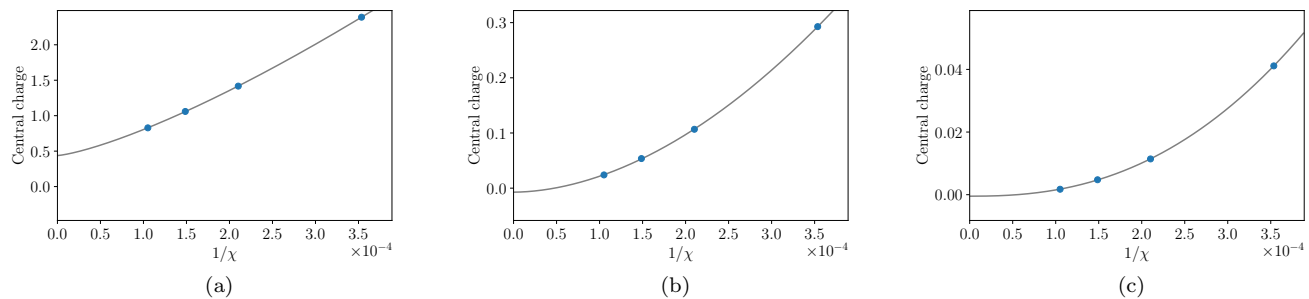

FIG. S22. Central charge vs $1/\chi$, calculated by scaling with bond dimension, for **(a)** $U/t = 9$, **(b)** $U/t = 10$, and **(c)** $U/t = 13$. The latter two clearly extrapolate to 0, suggesting that both the intermediate and high-$U$ phases are gapped. The extrapolation at $U/t = 9$ still appears to go to a nonzero value (approximately $1/2$), but this is likely still a finite bond dimension effect.

### E. Entanglement spectrum degeneracy

As demonstrated in Figure 5(c) in the main text, the entanglement spectrum of the ground state on the YC4 cylinder acquires an exact two-fold degeneracy when entering the intermediate phase, at $U/t \approx 8.4$. Here we offer some further details. To begin, it is helpful to see the full low-lying entanglement spectrum as a function of $U/t$, shown in Figure S23.

The two lowest-lying levels appear to come together somewhere in the vicinity of $U/t = 8$, and then pairs of levels come together at $U/t = 10.6$. This onset of four-fold degeneracy from two-fold degeneracy at $U/t = 10.6$ is visually obvious in the figure: each pair of lines that come together at that point do so at a sharp angle, so that the slope of the entanglement spectrum lines appears discontinuous at that point. The precise location of the first transition, from a nondegenerate entanglement spectrum to two-fold degeneracy, is not clearly visible in the same way.

To more precisely find where the two-fold degeneracy onsets, we take all of the entanglement levels for a given value of $U/t$ and a given bond dimension, and group them into adjacent pairs, with the lowest two levels together, the third and fourth together, and so forth. We then find the separation within each pair, and average the separation over the lowest $N$ pairs, for some large $N$. (We do not average over all pairs because the highest ones will be inaccurate for any finite bond dimension.) The logarithm of this average can be plotted vs $U/t$ for each bond dimension, which is shown for $N = 1000$ in Figure 5(c) of the main text. The curves for different bond dimensions all sit roughly on top of one another until around $U/t = 8$, where they start to deviate. For each bond dimension the separation drops towards 0 before flattening off at some finite average separation; as bond dimension increases, this flattening out happens at successively smaller separations. It is still difficult to identify the exact onset of degeneracy in the infinite bond dimension limit, but it appears to be somewhere within the region highlighted by the vertical gray bar, from $U/t = 8.2$ to $8.6$.

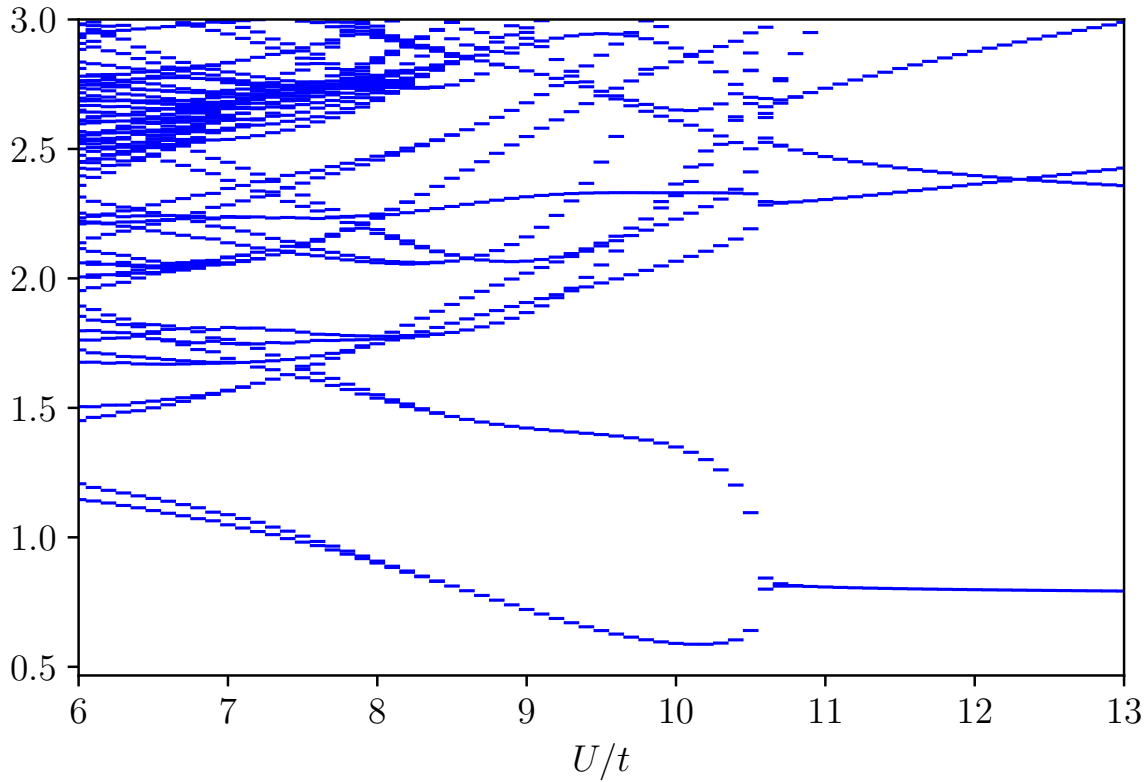

FIG. S23. Entanglement spectrum in the ground state of the YC4 cylinder, as calculated for bond dimension $\chi = 8000$.

For confirmation that this is indeed what a finite bond dimension approximation to an exact degeneracy should look like, we have also followed the same procedure with entanglement levels divided into groups of four, plotting the average separation between the highest and lowest levels in each group, which should go to 0 at the onset of four-fold degeneracy. This is shown in Figure 5(d) of the main text. This indeed shows essentially the same behavior at $U/t = 10.6$ as does the average pair splitting at $U/t \approx 8.4$, so we believe this is a valid and relatively rigorous way to locate the onset of degeneracy.

We also briefly comment on the reason for the observed degeneracy in the entanglement spectrum. The two-fold degeneracy in the intermediate region is expected for the semion sector of the chiral spin liquid as explained in the main text. In the high-$U$ region, the ground state in the full two-dimensional model is the 120°-order magnetic phase, but on the finite circumference cylinder it becomes a symmetry protected topological phase where the relevant symmetries are the space-group symmetries of the cylinder. The degeneracy in the entanglement spectrum then arises because the edge states (and therefore the entanglement spectrum) carry projective representations of the space group. The resulting degeneracy depends on the particular cylinder geometry; for example, the YC4 cylinder as shown here exhibits a four-fold degeneracy, while the YC6 cylinder does not. While this physics is quite interesting, it is not directly relevant to the chiral spin liquid phase in which we are interested, and thus we do not explore the details further.

### F. Chiral Domain Wall

As there is a two-fold degeneracy arising from the two possible chiralities, at nonzero temperature there will be regions of each chirality with domain walls between them. We characterize these domain walls by considering a finite cylinder segment, assuming an infinitely repeating MPS on each end; these half-infinite MPSs are taken from the ground states previously found for each of the two chiralities, with opposite chiralities at the two ends. Within the finite cylinder, we optimize the MPS tensors to minimize the energy as usual, resulting in the minimal energy

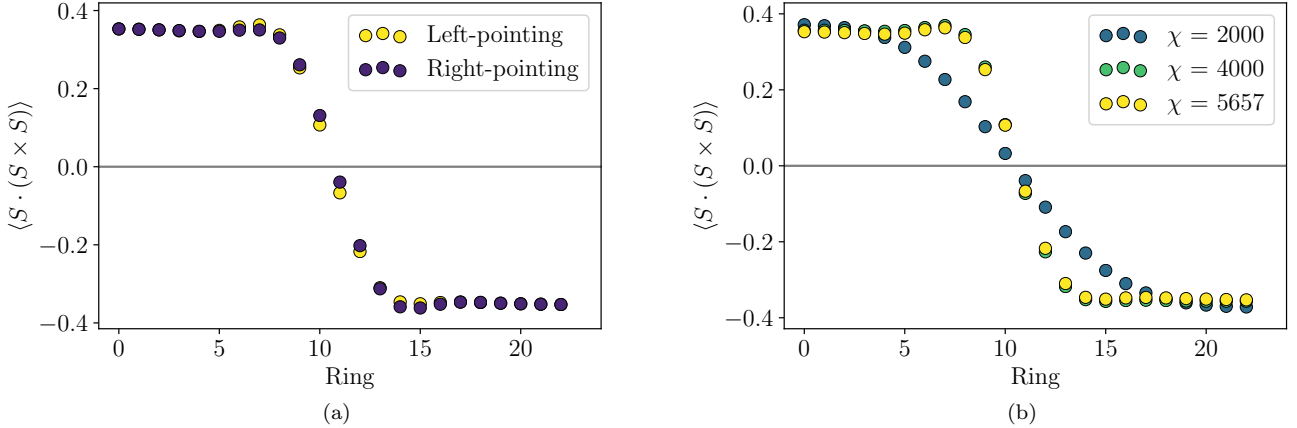

FIG. S24. (Color online) Shape of the domain wall between regions of the two different chiralities, as measured by the chiral order parameter for local triangles on each ring of a finite cylinder segment: **(a)** computed for $\chi = 5657$ for both triangle orientations, and **(b)** for the left-pointing triangles at three different bond dimensions, showing convergence by $\chi \approx 4000$.

configuration that interpolates between the two chiralities. This is the same method used in reference [6].

Here we show results for the YC4 cylinder at $U/t = 10$, deep in the chiral phase. The shape of the domain wall can be observed by calculating the chiral order parameter $\langle \mathbf{S} \cdot (\mathbf{S} \times \mathbf{S}) \rangle$ on successive triangles proceeding along the finite segment. On each ring of the cylinder, there are two inequivalent triangles, one which points left, and one which points right (see Figure 1 of the main text), and we calculate the order parameter on both. Note that all triangles in a particular ring with a given orientation are necessarily equivalent in our mixed real- and momentum-space basis. Figure S24a shows the evolution of the chiral order parameter across a finite segment of length 24 rings, with both triangle orientations; note that at each end the chiral order parameters for the two triangle orientations converge to a single constant value, the same one shown in Figure 3(e) of the main text. Figure S24b shows the dependence of the domain wall shape on the MPS bond dimension; although the behavior is quite different comparing the low bond dimension of 2000 with the higher ones, it appears to be essentially converged by $\chi = 4000$.

We can additionally calculate the energy cost of creating the domain wall by comparing, for each bond dimension, the minimized energy with the domain wall present with the minimized energy at the same bond dimension with just a single domain. The total energy cost of the domain wall on the YC4 cylinder, which is nearly converged for $\chi \gtrsim 4000$, is approximately $0.026t$, giving a domain wall tension of $0.0065t/a$. Using estimated values from the literature[18,19], this gives for the organic crystal $\kappa$-(BEDT-TTF)$_2$Cu$_2$(CN)$_3$ a tension of about $(4K) \times k_B$.

## G. Flux insertion data

In the main text, we show in Figure 6 the evolution of several quantities with flux insertion, including the chiral order parameter, the peak height of the spin structure factor, and the inverse correlation length from the MPS transfer matrix spectrum for operators with quantum numbers $(q, S_z) = (0, 0)$, which gives a qualitative estimate of the spin singlet gap. We also show the spin Hall effect in the chiral spin liquid (CSL) phase via spin pumping with flux insertion in Figure 13(d) and the spin- and momentum-resolved entanglement spectrum in the two topological sectors of the CSL in Figure 13(b). Here we provide some additional data on the evolution of the system with flux insertion, including fidelity, inverse correlation lengths in other quantum number sectors, and the evolution of the entanglement spectrum with flux insertion.

In Figure S25 we show a variety of quantities versus $U/t$ and flux insertion $\theta$. In particular, we show (a) the inverse correlation length for excitations with $q = 0$, $S_z = 1$, a proxy for the spin triplet gap that shows the distinction between the two high-$U$ phases; (b) the inverse correlation length for excitations with $q = 1$, showing the metal-insulator transition; (c) the fidelity with flux insertion, showing some phase transitions clearly; (d) the entanglement between rings; and (e) the truncation error in the last sweep of the DMRG simulation, showing that the accuracy of the simulation is essentially constant with flux insertion within each phase. All quantities clearly delineate the chiral phase.

We can also further examine the spin pumping from flux insertion at $U/t = 10$ shown in Figure 13(d) of the

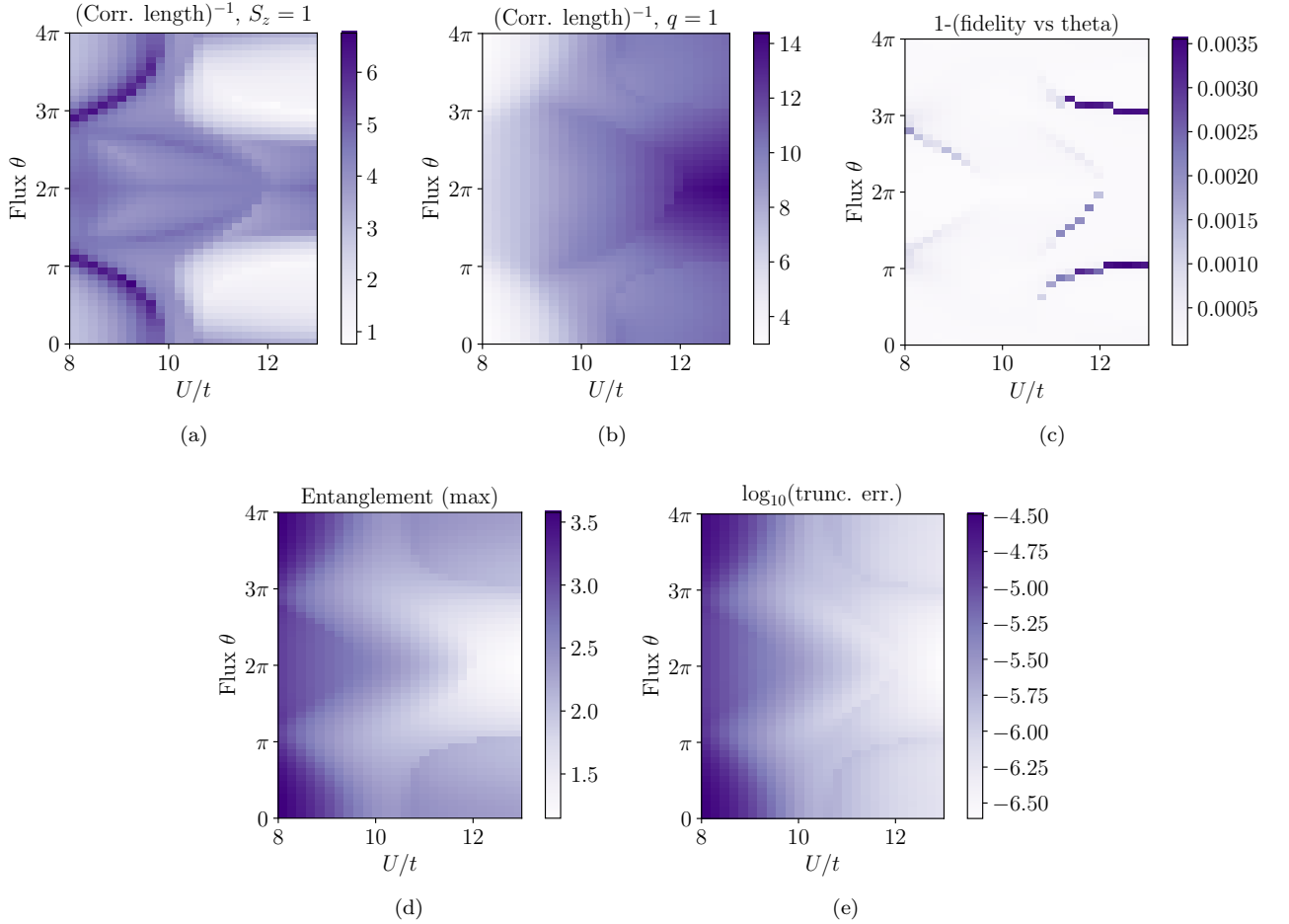

(a)

(b)

(c)

(d)

(e)

FIG. S25. (Color online) Various quantities for the YC4 cylinder vs $U/t$ and flux insertion $\theta$, with $\chi = 4000$. **(a)** Inverse correlation length for operators with $S_z = 1$, giving an estimate of the spin triplet gap. **(b)** Inverse correlation length for operators with $q = 1$, giving an estimate of the charge gap. **(d)** Fidelity of the flux insertion—in particular, we show $1 - |\langle\psi(\theta)|\psi(\theta + \pi/12)\rangle|^2$, so that a very large value would indicate a loss of adiabaticity. We see phase boundaries but no such non-adiabatic evolution. **(d)** Entanglement between two rings of the cylinder. **(e)** Log base 10 of the truncation error at the last step of the DMRG simulation.

main text. Compared with the spin pumping for YC6 (Figure 13(e)), there is substantially more deviation from a straight line, which arises because of the large shifts in the phase boundaries with flux insertion. In particular, the spin is pumped more quickly when the chiral order parameter is larger, and less quickly when it is smaller. This phenomenology can be seen clearly by looking at spin pumping at smaller and larger values of $U$ which clearly cross phase boundaries; in Figure S26 we show the spin pumping for $U/t = 9.2$ and $U/t = 11$. At $U/t = 9.2$, the system is in the chiral phase until shortly before $2\pi$ flux, where it enters the left-most nonchiral region seen in Figure 6(a) of the main text, at which point the pumped spin plateaus at exactly $1/2$. The spin pumping continues upon reentering the chiral phase. (Note that for smaller $U$, since the sign of the chirality is randomly opposite upon reentering that phase, the spin pumps back down to 0 at $4\pi$ flux instead of up to 1.) At $U/t = 11$, there is initially no spin pumping, but it begins after entering the chiral phase; the pumping stops again once the system once more enters the nonchiral phase after $3\pi$ flux.

## H. Transfer matrix and the two-dimensional excitation spectrum

In section IV above, we discuss how the MPS transfer matrix spectrum gives the excitation spectrum of the system. In particular, for a given $U$, flux insertion allows us to trace out a cut through the minimum of the excitation spectrum as a function of $k_x$ and $k_y$. In Figure S27 we show the transfer matrix spectrum for spin 1 excitations at $U/t = 10$

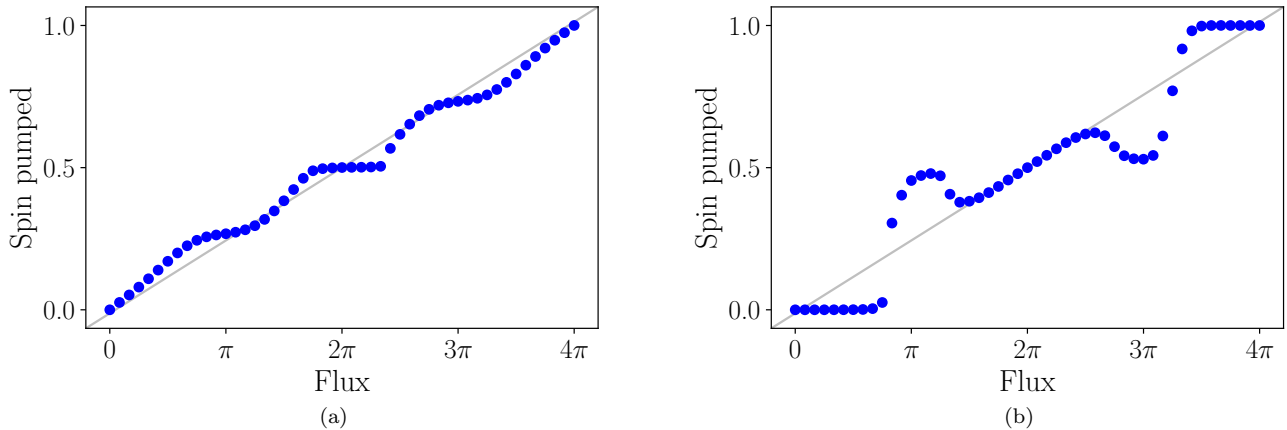

FIG. S26. (Color online) YC4 spin pumping: **(a)** $U/t = 9.2$ and **(b)** $U/t = 11.0$.

versus flux insertion, $k_x$, and $k_y$, at bond dimensions 2000, 4000, and 8000. Note that for this data, flux insertion was not performed adiabatically, but rather the DMRG was performed independently for each $\theta$; however, adiabatic flux insertion at $\chi = 4000$ gives a nearly identical result.

This data is somewhat difficult to interpret due to the shifting of phase boundaries with flux insertion. Another possibility is to, for each flux $\theta$, take the transfer matrix spectrum corresponding to the $U/t$ with the greatest chirality in order to find the transfer matrix within a single phase. This is shown (for $\chi = 4000$) in Figure S28.

### 1. Comparison with Dirac spin liquid

One purpose of this analysis is to check whether the Dirac spin liquid (DSL) is another candidate state for the two-dimensional model, which seems plausible given that the DSL can become a chiral spin liquid when the Dirac cones are gapped out. A DSL on the triangular lattice can be constructed with fermionic partons with staggered $\pi$ flux. There are several valid gauge choices, one of which is the model shown in Figure S6a above; the three inequivalent gauge choices give Dirac cone pairs at $(k_x, k_y) = \pm(\pi/\sqrt{3}, 0)$, $\pm(\pi/2\sqrt{(3)}, \pi/4)$, and $\pm(\pi/2\sqrt{(3)}, -\pi/4)$. The momenta of the corresponding transitions are indicated by vertical lines in Figures S27 and S28.

When plotted at constant $U/t = 10$, the minima in the spectrum do not appear to line up with the transitions between expected locations of the Dirac cones. When plotted instead at the approximate center of the chiral phase, the minima are in fact at $(\Delta k_x, \Delta k_y) = \pm(2\pi/\sqrt{3}, 0)$, which corresponds to one of the Dirac cone pairs. However, this result is at a single bond dimension, and these minima do not extend far below the rest of the spectrum, so it is difficult to reach a clear conclusion.

## VI. YC6 ADDITIONAL DATA AND ANALYSIS

### A. Metal to CSL transition in $k = 0$ sector of YC6 cylinder

For the YC4 cylinder, we used finite entanglement scaling to show that the metallic phase is gapless with $c = 6$ and that the intermediate and high-$U$ phases on that cylinder are gapped. For the YC6 cylinder, much larger bond dimensions (about $16\times$ larger) are needed to achieve the same accuracy, so we cannot estimate the central charge accurately. However, if we plot the central charge as estimated using equation (13) for pairs of bond dimensions (as described in section V D) as a function of $U/t$, we can still observe a qualitative change in the behavior of the system at one particular point, as shown in Figure S29a. This behavior is consistent with a gapless metallic phase at low $U$ and a gapped phase at intermediate $U$.

The transition can also be observed in the entanglement spectrum. Metallic phases charactistically have very densely spaced levels, which as shown in Figure S29b is true for the YC6 cylinder when $U/t \lesssim 8$ but is no longer true beyond that point. Just to the left of that same point, there is a corresponding rapid increase in the separation of the

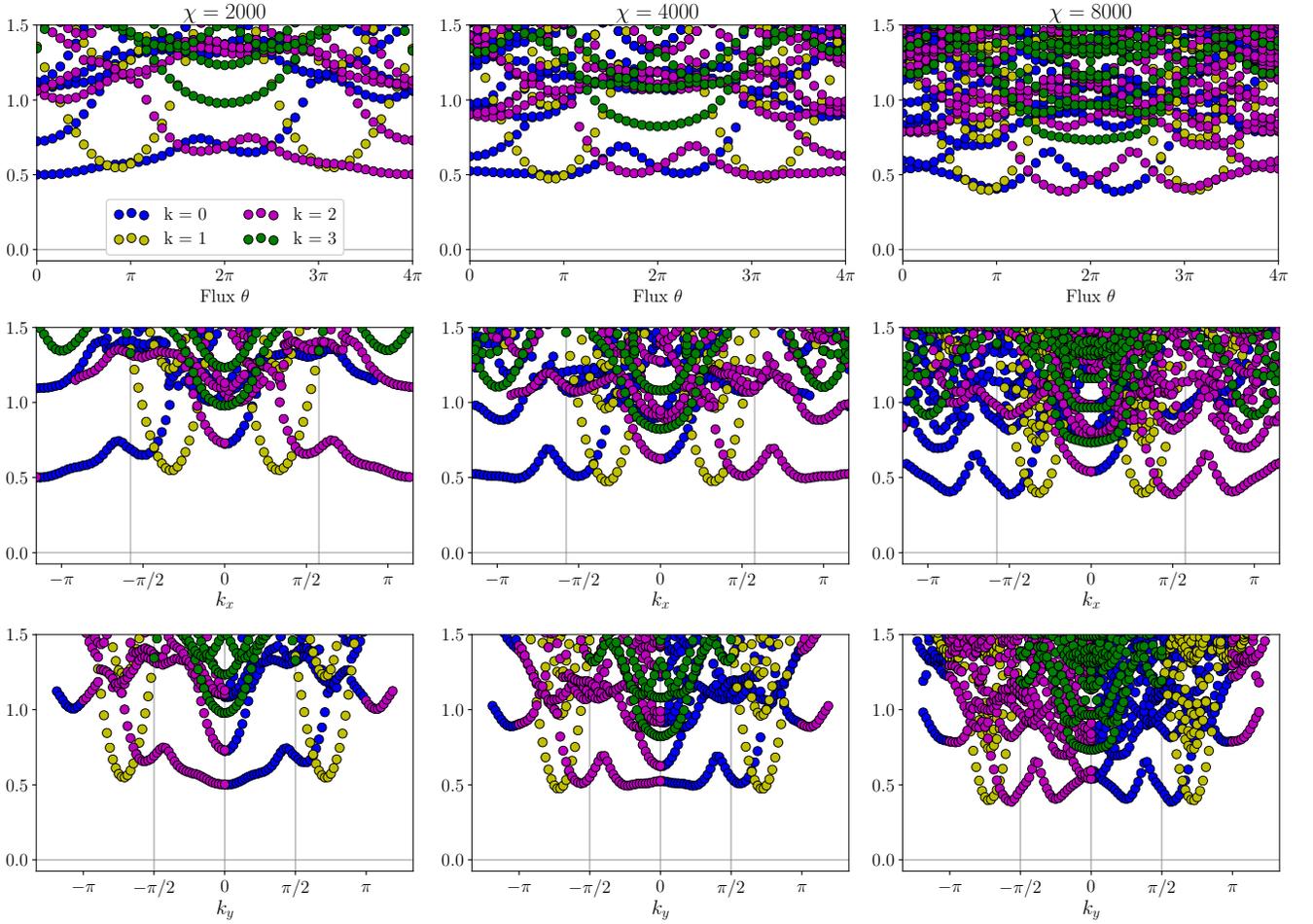

FIG. S27. (Color online) Low-lying part of the $S_z = 1$ part of the transfer matrix spectrum for $U/t = 10$ on the YC4 cylinder at $\chi = 2000$ (left column), 4000 (middle column), and 8000 (right column), plotted versus flux inserted (top row), $k_x$ (middle row), and $k_y$ (bottom row). (The lattice constant is set to 1.) Color corresponds to the momentum quantum number of the excitation as given in the panel in the upper left. Vertical lines correspond to momentum transfers between possible Dirac cone locations in the Dirac spin liquid; they do not clearly correspond to features observed in our data.

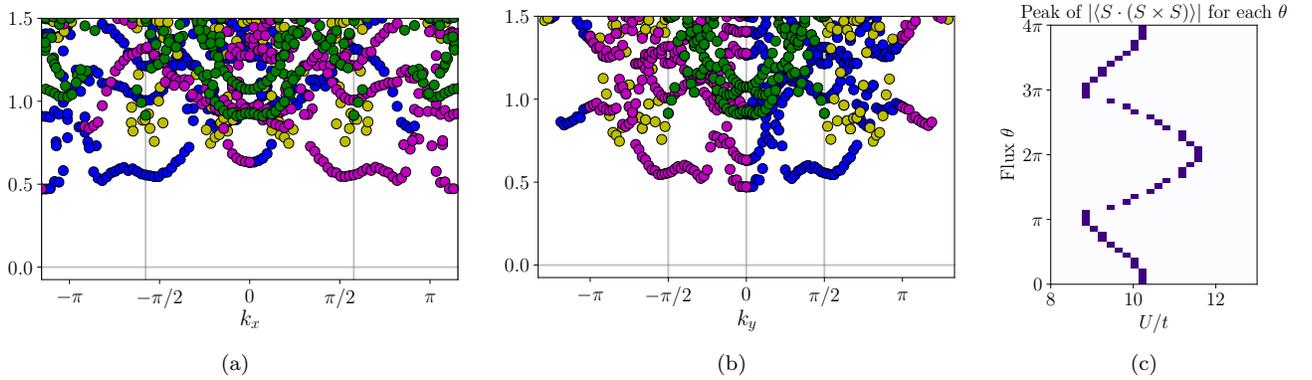

(a)                                      (b)                                      (c)

FIG. S28. (Color online) Low-lying transfer matrix spectrum where, for each amount of flux inserted, $\theta$, we use the $U/t$ with the largest chiral order parameter to "track" the chiral phase. **(a)** Spectrum plotted versus $k_x$. **(b)** Spectrum plotted versus $k_y$. **(c)** Location of the maximum $U/t$ used for each $\theta$.

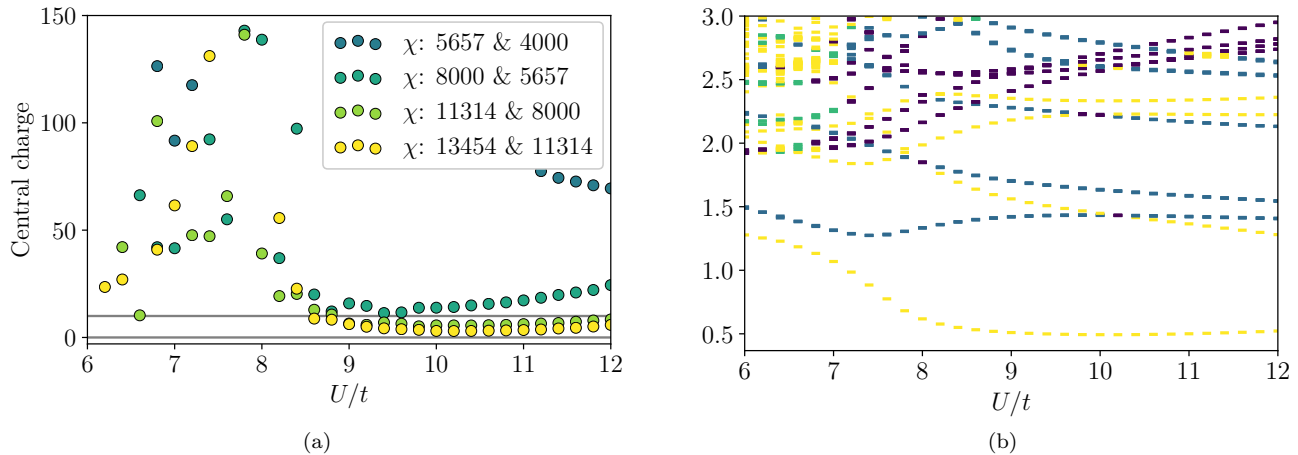

FIG. S29. (Color online) Evidence for metal to spin liquid transition in the $k = 0$ sector for the YC6 cylinder: **(a)** At $U/t \approx 8.5$, there is a qualitative change in the behavior of the finite entanglement scaling. To the left the scaling of entanglement with bond dimension appears chaotic, which is not surprising for a gapless system when the bond dimension is very small compared with the size of the Hilbert space; to the right the behavior becomes systematic, because the DMRG converges much more accurately even at low bond dimensions when the system is gapped. **(b)** The same transition is also visible in the entanglement spectrum, plotted here for $\chi = 8000$. The dense levels in the upper left are characteristic of a metallic state. At $U/t \approx 7.5$ there is a large increase in the separation between the lowest levels, and the low-lying levels become much more sparse in general, showing a transition into a non-metallic state. (Note that the coloring indicates degeneracy of the levels: yellow is non-degenerate and blue is 3x degenerate).

lowest levels; on its own, that feature would not be sharp enough to identify a transition, but in combination with everything else, it provides some additional evidence for the location of the transition.

The two indicators of a transition are slightly displaced from each other, but both are in the vicinity of $U/t \approx 8$, so we identify that region as the approximate location of the transition.

## B. Four-fold ground state degeneracy

As shown in Figure 7(a) of the main text, for the YC6 cylinder we find low-lying states in two different sectors of momentum around the cylinder per ring. In addition to this near-degeneracy, which as explained in the main text is between the trivial and semion topological sectors, we also observe an additional two-fold degeneracy between the two different possible chiralities.

When finding the ground state using the DMRG method, one begins with some initial matrix product state; if the ground state is not degenerate and the algorithm does not get stuck in a metastable state, the final wave function should be approximately independent of the initial state. If, on the other hand, there are multiple degenerate ground states, the algorithm will converge to one or another of them depending on the initial state used. (It will also tend to converge to minimally entangled states within the ground state manifold and not superpositions of them.[6])

In our case, over a wide range of $U/t$ for a bond dimension $\chi = 8000$, we initialized the DMRG with ten different random product states. In the center of the CSL phase, the energies of the final states within each momentum sector varied by up to about 0.01%, meaning that none of the final states were metastable. Although none of the states were numerically identical, they can be separated into two groups within which they are essentially the same, with an overlap per ring of about 0.99998; the overlap between states in opposite groups is about 0.22 per ring. That these two groups correspond to the two possible chiralities of the time-reversal symmetry-breaking phase can be seen from the momentum-resolved entanglement spectra, shown in Figure S30 for representative final states in each of the two groups for each topological sector for $U/t = 9$: the spectra are almost precisely related by $k \to -k$. (Note that parts (a) and (c) are essentially the same as the left and right respectively of Figure 13(a) of the main text; the main text figures were generated by using the ground states shown here as the initial states for DMRG optimization with a larger bond dimension.)

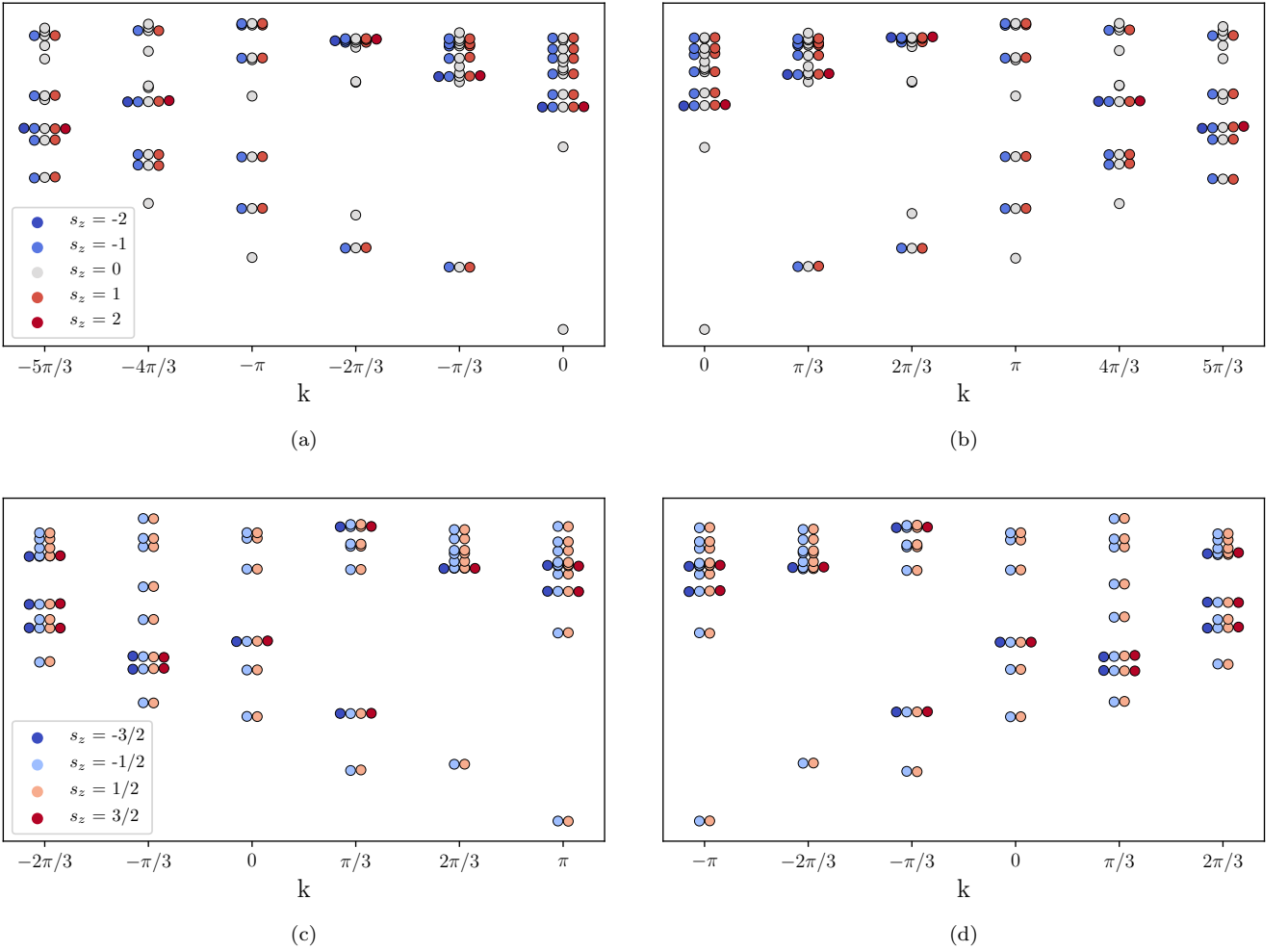

FIG. S30. (Color online) Spin- and momentum-resolved entanglement spectra for the four degenerate ground states of the YC6 cylinder with $U/t = 9$. **(a)** and **(b)** show the two chiralities for the $k = 0$ (trivial) sector, and **(c)** and **(d)** show the $k = \pi$ (semion) sector.

## C. Flux insertion data

In Figure 8 of the main text, we show the chiral order parameter, peak height of the spin structure factor, and inverse correlation length for excitations with $S_z = 0$, giving an estimate of the spin singlet gap. As for the YC4 cylinder in section V G above, we show here in Figure S31 a number of additional quantities versus $U/t$ and inserted flux, $\theta$, with bond dimension $\chi = 8000$. The relative behavior of the various quantities is consistent with what we found for YC4.

Unlike for YC4, here we actually observe a failure of the adiabatic flux insertion, for $U/t = 7$ and 7.2; this is visible in all quantities and is especially clear from the measurement of the overlap between the ground states with successive values of $\theta$, Figure S31c. The overlaps in this case are much smaller than those corresponding to phase boundaries, as seen at higher $U$; this is quite useful because it confirms that those phase boundaries do *not* indicate a loss of adiabaticity. The fact that the chiral phase gives way to a (lower energy) non-chiral phase near $2\pi$ flux might be concerning, but in fact the drop in energy is already much smaller for $U/t = 7.2$ than for $U/t = 7$, and it is likely that this is just the right-most edge of the metallic phase. So this could correspond to the nonchiral region extending to about $U/t = 10$ for YC4, and it is simply easier to maintain adiabaticity with flux insertion for YC6 at $\chi = 8000$ than YC4 with $\chi = 4000$ even with the much larger $\theta$ step, so that the transition into this phase happens later than it ought to. (Additionally, at $4\pi$ flux the nonchiral phase is much higher in energy than the chiral one, again supporting the fact that adiabatic evolution is easier to achieve here.)

We also briefly discuss the evolution of the entanglement spectrum. In Figure 13(a) of the main text, we show

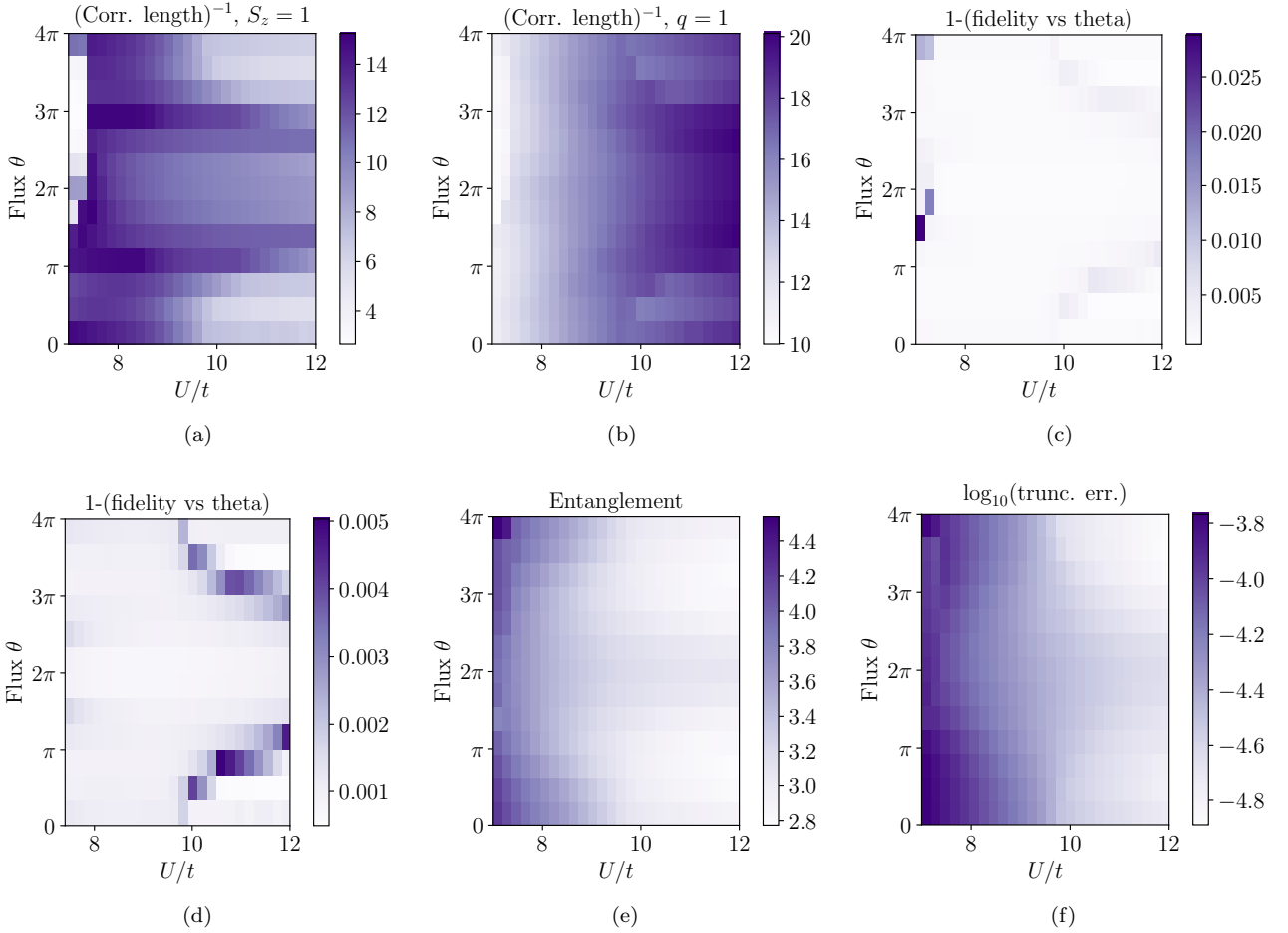

(a)

(b)

(c)

(d)

(e)

(f)

FIG. S31. (Color online) Various quantities for the YC6 cylinder vs $U/t$ and flux insertion $\theta$. **(a)** Inverse correlation length for operators with $S_z = 1$, giving an estimate of the spin triplet gap. **(b)** Inverse correlation length for operators with $q = 1$, giving an estimate of the charge gap. **(c)** Fidelity of the flux insertion, measued by $1 - |\langle \psi(\theta)|\psi(\theta + \pi/3)\rangle|^2$. For the two smallest values of $U$ there is a loss of adiabaticity. **(d)** Fidelity with the left two columns removed to more clearly show the results in the rest of the phase diagram. **(e)** Entanglement between two rings of the cylinder. **(f)** Log base 10 of the truncation error at the last step of the DMRG simulation. The error is quite large even at high $U$, indicating that these are the least converged of our results; this is why the chiral phase appears to extend to high $U$ near $2\pi$ flux.

the spin- and momentum-resolved entanglement spectrum in the trivial and semion sectors; these two sectors are the respective ground states at $U/t = 9$ in the $k = 0$ and $k = \pi$ momentum sectors. Alternatively, we can observe both sectors using flux insertion, as we did for the YC4 cylinder (main text Figure 13(b)). Beginning with the ground state with periodic boundaries in the $k = \pi$ sector, with entanglement spectrum given in Figure S30c, and adiabatically inserting $2\pi$ flux indeed produces an entanglement spectrum consistent with the trivial sector of the CSL, as shown in Figure S32a. An additional $2\pi$ flux converts back to the semion sector, as shown in Figure S32b. The spin multiplets in the spectrum are slightly less well converged after the flux insertion, but the qualitative behavior is exactly the same as at 0 flux.

### D. Transfer matrix and the two-dimensional excitation spectrum

As with YC4, we compute the transfer matrix spectrum for spin 1 excitations versus flux $\theta$, $k_x$, and $k_y$. In particular, we show in Figure S33 the spectrum for $U/t = 9$, deep in the chiral phase, at bond dimension 8000. The relatively low bond dimension given the circumference makes it difficult to reach any meaningful conclusions as to whether the observed state is consistent with a chiral spin liquid formed by gapping out a Dirac spin liquid. Note however that the observed minima at this bond dimension are not consistent with the expected momenta of transitions between

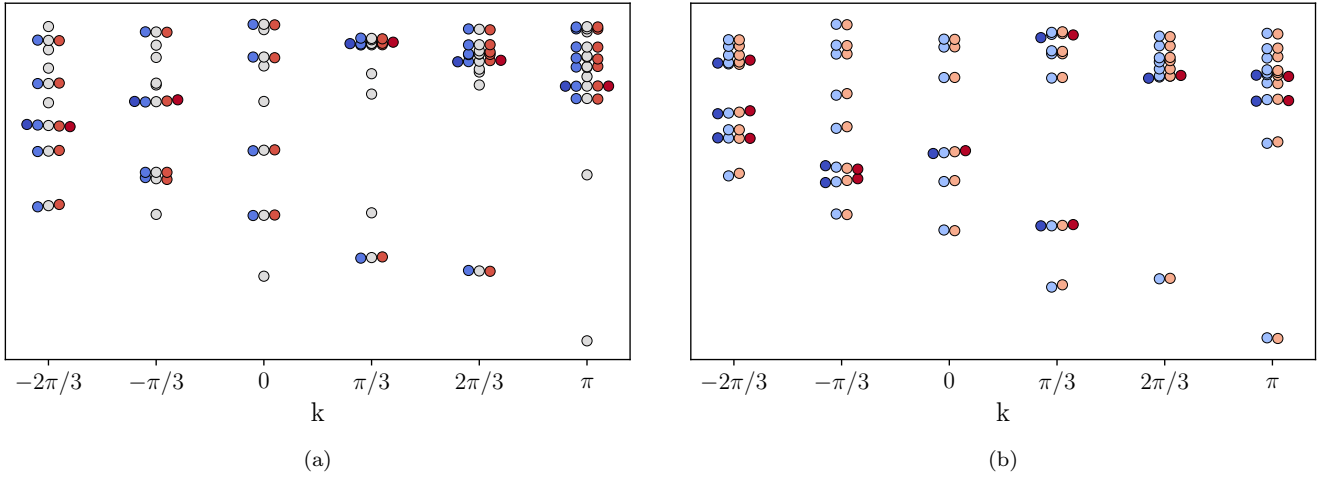

FIG. S32. (Color online) The transformation of the YC6 spin- and momentum-resolved entanglement spectrum with flux insertion for $U/t = 9$, beginning in the semion sector (as shown in Figure S30c): **(a)** $2\pi$ flux converts the system to the trivial sector (compare with Figure S30a) and **(b)** an additional $2\pi$ flux, or $4\pi$ total, returns to the initial state.

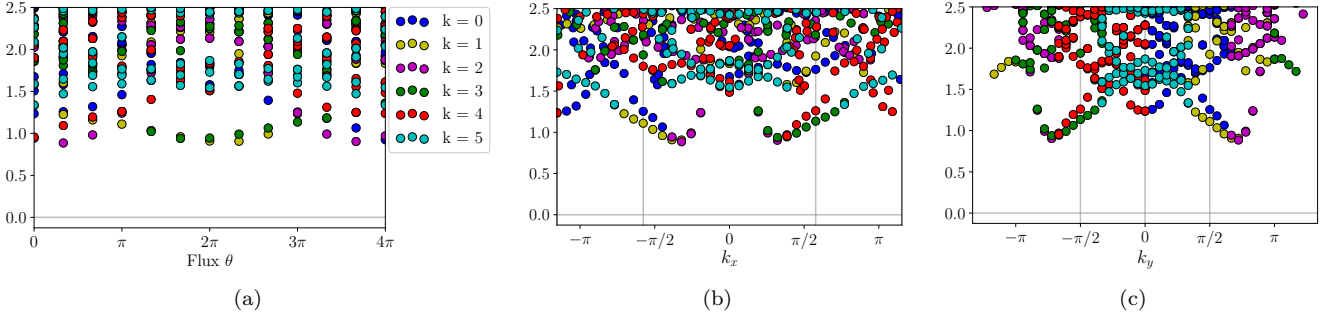

FIG. S33. (Color online) $S_z = 1$ part of the transfer matrix spectrum for the YC6 cylinder with $U/t = 9$, with $\chi = 8000$. Color indicates the momentum quantum number of each transfer matrix eigenvalue. Vertical lines correspond to possible locations of Dirac cones for a Dirac spin liquid. **(a)** Plotted versus flux inserted, $\theta$. **(b)** Plotted versus $k_x$. **(c)** Plotted versus $k_y$.

Dirac cones, though there are local minima at $(k_x, k_y) = \pm(2\pi/\sqrt{3}, 0)$.

## VII. YC5 ADDITIONAL DATA AND ANALYSIS

### A. Flux insertion data

In Figure 9 of the main text, we show the chiral order parameter, peak height of the spin structure factor, and inverse correlation length for excitations with $S_z = 0$, giving an estimate of the spin singlet gap. Here we show in Figure S34 a number of additional quantities versus $U/t$ and inserted flux, $\theta$, with bond dimension $\chi = 8000$. In the chiral phase, the relative behavior of the various quantities is consistent with what we found for YC4 and YC6.

As with YC6, adiabatic flux insertion leads to metastable states at low $U$, as seen by the asymmetry around $2\pi$ flux at low $U$; there is a sudden jump back to the ground state between $3\pi$ and $4\pi$ flux as seen by the line of low fidelity in Figure S34c.

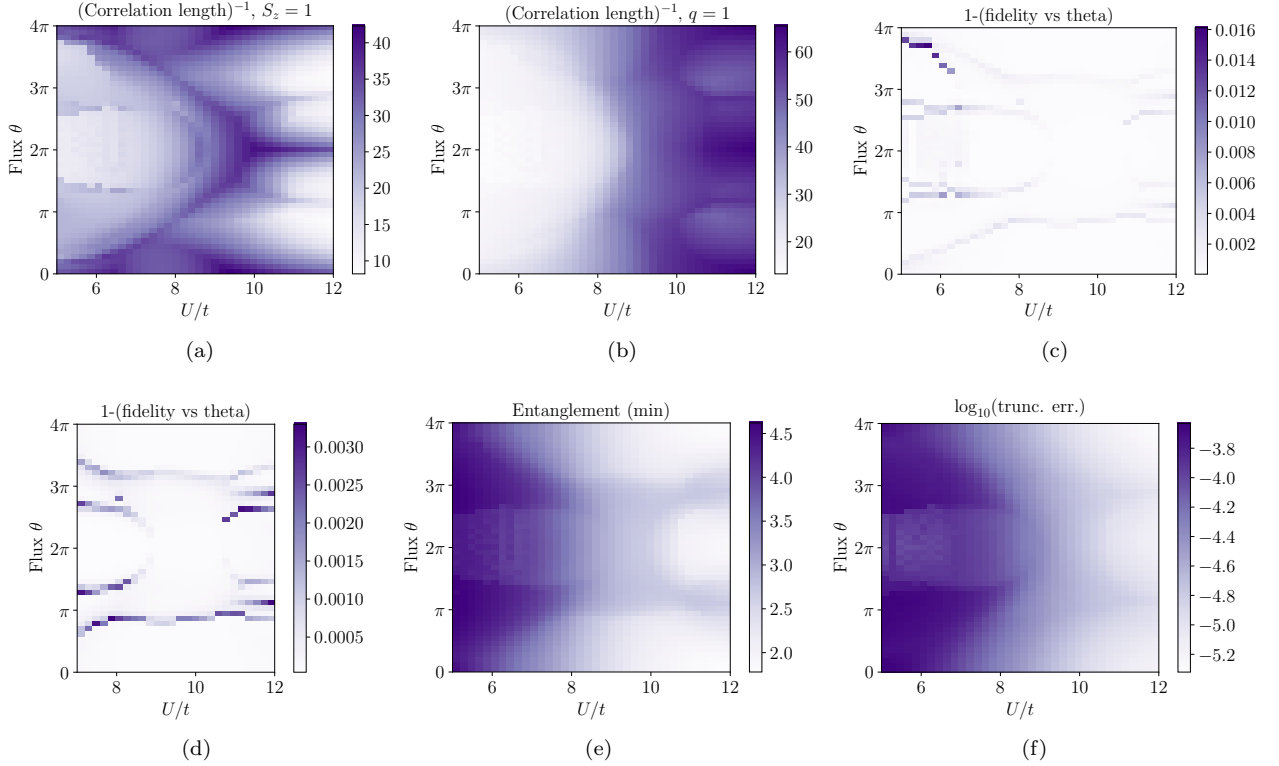

(a)  (b)  (c)

(d)  (e)  (f)

FIG. S34. (Color online) Various quantities for the YC5 cylinder vs $U/t$ and flux insertion $\theta$. **(a)** Inverse correlation length for operators with $S_z = 1$, giving an estimate of the spin triplet gap. **(b)** Inverse correlation length for operators with $q = 1$, giving an estimate of the charge gap. **(c)** Fidelity of the flux insertion, measued by $1 - |\langle\psi(\theta)|\psi(\theta + \pi/12)\rangle|^2$. At small $U$ there is a loss of adiabaticity between $3\pi$ and $4\pi$ flux. **(d)** Fidelity with the leftmost columns removed to more clearly show the results in the rest of the phase diagram. **(e)** Entanglement between two rings of the cylinder. There is symmetry-breaking between the two rings of the unit cell; here we show the smaller of the two entanglements—between the two rings in the unit cell and between unit cells. **(f)** Log base 10 of the truncation error at the last step of the DMRG simulation.

### B. Transfer matrix and the two-dimensional excitation spectrum

We again compute the transfer matrix spectrum for spin 1 excitations versus flux $\theta$, $k_x$, and $k_y$. In particular, we show in Figure S35 the spectrum for $U/t = 10$ at bond dimensions 2000, 4000, 8000, and 11314.

As with YC4 and YC6, above, we investigate how the transfer matrix spectrum does or does not correspond to a possible Dirac spin liquid (DSL). In this case, our particular goal is to check the possibility that the non-chiral state found near 0 flux could be a DSL, which becomes gapped out near $2\pi$ flux, thus becoming a chiral spin liquid instead. We see no evidence for this conclusion.

There are minima in the excitation spectrum at the values of $\theta$ where the transition between chiral and non-chiral states occurs, and these points will plausibly become gapless in the infinite bond dimension limit; however, the momenta corresponding to these gapless points are completely different from the transitions between expected locations of Dirac cones for the DSL, indicated by the vertical lines in Figure S35 (also at the edges of the plotted region in $k_x$).

Instead of a transition between DSL and CSL, the most plausible explanation is that there is simply a gap closing at a second-order transition between the chiral spin liquid and a bond-ordered state (see main text).

### C. Appearance of chirality with hopping anisotropy

In the main text we argue that the nonchiral state at intermediate $U$ near zero flux may be prevented from becoming the chiral spin liquid due to a large degree of anisotropy in the spin interactions. To test this, we explicitly introduce a hopping anisotropy in the model, weakening the bonds around the cylinder by 10%, thus producing more isotropic

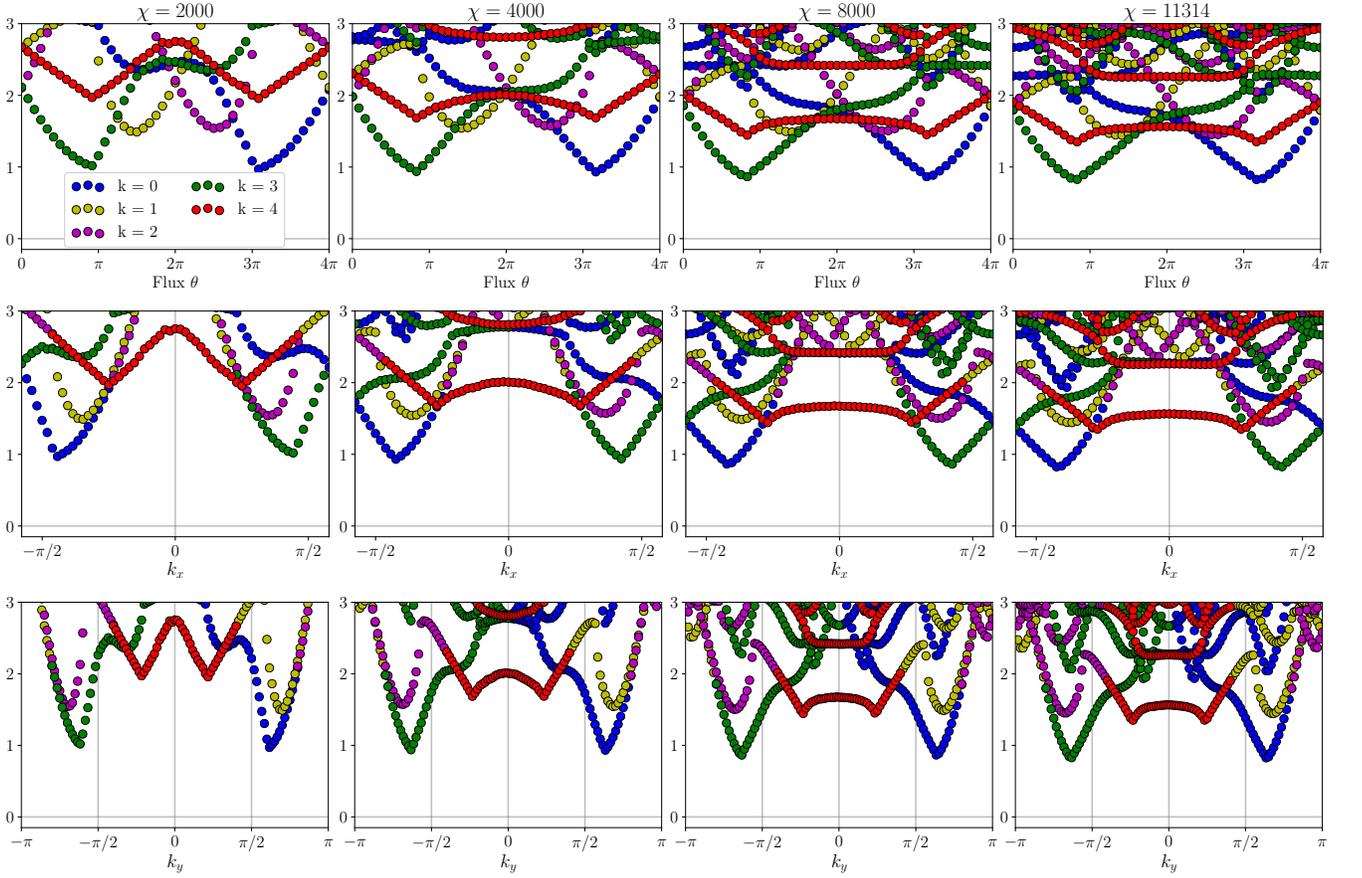

FIG. S35. (Color online) Low-lying part of the $S_z = 1$ part of the transfer matrix spectrum for $U/t = 10$ on the YC5 cylinder at $\chi = 2000$ (left column), 4000 (second column), 8000 (third column), and 11314 (right column), plotted versus flux inserted (top row), $k_x$ (middle row), and $k_y$ (bottom row). (The lattice constant is set to 1.) Color corresponds to the momentum quantum number of the excitation as given in the panel in the upper left. Vertical lines correspond to possible Dirac cone locations in the Dirac spin liquid; they do not clearly correspond to features observed in our data.

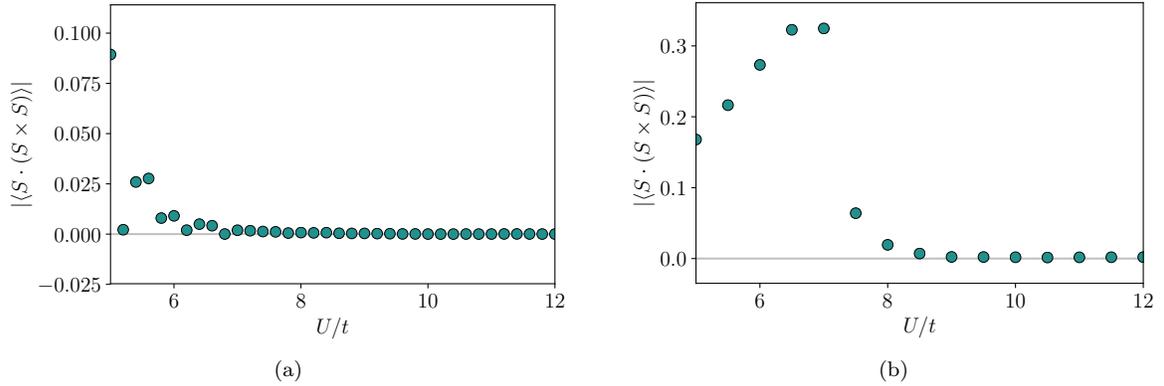

FIG. S36. (Color online) Chiral order parameter versus $U/t$ at zero flux for $\chi = 4000$. **(a)** With isotropic hopping in the Hubbard model, there scalar chiral order parameter is zero. It appears nonzero at low $U$, but this is not systematic and is likely a finite bond dimension effect. **(b)** When hopping on bonds around the cylinder is weakened by 10%, the chirality becomes large, of the same order as that observed for YC4.

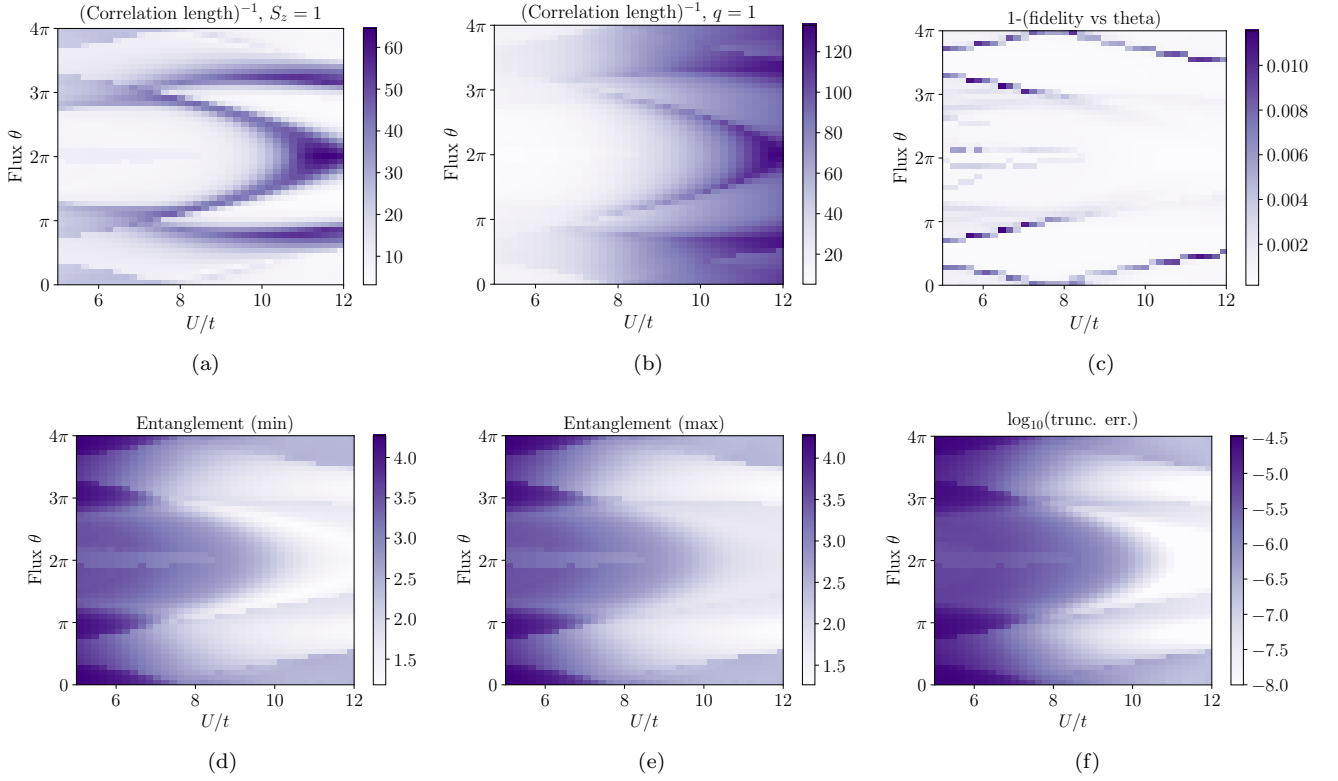

FIG. S37. (Color online) Various quantities for the YC3 cylinder vs $U/t$ and flux insertion $\theta$. **(a)** Inverse correlation length for operators with $S_z = 1$, giving an estimate of the spin triplet gap. **(b)** Inverse correlation length for operators with $q = 1$, giving an estimate of the charge gap. **(c)** Fidelity of the flux insertion, measued by $1 - |\langle\psi(\theta)|\psi(\theta + \pi/12)\rangle|^2$. **(d)** Entanglement between two rings of the cylinder. The unit cell is four rings, and here we show the smallest entanglement across any of the four distinct cuts. **(e)** Here we show just the largest of the four entanglements; it is nearly identical to the smallest. **(f)** Log base 10 of the truncation error at the last step of the DMRG simulation.

spin interactions that are therefore more consistent with the symmetries of the two-dimensional model. This indeed gives rise to nonzero scalar chirality at zero flux, as shown in Figure S36.

## VIII.   YC3 ADDITIONAL DATA AND ANALYSIS

### A.   Flux insertion data

In Figure 11 of the main text, we show the chiral order parameter, peak height of the spin structure factor, and inverse correlation length for excitations with $S_z = 0$, giving an estimate of the spin singlet gap. Here we show in Figure S37 a number of additional quantities versus $U/t$ and inserted flux, $\theta$, with bond dimension $\chi = 4000$. The relative behaviors of the various quantities do not match what we observed for the larger cylinders, so it is unclear whether the region with nonzero chirality corresponds to a chiral spin liquid.

## IX.   XC4 ADDITIONAL DATA AND ANALYSIS

### A.   Flux insertion data

In Figure 12 of the main text, we show the chiral order parameter, peak height of the spin structure factor, and inverse correlation length for excitations with $S_z = 0$, giving an estimate of the spin singlet gap. Here we show in Figure S38 a number of additional quantities versus $U/t$ and inserted flux, $\theta$, with bond dimension $\chi = 4000$. Again the relative behaviors of the various quantities do not match what we observed for the larger cylinders, and also the

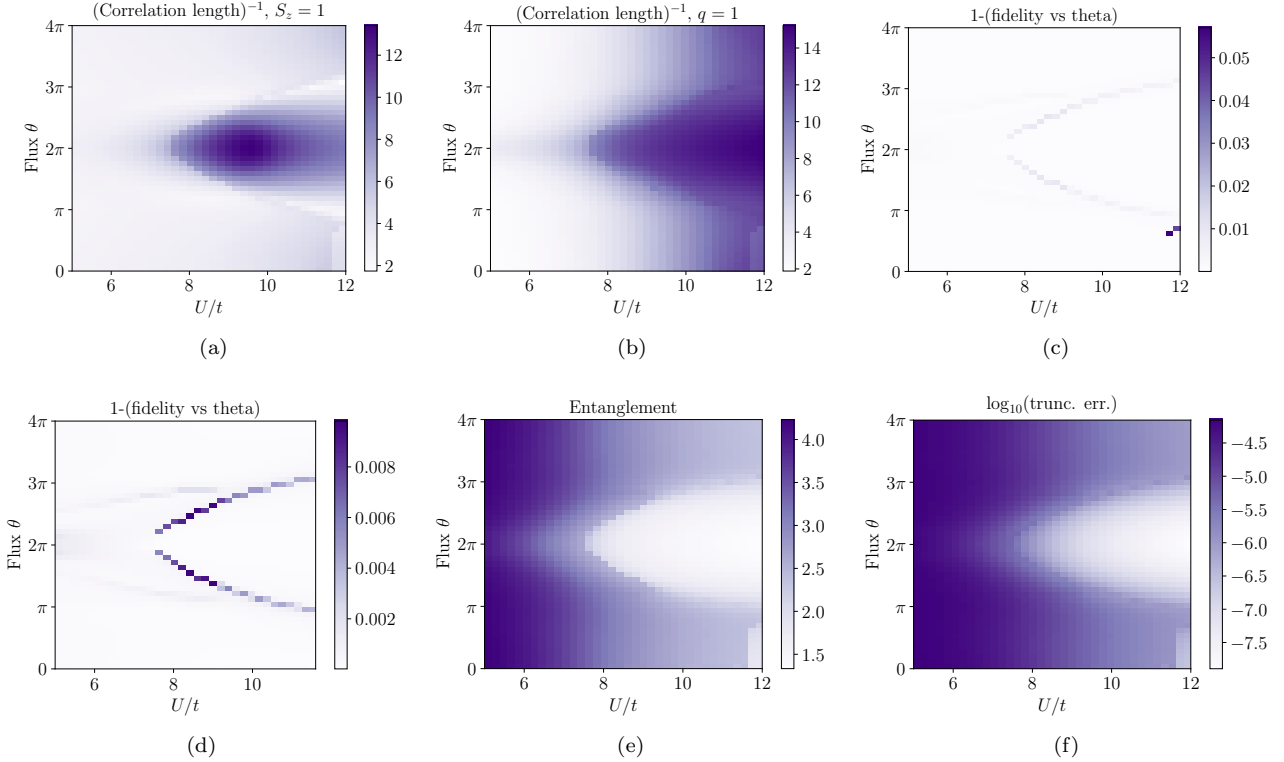

FIG. S38. (Color online) Various quantities for the XC4 cylinder vs $U/t$ and flux insertion $\theta$. **(a)** Inverse correlation length for operators with $S_z = 1$, giving an estimate of the spin triplet gap. **(b)** Inverse correlation length for operators with $q = 1$, giving an estimate of the charge gap. **(c)** Fidelity of the flux insertion, measued by $1 - |\langle \psi(\theta)|\psi(\theta + \pi/12)\rangle|^2$. **(d)** Fidelity with the rightmost columns removed to more clearly show the results in the rest of the phase diagram. **(e)** Entanglement between two rings of the cylinder. **(f)** Log base 10 of the truncation error at the last step of the DMRG simulation.

chirality at zero flux likely goes to zero with increasing bond dimension (as shown in Figure 12(a) of the main text), so it is unclear whether the region with nonzero chirality corresponds to a chiral spin liquid.

Note that as with YC6 and YC5, there is some nonadiabaticity in the flux insertion, in this case at $U/t = 11.8$ and 12. The state at near zero flux is metastable, about 1% higher in energy than the state near $4\pi$ flux.



\end{document}